\documentclass[]{aa}  

\usepackage{graphicx}
\usepackage{txfonts}
\usepackage{lipsum}
\usepackage{float}
\usepackage{subcaption}         
\usepackage{lscape}             
\usepackage{placeins}          
\usepackage[colorlinks=true,linkcolor=blue,citecolor=blue]{hyperref}
\usepackage{lipsum,multicol}
\usepackage{graphicx}
\usepackage{array}      
\usepackage{multirow}    
\usepackage{booktabs} 
\usepackage[utf8]{inputenc}
\usepackage{rotating}
\usepackage{xcolor}
\usepackage{longtable}
\usepackage{makecell}

\bibpunct{(}{)}{;}{a}{}{,} 
\usepackage{txfonts}
\usepackage{caption}
\usepackage{soul}
\usepackage{pdflscape}
\usepackage{capt-of}

\begin{document}

\title{Validation of Ariel candidates orbiting K-M stars}

%
%
%

   \title{Validation of Ariel candidates orbiting K-M stars}

   \author{Giulia Piccinini\inst{1,2}\corrauth{\texttt{giulia.piccinini@inaf.it}} 
   \and
           Claudia Di Maio\inst{1}\,
           \and
           Giuseppina Micela\inst{1}\,
           \and
           Jonathan Brande\inst{3}\, 
           \and  \\
           David R. Ciardi\inst{4}\, 
           \and 
           Catherine A.\ Clark\inst{4}\,
           \and
           Ian J.\ M.\ Crossfield\inst{5} \and
           Julian C. van Eyken\inst{4}\,
           \and \\
           Miranda Felsmann\inst{4}\,
           \and
           Emily A.\ Gilbert\inst{4}\,
           \and 
           Michael B. Lund\inst{4}\, 
           \and
           Rachel A.\ Matson\inst{6}\, 
           \and \\
           Elisabeth C. Matthews\inst{7}\,
           \and 
           Alex S. Polanski\inst{8}\,\thanks{Percival Lowell Fellow}\,
           \and
           Joshua E.\ Schlieder\inst{9}\,
          }

    \institute{INAF -- Astronomical Observatory of Palermo, Piazza del Parlamento 1, 90134, Palermo, Italy \and
    Department of Physics and Chemistry, University of Palermo, Via Archirafi 36, 90123, Palermo, Italy \and
    Department of Astronomy, University of Maryland, PSC 1113, 4296 Stadium Dr., College Park, MD 20742, USA \and
    NASA Exoplanet Science Institute, Caltech/IPAC, Pasadena, CA 91125, USA \and
    Department of Physics and Astronomy, University of Kansas, Lawrence, KS 66045, USA \and
    U.S.~Naval Observatory, 3450 Massachusetts Avenue NW, Washington, DC 20392, USA \and
    Max-Planck-Institut für Astronomie, Königstuhl 17, 69117 Heidelberg, Germany \and
    Lowell Observatory, 1400 W Mars Hill Road, Flagstaff, AZ 86001, USA \and
    NASA Goddard Space Flight Center, Exoplanets and Stellar Astrophysics Laboratory, 8800 Greenbelt Road, Greenbelt, MD 20771, USA}

   \date{Received 15 May 2026, Accepted 14 November 2026}

  \abstract
  {One of the key goals of planetary population studies is to achieve broad coverage of the parameter space, which is not yet fully represented by confirmed planets. This limitation can be mitigated by including planets that have been statistically validated through a vetting procedure, even if they are not fully confirmed. In preparation for the Ariel mission, a Mission Candidate Sample has been assembled, comprising both confirmed planets and several unvalidated candidates. In this context, we apply a vetting analysis to a selected subsample of these candidates. Our methodology is based on the \texttt{TRICERATOPS} tool, aimed at confirming the presence of transiting planets or ruling out false-positive scenarios. We applied \texttt{TRICERATOPS} to a sample of 318 targets drawn from the last three Ariel TESS Planet Candidates catalogues. The analysis includes evaluation of statistical probabilities, as well as checks on stellar binarity indicators such as the Renormalized Unit Weight Error (RUWE) and the \texttt{IPD\_frac\_multi\_peak} values from {\it Gaia} Data Release 3. Our analysis validated 56 candidates as planets, identified 29 false positives, and classified 35 as likely planets. These results are intended to support the Ariel mission by providing a probabilistic priority ranking to guide follow-up observations and optimise targets for atmospheric characterisation studies.
    }

\maketitle
\nolinenumbers

\section{Introduction}
\label{sec:introduction}

The Atmospheric Remote-Sensing Infrared Exoplanet Large-survey \citep[Ariel;][]{ariel1,ariel} Mission has as its primary objective the realisation of the first large-scale, homogeneous spectroscopic survey of exoplanet atmospheres. By studying approximately 1000 exoplanets through transit, secondary eclipse spectroscopy or phase curves, Ariel aims to address three fundamental scientific questions: the physical processes shaping planetary atmospheres, the chemical diversity of exoplanets, and the formation and evolution mechanisms of planetary systems. To support these goals, the mission employs a 1-meter class telescope providing simultaneous spectral coverage from 0.5 to 7.8$\,\mu m$. The onboard payload comprises three scientific instruments: the Fine Guidance System, with three photometric channels operating between 0.5 and 1.1$\,\mu m$; the Near Infrared Spectrometer, operating between 1.1 and 1.95 $\,\mu m\,$ with a spectral resolution of R$\,\sim\,$10; and the Ariel InfraRed Spectrometer (AIRS), split into the AIRS-CH0 (1.95–3.9$\,\mu m\,$ at R\,=\,100–200) and AIRS-CH1 (3.9–7.8$\,\mu m\,$ at R\,=\,30–60). This broad wavelength range captures most of the thermal emission from warm and hot planets, enabling accurate estimates of atmospheric heat redistribution, Bond albedo, molecular abundances, and vertical thermal structures \citep{ariel1,Charnay22}.
To optimise observing time and maximise scientific return, Ariel adopts a tiered observing strategy. Tier 1 (Reconnaissance Survey) will provide an initial, low-resolution characterisation of approximately 1000 planets, targeting the presence of primary atmospheres, clouds, and hazes. A subset of around 500 targets will then be studied in Tier 2 (Deep Survey) at higher resolution (R$\,\sim\,$100 for AIRS-CH0, R$\,\sim\,$30 for AIRS-CH1) to constrain chemical composition, cloud properties, and temperature profiles. Tier 3 (Benchmark Survey) will be dedicated to the most favourable targets (roughly 50 objects) observed repeatedly at the native instrument resolution to monitor temporal and spatial variability in chemistry and cloud coverage. Finally, Tier 4 will accommodate complementary science outside this framework, including phase curve observations and in-depth studies of smaller planets that may host secondary atmospheres \citep{targetlist}. The mission will systematically explore a wide parameter space, analysing objects ranging from rocky planets to inflated Jupiters ($1.8\,R_{\oplus} \lesssim R_p \lesssim 26\,R_{\oplus}$), primarily at equilibrium temperatures between 500 and 2500\,K\, orbiting host stars from M-dwarfs to A-type stars \citep{ariel1}.
The selection of the Mission Reference Sample is designed to maximise both observational efficiency and sample diversity. For each candidate, the number of transits or eclipses required to achieve signal-to-noise\,$>$\,7 at the native spectral resolution of the Ariel instruments is estimated assuming an observing window of 2.5 times the transit or eclipse duration. Targets requiring up to five observations to meet Tier~1 goals, or up to 20 for Tier~2, are considered suitable for inclusion, even though the boundary is not strictly defined and some targets may be assessed on a case-by-case basis. Although over 2000 candidates already satisfy the Tier~1 requirement, the Mission Reference Sample will comprise approximately 1000 targets, exploiting this surplus to ensure scheduling flexibility and avoid observational redundancy. Target classification is currently based on stellar effective temperature, planetary radius, and planetary equilibrium temperature; future selections could additionally incorporate planetary mean density and host star metallicity \citep{Edwards2019}, five properties with direct and well-established impact on atmospheric chemistry and evolution. Among the scientific objectives of the mission, the study of planets smaller than $3.5\,R_{\oplus}$, which occupy the transition zone between rocky and gaseous worlds, will be of particular interest. Understanding whether their atmospheres are primordial or more evolved is key to constraining formation models and clarifying evolutionary processes such as hydrogen escape or interaction with the interior composition. Since the catalogue of potential targets continues to grow with new discoveries, Ariel's final target list will remain dynamic until launch, currently scheduled for 2031, ensuring the most diverse and statistically significant sample possible \citep{targetlist}.

An initial sample of potential targets has already been identified \citep{2018Zingales,Edwards2019,arielrad,targetlist}. This sample, referred to as the Mission Candidate Sample (MCS), is divided into two catalogues: the {\it Known} catalogue, consisting of confirmed planets that meet Ariel's requirements for atmospheric characterisation within a reasonable number of transits or eclipses, and the TESS Planet Candidates (TPCs) catalogue, which includes planet candidates detected by TESS that still require further characterisation. In this context, the planetary mass represents a key parameter for accurately constraining atmospheric properties \citep{Batalha2019,2020Changeat,DiMaio2023}. It is therefore essential that planet candidates are properly vetted, validated, and subsequently confirmed through follow-up observations. Several radial velocity campaigns are dedicated to this purpose, i.e. the follow-up of TESS candidates \citep{Chontos2022}, relying on HIRES/Keck \citep{hires} for precise mass measurements and more recently the Ariel Mass Survey \citep{dimaio2026}, carried out with the HARPS-N spectrograph \citep{HARPSN} at the Telescopio Nazionale Galileo, which specifically focuses on determining the masses of planets in the transition regime between super-Earths and sub-Neptunes.

Before conducting follow-up campaigns, which often demand a significant number of observations for mass characterisation, it is essential to perform a rigorous validation procedure to identify and exclude false positive candidates. False positives correspond to detections in which the observed transit is caused by another astronomical phenomenon or by instrumental effects. In the last years, vetting techniques have attracted increasing attention due to their crucial role in selecting the most promising candidates for follow-up observations, with several recent studies focusing on the application of these methods to large samples of targets; i.e. \citet{Hord2024}, \citet{kostov}, and, more recently, \citet{RAVEN}, who applied the \texttt{RAVEN} pipeline, a machine learning-based classification tool, to TESS data up to Sector 55. This effort resulted in the validation of roughly one hundred candidates and the identification of new objects of interest that had not been previously detected. 
This work evaluates the three most recent Ariel TPCs catalogues (released on 09-07-2024, 18-07-2025, and 18-08-2025) by implementing a rigorous vetting analysis and statistical validation. The primary objective is to identify and prioritise high-quality candidates for subsequent confirmation and atmospheric characterisation, ensuring alignment with Ariel’s mission criteria. Specifically, we performed a vetting analysis on a selected subset of the TPCs sample to assess whether to validate or reject them. Our goal is therefore not to definitively determine the nature of the candidates, but rather to define a priority list identifying the targets that should be observed and analysed first. In this way, once these validated planets have been observed, their mass measured and they have been confirmed they will help expand the parameter space currently covered by the {\it Known} catalogue, in particular with respect to equilibrium temperature, planetary radius, and orbital period. In this work, we prioritises targets where detecting low-mass planets is more favourable, which is of special interest for Ariel's investigation of small planets in the transition regime between rocky and gaseous bodies. To achieve this, we adopt a stellar effective temperature cutoff at $T_{\rm eff} = 5000\,\text{K}$, explicitly separating low-mass M- and K-type dwarfs from more massive hosts. This selection does not exclude the possibility to find massive planets or brown dwarfs.

The article is organised as follows: in Section~\ref{sec:analysis}, we describe the methodology employed in our analysis and we present the criteria and procedure for target selection; in Section~\ref{sec:Results}, we report the results of our statistical validation; in Section~\ref{sec:discussion}, we discuss our findings, and in Section~\ref{sec:summary}, we summarise our conclusions.

\section{Analysis}
\label{sec:analysis}

\subsection{Method}
The vetting procedure consists of a detailed analysis of photometric data aimed at identifying and examining periodic, transit-like signals from target stars, with the goal of classifying them as either genuine planet candidates or false positives. This approach enables the statistical rejection of astrophysical false positive scenarios with a high degree of confidence.
Our analysis is based on the \texttt{TRICERATOPS} tool, whose theoretical framework is described in detail in \cite{triceratopscode} and in \cite{2021giacalonevetting}. Among the various tools available for the statistical validation, including \texttt{BLENDER} \citep{blender} and \texttt{VESPA} \citep{vespa}, \texttt{TRICERATOPS} is specifically designed to vet TESS planet candidates and is continuously maintained and updated \citep{2023Morton}. It evaluates the probabilities of multiple transit-producing scenarios and provides estimates of both the False Positive Probability (FPP) and the Nearby False Positive Probability (NFPP). 
The tool requires as input the orbital period, time of conjunction, transit depth and duration, and  star mass, radius, and temperature, which were obtained from the Exoplanet Follow-up Observing Program (ExoFOP\footnote{\url{https://exofop.ipac.caltech.edu/tess/}}), along with the TESS light curves.  In this work, we only considered the 2-minute cadence data light curves, corrected from any instrumental systematics using the Presearch Data Conditioning Simple Aperture Photometry flux \citep{2012Stumpe,Stumpe2014,Smith2012}. When available, we incorporated contrast curves from high-resolution imaging technique. As described in \citet{2021giacalonevetting}, these data allow us to determine the angular separation beyond which a simulated secondary or fore- or background star would have been detectable. As input to the tool, we provided information on the instruments and filters used, as well as the corresponding publicly available imaging data from ExoFOP, all of which had been uploaded more than one year prior to the time of writing (February 2026). We employed contrast curves from several facilities, with additional details on the observing sites provided in Appendix \ref{app:app_imaging}.
If no high-resolution imaging data are available, an angular separation of 2.2'' is assumed \citep{GDR2,2021giacalonevetting}. From the simulations of the scenarios, two key probabilities are calculated for each candidate, providing a statistical classification of the transit signal.

 \begin{figure}
    \centering
    \includegraphics[width=\hsize]{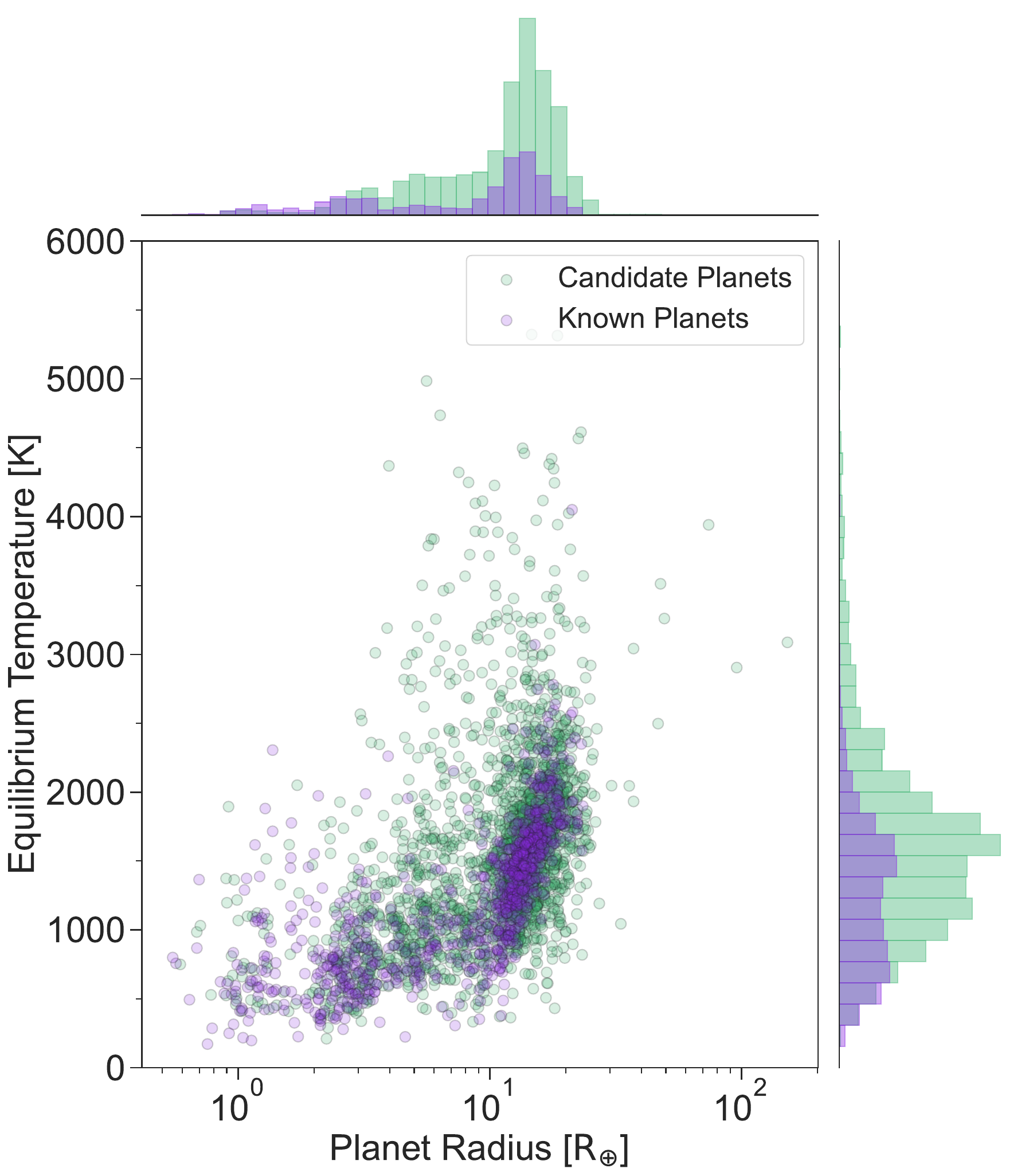}
    \caption{Comparison between TESS planet candidates catalogue (09-07-2024, 18-07-2025, 18-08-2025), in green, and {\it Known} catalogue (18-08-2025), in violet.}
    \label{fig:KnownTPC_radiustemp}
\end{figure}

 \begin{figure}
    \centering
    \includegraphics[width=\hsize]{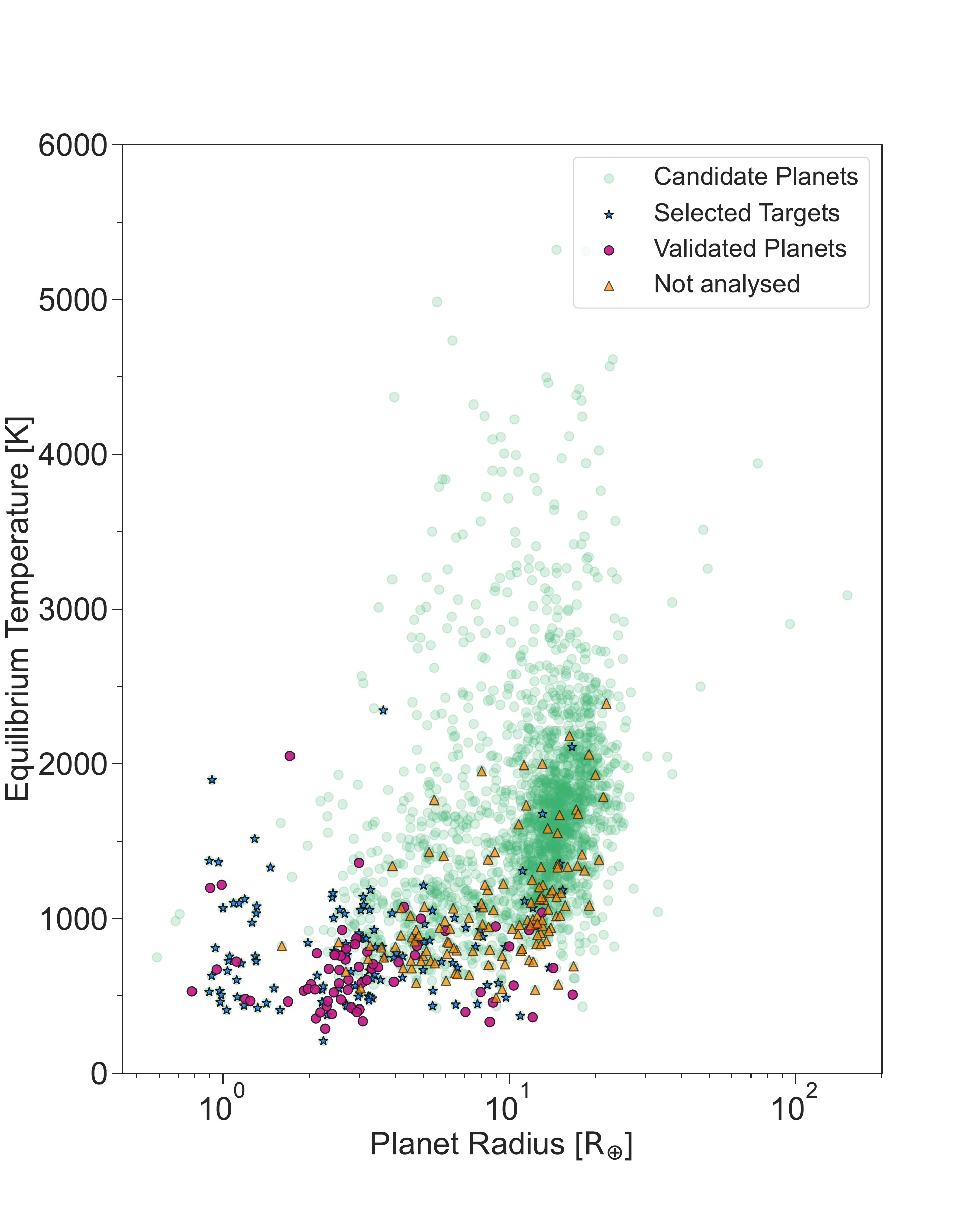}
    \caption{Comparison between candidate planets and analysed sample. All elements from the last three TPC catalogues (green circles), the selected sample analysed in this work (blue stars), validated candidates including both newly validated targets and those previously validated in the literature (magenta circles), and targets that could not be analysed (yellow triangles).}
    \label{fig:radiustemp}
\end{figure}

The first is the FPP, defined as 

\begin{equation}
    \mathrm{FPP} = 1 - (\mathcal{P}_\mathrm{TP} + \mathcal{P}_\mathrm{PTP} + \mathcal{P}_\mathrm{DTP}),
\end{equation}
which represents the probability that the observed transit signal is caused by a phenomenon other than a planet transiting the target star. The terms in parentheses represent three scenarios in which a planet transits at the reported orbital period: a transit around the target star with no unresolved stellar companion (TP), a transit around the target star with an unresolved bound companion (primary TP, PTP), and a transit around the target star in the presence of an unresolved fore- or background star along the line of sight (diluted TP, DTP).

The second probability is the NFPP, calculated as

\begin{equation}
    \mathrm{NFPP} = \sum(\mathcal{P}_\mathrm{NTP} + \mathcal{P}_\mathrm{NEB} + \mathcal{P}_\mathrm{NEBx2P}),
\end{equation}
and quantifies the probability that the observed transit signal originates from a resolved nearby star rather than the target star.

This includes scenarios in which the transit is caused by a planet with the reported orbital period around that nearby star (nearby TP, NTP), an eclipsing binary (EB) with the reported orbital period around that nearby star (nearby EB, NEB), or an eclipsing binary with twice the reported orbital period (NEBx2P).

For the classification of the candidates, we adopted the following thresholds, as recommended by \citet{2021giacalonevetting}:
\begin{itemize}
    \item[-] Validated planets: FPP $<$ 1.5\% and NFPP $<$ 0.1\%
    \item[-] Likely planets: FPP $<$ 50\% and NFPP $<$ 0.1\%
    \item[-] False Positives: NFPP $>$ 10\%
\end{itemize}

Additionally, we included candidates that fall outside these defined thresholds but for which the transiting planet scenario remains the most likely. For these cases, we introduced a separate classification: Other. Notably, \texttt{TRICERATOPS} does not account for instrumental systematics or stellar variability and therefore cannot evaluate false alarm scenarios. Moreover, if the transit signal is affected by stellar variability, this may introduce a bias in the probability estimates, potentially leading to misclassification: for instance, a candidate that would otherwise be validated may instead be classified as a likely planet. This risk is mitigated when multiple TESS sectors are available, as the increased photometric baseline reduces the relative contribution of stellar variability to the overall signal.

As the vetting procedure is statistical in nature, we performed 10 runs for each target and report the mean values and standard deviations of the FPP and NFPP results. We opted for 10 runs rather than 20 recommended by \cite{2021giacalonevetting}, primarily for computational efficiency given the large size of our sample. Moreover, for a significant fraction of targets, additional contrast curves were available, which effectively increases the constraining power of each individual run. To verify that this choice did not compromise the robustness of our results, we assessed the statistical convergence of the outputs, finding that after 10 runs the standard deviations of both the FPP and the NFPP are already very small. The stability of the mean values indicates that doubling the number of iterations to 20 would not alter the final classification of any candidate.

\subsection{Target selection}
\label{TargetSelection}

The sample selection was based on the last three TPCs catalogues (09-07-2024, 18-07-2025, 18-08-2025), which are publicly available\footnote{\url{https://github.com/arielmission-space}}. We compared these versions because, although some objects were removed from the latest TPC as they no longer satisfied the Ariel selection criteria, they may still remain valuable for other studies.
Figure~\ref{fig:KnownTPC_radiustemp} shows the distribution of the Ariel targets in the {\it Known} catalogue (18-08-2025) compared to the three TPC catalogues. The figure highlights the regions of parameter space that are currently under-represented in the {\it Known} catalogue and that could be populated through the vetting, validation, and confirmation of TPC targets. In particular, a significant expansion is expected at equilibrium temperatures above 1000\,K and planetary radii larger than $10\,\mathrm{R_{\oplus}}$. We focused on candidates hosted by stars cooler than 5000 K to specifically prioritise low-mass stars that are more probable to host small planets. The initial sample comprises 320 candidates, of which 272 are hosted by early-M to mid-K stars (3500\,K $<T_{\rm eff}<$\,5000\,K) and 48 by M-type stars ($T_{\rm eff} <$ 3500\,K).
The effective temperature thresholds were chosen according to key structural transitions. As reported in \citet{Chabrier1997} and in \citet{PecautMamajek2013}, a temperature of 5000\,K, corresponding to a stellar mass of approximately 0.9\,$M_{\rm \odot}$, separates solar-type stars from lower-mass dwarfs, while 3500\,K (roughly 0.40\,$M_{\rm \odot}$) identifies the transition between partially and fully convective stars.

We considered candidates with planetary radii between 0.5 and 22\,$\mathrm{R_{\oplus}}$, including inflated Jupiter-sized objects. We did not adopt the radius cut of \citet{2021giacalonevetting}, which excludes planets larger than 8\,$R_{\rm \oplus}$, in order to retain giant planets and potential brown dwarfs (BD). \citet{2021giacalonevetting} advise against applying the vetting procedure to targets with ${R_{\rm p}} >$\,8\,$R_{\rm \oplus}$, as giant planets, BDs, and low-mass stars cannot be reliably distinguished based on radius alone. For these reasons, we performed additional checks, rather than relying solely on the statistical vetting procedure, as better described in Section \ref{sec:Results}. Transiting BDs are rare, with no more than 60 reported to date \citep[e.g.][]{Vowell25BD,browndwarf}, and their confirmation and atmospheric characterisation would provide valuable information for Ariel complementary science. Consequently, we excluded 2 candidates from the early-M to mid-K sample with radius $>$\,22\,$R_{\rm \oplus}$, resulting in a final total selection of 318 targets. Figure \ref{fig:radiustemp} shows all the candidates reported in TPCs catalogue 18-08-2025, together with the final sample analysed in this work. Among them, several targets had already been vetted in previous studies, particularly by \citet{vatestI,vatestII,vatestIII}, and \citet{Hord2024}, using different methodology or thresholds. \citet{vatestI,vatestII,vatestIII} describe the Validation of Transiting Exoplanets using Statistical Tools project, in which the \texttt{LATTE} tool \citep{latte} is initially employed as a diagnostic test to provide a preliminary indication of whether the detected signal originates from the target star or from nearby sources. This step is followed by candidate validation with \texttt{TRICERATOPS}, with ground-based follow-up photometry incorporated into the vetting process. However, since \texttt{LATTE} provides only approximate information and does not offer a high level of confidence regarding the true origin of the signal, as also noted in \cite{vatestII}, we opted not to use it and instead assessed the origin of the false positive results a posteriori. \citet{Hord2024} employ both \texttt{VESPA} and \texttt{TRICERATOPS} for vetting analysis. According to the authors, for a candidate to be considered validated, it must satisfy the thresholds of both tools: FPP $<$ 10\% for \texttt{VESPA}, and FPP $<$ 1.5\% and NFPP $<$ 0.1\% for \texttt{TRICERATOPS}. These objects were reanalysed to assess the robustness of our results.

\subsection{Terminology adopted}
\label{Terminology}

The terminology used in the literature to describe the different stages of planet candidate characterisation is not fully standardised, with terms such as false positive, planet candidate, vetted candidate, validated planet, and confirmed planet often used inconsistently across studies. To avoid ambiguity, we here define the nomenclature adopted throughout this work, drawing from a survey of the existing literature and from the conventions established by the statistical tools employed.

The classification of a transit signal depends strongly on the tool and methodology used. For instance, \texttt{TRICERATOPS} distinguishes three categories — false positives, likely planets, and validated planets — based on the values of the FPP and NFPP computed from the relative probabilities of a set of astrophysical scenarios. A different scheme is adopted by \texttt{LEO-Vetter} \citep{LEOvetter}, a tool designed to vet TESS Threshold Crossing Events (TCEs), which identifies three categories: false alarms, false positives, and planet candidates. In that framework, false alarms encompass stellar variability, instrumental noise, and scattered light; false positives indicate transit-like signals not attributable to a planet orbiting the target star, such as eclipsing binaries; and planet candidates are TCEs that pass a set of 13 flux-level tests against false alarms and five against false positives. Under this definition, a planet candidate and a vetted candidate are effectively synonymous, and in the context of \texttt{TRICERATOPS}, a likely planet falls within the same category. Similarly, \citet{RAVEN} distinguish between validated planets, which satisfy all their adopted criteria (widely described in their paper), and vetted candidates, which do not meet the full set of requirements.

Drawing from these conventions, we adopt the following definitions for this work:

\begin{itemize}
    \item False alarm: a TCE caused by stellar variability, instrumental systematics, or other non-transiting phenomena that can mimic a transit-like signal;
    
    \item False positive: a transit-like signal that is not produced by a planet transiting the target star, including eclipsing binaries, hierarchical systems, and transits occurring around unresolved background or nearby stars;
    
    \item Likely planet/vetted candidate/planet candidate: these terms are used interchangeably throughout this paper to refer to a TCE for which the considered false alarm and false positive scenarios can be ruled out with high statistical confidence. A vetted candidate may subsequently undergo formal statistical validation;
    
    \item Validated planet: a planet candidate for which a statistical validation analysis has been performed and whose planetary scenario probability satisfies the criteria of the adopted method (e.g., \texttt{TRICERATOPS} or \texttt{RAVEN}). Statistical validation relies primarily on photometric and imaging data and does not provide a dynamical mass measurement;
    
    \item Confirmed/Known planet: a planet for which a mass has been measured through an appropriate follow-up technique, such as radial velocity. A confirmed planet may or may not have previously undergone a formal vetting or statistical validation procedure.
\end{itemize}

For completeness, we include the false alarm category in our definitions, even though it represents a prior step with respect to the vetting analysis discussed in this work. For consistency with the \texttt{TRICERATOPS} terminology, we refer to vetted candidates as likely planets throughout the analysis, while stressing that the two terms are treated as fully synonymous.

\section{Results}
\label{sec:Results}
\subsection{Statistical validation}

Our analysis resulted in the validation of 56 planets (VP). We also identified 29 false positives (FP), 35 likely planets (LP), and 47 objects classified as Other, with the TP scenario being the most probable (OT). We also classified a small group of objects as eclipsing binaries (EB) candidates, as well as targets with ambiguous results. The six EB candidates are the ones with FPP and NFPP values that do not fall within the FP thresholds, but for which an EB scenario is still the most probable. We identified seven targets with ambiguous results (AR), as their classifications varied depending on the contrast curves used.
A total of 153 objects could not be analysed. Specifically, for 133 targets the 2-minute cadence TESS light curves were unavailable, the transit signal fell within observation gaps, or the light curve was too noisy to identify a reliable transit. The remaining 20 candidates were missing at least one key stellar parameter (mass, radius, or parallax). For these objects, we investigated whether the missing parameters could be independently derived. Specifically, for those targets with an available parallax measurement in \textit{Gaia} Data Release 3 \citep[DR3;][]{Gaia2023}, since stellar mass and radius are not reported in the TESS Objects of Interest Catalog \citep{Guerrero_2021}, we estimated the stellar radius and mass by interpolating the empirical tables of \citet{PecautMamajek2013}. This approach allowed us to recover 15 additional targets, reducing the number of Not Classified (NC) targets to 138. The analysis of these targets is described in Section~\ref{sec:derived_param}. Figure~\ref{fig:radiustemp} shows the distribution of the TPC candidates in our selected sample, together with the validated planets and the targets that could not be classified. The validated sample is found to preferentially populate the small-planet regime below $4\,\mathrm{R_{\oplus}}$.

A summary of the results of the entire sample is reported in Table \ref{tab:summary}. We represented in Figures~\ref{fig:Sample_KM} and \ref{fig:Sample_M} the analysed sample compared to the TPCs catalogues and in Figures~\ref{fig:KM_KP} and \ref{fig:M_KP} the analysed sample compared to the {\it Known} catalogue dated 18-08-2025.
In Tables \ref{tab:validated_new} and \ref{tab:prop_vetted} we report the planetary radius, orbital period, stellar mass, and magnitude, along with the Transmission Spectroscopy Metrics \citep[TSM;][]{Kempton_2018} retrieved from the ExoFOP database. The TSM, which scales with the planet’s radius and mass, equilibrium temperature, and stellar radius, quantifies the suitability of a target for atmospheric characterisation with JWST \citep{JWST} and the future Ariel mission. Planetary masses were estimated using the \cite{Chen2017} forecasting relation; however, as noted by \cite{2024polanskitess}, this approach underestimate masses by approximately 24\% compared to the masses obtained via radial velocity measurements, potentially resulting in overestimated TSM values. This systematic offset is less pronounced for planets smaller than 4$\,\mathrm{R_{\oplus}}$, where the underestimation reduces to approximately 10\%, suggesting that the impact on the derived TSM values is likely modest for this subset of targets. For planets larger than $4\,\mathrm{R_{\oplus}}$, the underestimation is more significant, potentially leading to a larger overestimation of the TSM. However, since more massive planets generally exhibit larger atmospheric scale heights, their atmospheric signals remain detectable even when the planetary mass is underestimated. Given that the majority of our analysed sample have radii smaller than $4\,\mathrm{R_{\oplus}}$, this systematic effect is not expected to significantly alter the overall priority ranking of our sample.
As additional check on the results, we provided in Tables \ref{tab:validated_new} and \ref{tab:prop_vetted} the Renormalized Unit Weight Error \citep[RUWE;][]{Lindegren2021} and the \texttt{IPD\_frac\_multi\_peak} \citep[$\mathrm{IPD_{FMP}}$;][]{halbwachs2023} values. These parameters, available in the {\it Gaia} catalogues \citep{Gaia2016,belokurov_binarity,gaia_IPD}, offer insight into the presence of potential unresolved binary companions. While \cite{StassunTorres} showed that even mild RUWE values in the range 1.0–1.4 are sensitive to photocentre motion and can indicate unresolved companions, values of $\text{RUWE} > 1.4$ are widely adopted as the standard threshold to identify severe astrometric noise due to multiplicity. The $\mathrm{IPD_{FMP}}$ parameter ranges from 0 to 100 and helps identify close companions, with values $\leq$ 2 typically indicating single stars. Since multiple peaks can arise from either background stars or physical companions, we considered both parameters, adopting the thresholds RUWE $>$ 1.4 and $\mathrm{IPD_{FMP}}$ $>$ 2. Moreover, we consulted the GDR3 Non-Single Stars (NSS) catalogue \citep{Holl2023,halbwachs2023} to assess each case individually. Candidates listed in the \textit{Gaia} DR3 NSS catalogue, or for which both RUWE and $\mathrm{IPD_{FMP}}$ exceed the adopted thresholds, are not considered as fully validated. Candidates for which only one of the two indicators exceeds the threshold are validated, but flagged with a cautionary note.
Finally, we note that while \texttt{TRICERATOPS} does not impose a lower limit on the signal-to-noise ratio (S/N) of the input lightcurves, the tool is known to perform more reliably when S/N\,$>$\,15. Targets falling below this threshold are flagged accordingly throughout the results.
In Table \ref{results_vetted} we listed the TESS sectors used for the analysis, the combined S/N of the resulting light curves, and the contrast curves employed for each target. Furthermore, the table reports the FPP and NFPP values calculated for each contrast curve. Finally, Table \ref{info_vetted} lists the characteristics of the high-resolution imaging data used, including observatory, instrument, and filters used, pixel scale and full width at half maximum (FWHM) of point spread function (PSF), as well as the Principal Investigator (PI) and acquisition date for each target.

\begin{table}[]
\fontsize{9.5}{12}\selectfont
    \centering
        \caption{Summary of the sample divided by host spectral type.}
    \newcolumntype{M}[1]{>{\centering\arraybackslash}m{#1}}
    \begin{tabular}{M{1.3cm}|M{2.3cm}|M{1.6cm}}
    \hline
    \hline
         & Middle-K to early-M stars & M stars\\
         \hline
        Total & 270 & 48 \\
        \hline
        VP & 42 & 15 \\
        FP & 21 & 8 \\ 
        LP & 31 & 4 \\ 
        OT & 38 & 9 \\ 
        EB & 4 & 2 \\
        AR & 4 & 2 \\ 
        NC & 130 & 8 \\ 
        \hline
        ${R_{\rm p}} >$\,8\,$R_{\rm \oplus}$ & 19 & 1 \\
        ${R_{\rm p}} <$\,8\,$R_{\rm \oplus}$ & 106 & 39 \\
        TDP & 15 & 0 \\
        \hline
    \end{tabular}
    \label{tab:summary}
    \\
\footnotesize\raggedright{{\textbf{Note.} The numbers represent the total analysed} sample, comprehensive of the three TPCs catalogues (09-07-2024, 18-07-2025, and 18-08-2025), before any additional checks. VP = Validated Planet, FP = False Positive, LP = Likely Planet, OT = Other, EB = Eclipsing Binary, AR = Ambiguous Result, NC = Not Classified, and TDP = Targets with Derived Parameters.}
    
\end{table}

\begin{figure}
    \centering
    \includegraphics[width=\hsize]{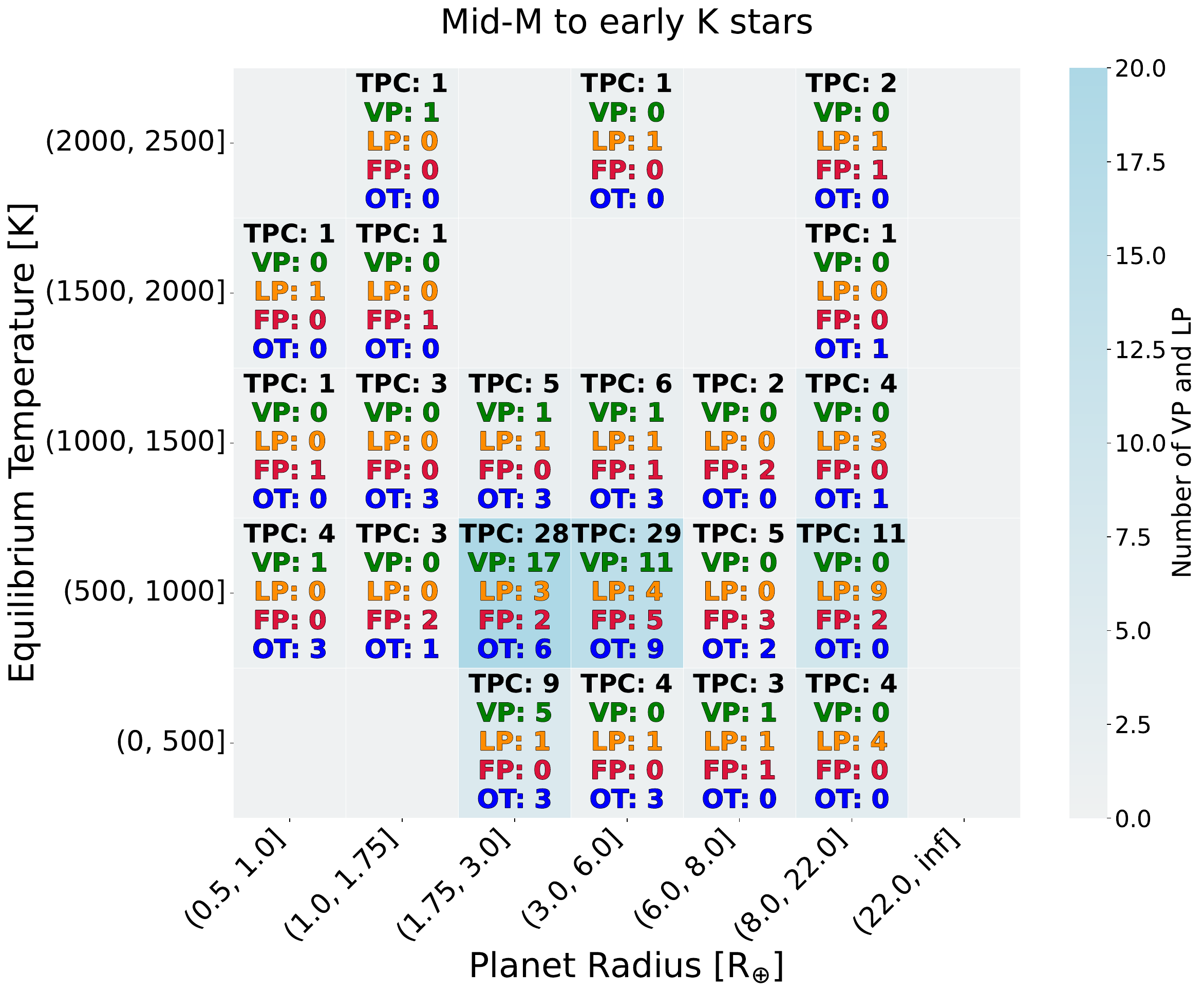}
    \caption{Coverage of the full early-M and middle-K star sample compared to the three TPC catalogues. The total number of candidates (TPC) around middle-M and early-K stars (3500 $< T_{\rm eff} <$\,5000\,K) is reported in black. The number of validated planets (VP) is green, the number of likely planets (LP) is orange, the number of false positives (FP) is red, and the number of Other (OT) is blue. Targets that were excluded from the validated sample following our checks (see Section~\ref{validated}) are not included in the VP count.}
    \label{fig:Sample_KM}
\end{figure}

\begin{figure}
    \centering
    \includegraphics[width=\hsize]{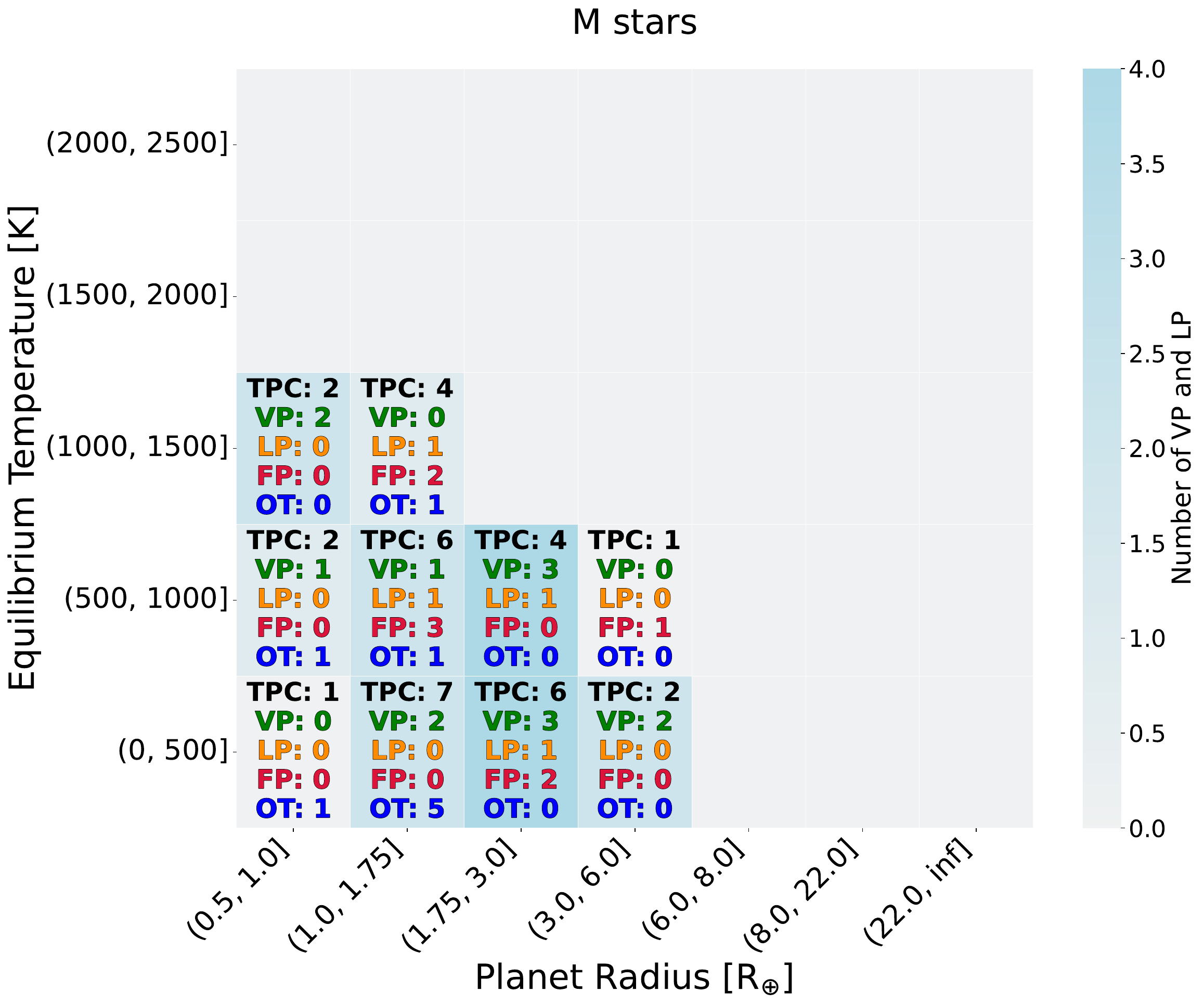}
    \caption{Coverage of the full M-dwarf star sample compared to the three TPC catalogues. The total number of candidates (TPC) orbiting M stars ($T_{\rm eff} <$\,3500\,K) is reported in black. The number of validated planets (VP) is green, the number of likely planets (LP) is orange, the number of false positives (FP) is red, and the number of Other (OT) is blue. Targets that were excluded from the validated sample following our checks (see Section~\ref{validated}) are not included in the VP count.}
    \label{fig:Sample_M}
\end{figure}

\begin{figure}
    \centering
    \includegraphics[width=\hsize]{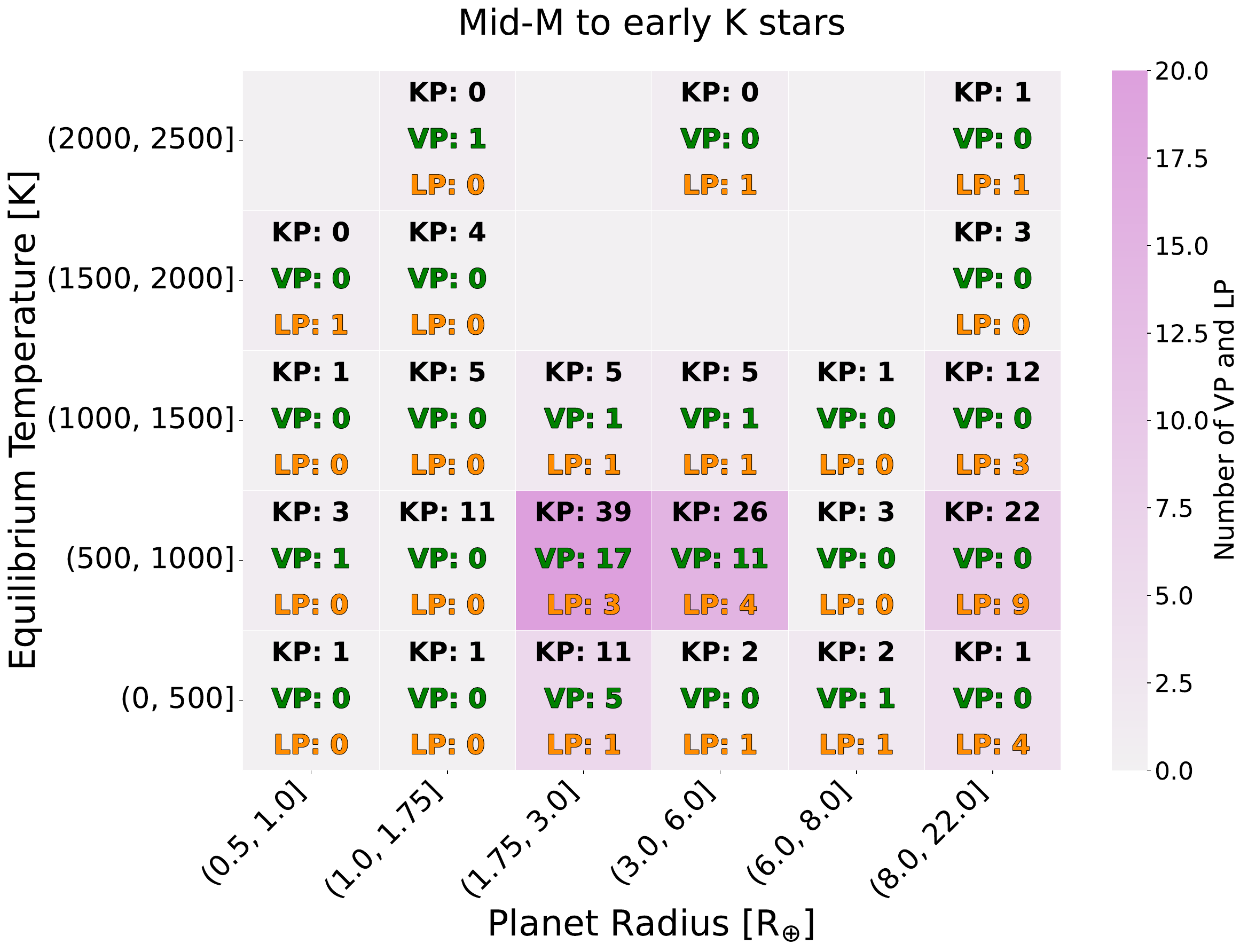}
    \caption{Coverage of the full vetted and validate candidates around early-M and middle-K star sample compared strictly to the latest {\it Known} catalogue (18-08-2025). The total number of known planets (KP) orbiting early-M and middle-K stars is reported in black. The number of validated planets (VP) is green and the number of likely planets (LP) is orange. Targets that were excluded from the validated sample following our checks (see Section~\ref{validated}) are not included in the VP count.}
    \label{fig:KM_KP}
\end{figure}

\begin{figure}
    \centering
    \includegraphics[width=\hsize]{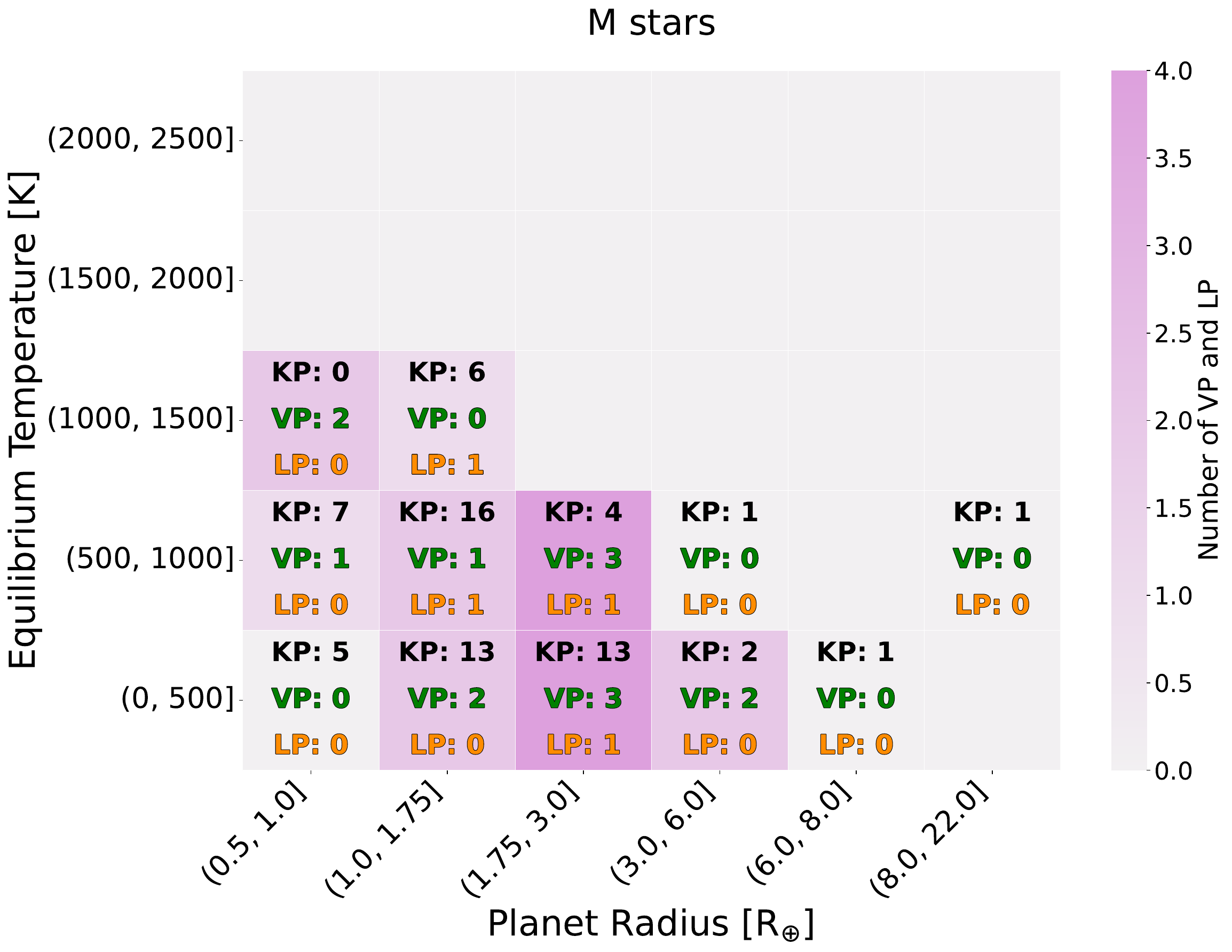}
    \caption{Coverage of the full vetted and validate candidates around M-dwarf star sample compared strictly to the latest {\it Known} catalogue (18-08-2025). The total number of known planets (KP) orbiting M stars is reported in black. The number of validated planets (VP) is green and the number of likely planets (LP) is orange. Targets that were excluded from the validated sample following our checks (see Section~\ref{validated}) are not included in the VP count.}
    \label{fig:M_KP}
\end{figure}

\renewcommand\arraystretch{1.8}
\begin{table*}[!t]
\fontsize{7.5}{9}\selectfont
\centering

\caption{Results of the vetted systems.}
\label{results_vetted}
\newcolumntype{M}[1]{>{\centering\arraybackslash}m{#1}}
\begin{tabular}{M{1.8cm}M{2.8cm}M{1.8cm}M{3cm}M{2cm}M{2cm}}

\hline
\hline
TOI&TESS sectors&S/N&Contrast curve&FPP&NFPP\\
\hline
\makecell[t]{133.01} &
\makecell[t]{1, 28, 68} &
\makecell[t]{47.8} &
\makecell[t]{Zorro (562\,nm) \\ Zorro (832\,nm)} &
\makecell[t]{0.457 $\pm$ 0.053 \\ 0.344 $\pm$ 0.092} &
\makecell[t]{(1.40 $\pm$ 0.24)$\times10^{-6}$ \\ (1.81 $\pm$ 0.32)$\times10^{-6}$} \\

210.01&1, 2, 3, 4, 5, 7, 8, 9, 10, 11, 12, 13, 27, 28, 29, 30, 31, 32, 33, 34, 35, 37, 38, 39, 61, 62, 63, 64, 65, 67, 68, 69&13.5&No&0.867 $\pm$ 0.299&0.339 $\pm$ 0.417\\

\makecell[t]{213 b} &
\makecell[t]{1, 2, 13, 27, 29, 67, 69} &
\makecell[t]{93.2} &
\makecell[t]{Zorro (562\,nm) \\ Zorro (832\,nm)} &
\makecell[t]{(0.64 $\pm$ 0.29)$\times10^{-6}$ \\ (0.94 $\pm$ 2.83)$\times10^{-6}$} &
\makecell[t]{(4.51 $\pm$ 0.89)$\times10^{-12}$ \\ (4.83 $\pm$ 1.49)$\times10^{-11}$} \\

\makecell[t]{233 b} &
\makecell[t]{2, 29, 42, 69, 70, 92} &
\makecell[t]{77.5} &
\makecell[t]{Zorro (562\,nm) \\ Zorro (832\,nm) \\ PHARO (Br$\gamma$)} &
\makecell[t]{(5.6 $\pm$ 9.8)$\times10^{-7}$ \\ (5.0 $\pm$ 5.6)$\times10^{-8}$ \\ (8.8 $\pm$ 9.3)$\times10^{-5}$} &
\makecell[t]{0.0 $\pm$ 0.0 \\ 0.0 $\pm$ 0.0 \\ 0.0 $\pm$ 0.0} \\
\hline
\end{tabular}
\\
\footnotesize\raggedright{{\textbf{Note.} The complete table is available in a machine-readable form. We reported the total S/N. FPP and NFPP values derive from the mean of the 10 runs and their errors are the standard deviations.}}
\end{table*}

\begin{figure}
    \centering
    \includegraphics[width=\hsize]{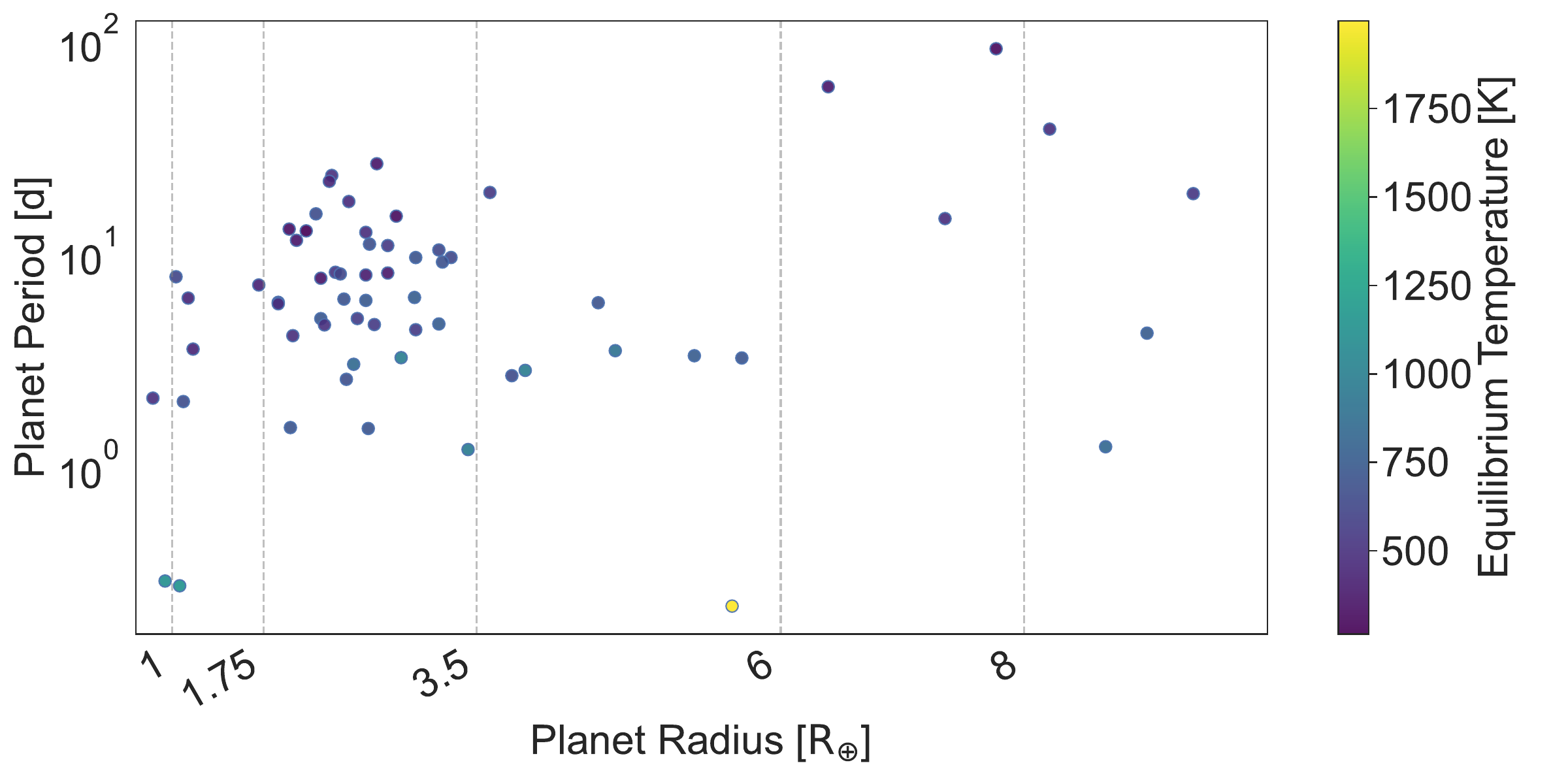}
    \caption{Representation of the validated candidates in a radius-period diagram. Vertical dotted lines delineate the different planetary categories, following \citet{Fressin2013} classification: Earths, super-Earths, small Neptunes, large Neptunes and Giant and the 8\,$\mathrm{R_{\oplus}}$ limit between planets and brown dwarfs.}
    \label{fig:allvalidated}
\end{figure}

\subsection{Validated planets}
\label{validated}
Of the 56 validated targets, 19 are newly validated and passed our stellar binarity checks, 15 have already been validated and 18 have already been classified as planets. We retain these previously classified candidates to validate our method and to enable a comparison with the results already published. Figure \ref{fig:allvalidated} presents the full set of validated candidates, including both newly validated targets and those previously reported in the literature (either validated or characterised). We present the 19 newly validated planets and the 15 candidates previously validated in the literature in Table~\ref{tab:validated_new}, separate from the full results, as these represent the most promising targets of our analysis for which we encourage follow-up observations to measure their masses and fully characterise their nature. It can be noted that the majority of the validated sample falls within the rocky planet regime. All validated planets have FPP and NFPP values well below the adopted thresholds, with tight and stable standard deviations, with the exception of two cases. For these targets, in order to improve the stability of the results and to better constrain the statistical nature of the candidates, we repeated the analysis using 20 runs. TOI-233.02 yields FPP values consistent with a validated planet for two of the available contrast curves; however, for the third curve we obtain a FPP of $1.3 \pm 0.3$\%. Upon repeating the analysis with 20 runs, this value increases to $2.0 \pm 0.3$\%, consistent with a likely planet disposition. We therefore moved this target to the Ambiguous Results category (Section~\ref{ambiguous}). TOI-475.01 shows a FPP of $1.4 \pm 0.3$\%; upon repeating the analysis with 20 runs, we obtain $1.3 \pm 0.2$\%, and we therefore retain it in the validated sample. Among the validated candidates, six stars have RUWE or $\mathrm{IPD_{FMP}}$ above the considered limits. TOI-233, TOI-663, TOI-1467 and TOI-5172 exhibit high RUWE values, yet none are included in the {\it Gaia} DR3 NSS catalogue, and all have planet candidates listed on ExoFOP. TOI-3494 exhibits high RUWE (3.73) and $\mathrm{IPD_{FMP}}$ (77) values, strongly suggesting stellar binarity, yet it is absent from the {\it Gaia} DR3 NSS catalogue and is listed as hosting a planet candidate on ExoFOP. TOI-7054 has the highest $\mathrm{IPD_{FMP}}$ value of our sample (88), indicating stellar binarity. It appears in the {\it Gaia} DR3 NSS catalogue with the AstroSpectroSB1 model, which is used for sources consistent with astrometric and single-line spectroscopic binaries. The period reported is $\sim$\,760\,days. On ExoFOP, a planet candidate disposition is reported. Even if TOI-1280 and TOI-3585 do not show either high RUWE or $\mathrm{IPD_{FMP}}$ values, they are included in the {\it Gaia} DR3 NSS catalogue with a nine-parameter variable-acceleration model. This model is used when the gravitational pull within the system changes noticeably during the mission, even though the full orbital period is still much longer than the observation period. On ExoFOP, both are listed as hosting planet candidates. 

For the {\it Gaia} processing analysis, acceleration solutions were retained for many binary stars with orbital periods shorter than the mission duration. While this choice may no longer be optimal for DR3 processing, \cite{halbwachs2023} considered it preferable to risk an inaccurate orbital period rather than provide only a polynomial fit for a trajectory that could be elliptical. This approach will be reconsidered in {\it Gaia} DR4, with the expectation that the NSS solutions will be improved and will supersede the original DR3 solutions.

We cross-checked our results with {\it Gaia} catalogues and, based on this comparison, we excluded TOI-1280, TOI-3494, TOI-3585, and TOI-7054 from the final validated sample, as they exhibit elevated RUWE or $\mathrm{IPD_{FMP}}$ values, or are listed in the \textit{Gaia} DR3 NSS catalogue.

Four targets in our validated sample have S/N below 15: TOI-521 (9.9), TOI-654 (12), TOI-1280 (14.4), and TOI-2431 (13.2). Among the validated candidates, several have previously appeared in the literature at various stages of characterisation. At the vetting level, TOI-233\,b was classified as a vetted candidate by both \citet{Cloutier2019} and \citet{RAVEN}. TOI-5735\,b was reported as a potential planet by \citet{Gomez_5706} (a classification equivalent to a vetted disposition) although \citet{Hord2024} later flagged it as a false positive. Additionally, 16 further targets had already been statistically validated prior to this work: TOI-2453\,b \citep{RAVEN}, TOI-406\,b, TOI-1683\,b \citep[with mass estimated from the mass–radius relation of][]{2023zhang}, and TOI-1806\,b \citep{Hord2024}, TOI-213\,b, TOI-435\,b, TOI-783\,b, TOI-1806\,c, TOI-2540\,b, TOI-3585\,b, TOI-4324\,b, TOI-6662\,b \citep{Collier2026}, TOI-1301\,b \citep{2025Crossfield,RAVEN}, TOI-2407\,b \citep{JanoMunoz2025}, TOI-5799\,b \citep{vatestIII,2025Crossfield}, and TOI-7166\,b \citep{barkaoui_7166,2026Jiang7166}.

Finally, 18 targets in our validated sample have already been confirmed through dynamical mass measurements. These are: TOI-521\,b and TOI-912\,b \citep{2026Lacedelli}, TOI-654\,b \citep{Ikuta_toi654}, TOI-663\,b and TOI-663\,c \citep{2024cointepas663}, TOI-782\,b \citep[mass upper limit;][]{2024Hori}, TOI-815\,b \citep{2024psaridi815}, TOI-1180\,b, TOI-1410\,b, and TOI-1467\,b \citep[the latter with a mass upper limit;][]{2024polanskitess}, TOI-1450\,b \citep{2024brady1450}, TOI-1803\,c \citep{2025Zingales}, TOI-2134\,b \citep{2024rescigno2134}, TOI-2431\,b \citep{HanTas2025}, TOI-4491\,b \citep{2025gerald4491}, TOI-4529\,b \citep[mass upper limit;][]{TOI4529}, TOI-5734\,b \citep{Filomeno_5734}, and TOI-6324\,b \citep{2025ALee}. One additional target, TOI-493\,b, was previously reported as a false positive by \citet{vatestII}; in this work, we extended its analysis by incorporating three additional TESS sectors and three contrast curves.

\subsection{False positives}
Initially, we identified 22 false positives, which may result from instrumental noise or other astrophysical sources. Some of these candidates have already been analysed in previous studies, sometimes classifying them as false positives and other times confirming their planetary nature. It is important to note that our probabilistic approach is intended to establish a priority ranking rather than provide a definitive determination of the nature of each signal. In particular, one has already been classified as false positive, while six have been validated or confirmed. Considering the RUWE and $\mathrm{IPD_{FMP}}$ values, TOI-3433 meets the criteria indicating a possible unresolved stellar binary, even if it is not listed in the {\it Gaia} DR3 NSS catalogue.

Ten targets in our false positive sample have S/N below 15: TOI-210 (13.5), TOI-756 (8.9), TOI-771 (14.3), TOI-2290 (10.4), TOI-4114 (7.4), TOI-4527 (7.7), TOI-4552 (6.1), TOI-5974 (8.6), TOI-6008 (7.3), and TOI-6258 (6.2). Nine of our false positive candidates have previously appeared in the literature, sometimes with discrepant classifications. TOI-2290.01 was classified as a vetted candidate by \citet{2022Giacalone}, while TOI-4527\,b was statistically validated by \citet{Hord2024} on the basis of additional ground-based photometry. Four candidates that we classify as false positives have since been confirmed as planets through mass measurements using TESS, ground-based photometry, high resolution imaging and radial velocity data: TOI-756\,b \citep{parc_toi756}, TOI-1448\,b \citep[mass upper limit;][]{2024Hori}, TOI-2015\,b \citep{2024Jones,2025Barkaoui}, and TOI-6008\,b \citep{2024Barkaoui}. Three further cases are particularly noteworthy: these targets were flagged as false positives both in our analysis and in previous works, yet were ultimately confirmed through dedicated follow-up observations. TOI-771\,b is classified as a false positive here and as a possible false positive by \citet{Hord2024}, but was subsequently validated by \citet{vatestIII} using ground-based photometry and later confirmed by \citet{2025Lacedelli}, who measured its mass. TOI-3568\,b is similarly classified as a false positive both in this work and by \citet{vatestII}, yet \citet{2024Martioli} derived a mass consistent with a planetary nature. Finally, TOI-4552\,b received conflicting classifications in \citet{Hord2024}, being validated by \texttt{VESPA} but flagged as a false positive by \texttt{TRICERATOPS}; a mass measurement for this target has since been reported by \citet{toi4552}.

\subsection{Likely planets}
We first identify 16 likely planets ($R_{\rm P}<\,8\,R_{\rm \oplus}$). Among them three have previously been classified as likely planets in the literature, one has been statistically validated, three have been confirmed through mass measurements, and one has been reported as a false positive. Notably, TOI-1224 and TOI-5706 exhibit high RUWE values, suggesting potential stellar binarity, even though they are absent from the {\it Gaia} DR3 NSS catalogue and are listed as hosting planet candidates on ExoFOP. Three targets have S/N below 15: TOI-864 (8.15), TOI-5706 (10.1), and TOI-7336 (5.9). The three candidates previously classified as likely planets are TOI-864.01 \citep{Hord2024}, TOI-2408.01\, and TOI-3913.01 \citep{vatestII}. TOI-1224\,b has been statistically validated by \citet{2024Thao}. Three targets have been confirmed as planets through dynamical mass measurements: TOI-1803\,b \citep{2025Zingales}, TOI-1184\,b \citep{2024polanskitess}, and TOI-5388\,b \citep[mass upper limit;][]{TOI4529}. TOI-5706.01 was classified as false positive using \texttt{TRICERATOPS+}\footnote{\url{https://github.com/JGB276/TRICERATOPS-plus}} together with ground-based photometry \citep{Gomez_5706}.

\subsection{Eclipsing binary candidates}

For completeness, we also report five targets that do not meet the FP thresholds, although an EB scenario appears to be the most likely explanation. For TOI-278.01, the most probable scenario is a background eclipsing binary with twice the reported orbital period around a companion (BEBx2P); \citet{toi278} reported it as likely stellar variability rather than a planet candidate. We inspected the TESS Exoplanet Vetter\footnote{\url{https://tev.mit.edu/}}. The DV pipeline identifies two planet candidates with identical orbital periods ($\sim$0.299 days) but epochs offset by approximately half the orbital period ($\Delta t \approx 0.146$ days $\approx P/2$). This configuration is the hallmark of an eclipsing binary system in which the primary and secondary eclipses are both detected as separate planet candidates, with the true orbital period being twice the reported one, consistent with the BEBx2P scenario returned by \texttt{TRICERATOPS}. For TOI-450.01, the most probable scenario is a secondary eclipsing binary with twice the reported orbital period (SEBx2P), consistent with the eclipsing binary classification reported by \citet{Tofflemire_450}. For TOI-1412.01, the most likely scenario is a secondary eclipsing binary (SEB), and ExoFOP recommends its retirement as a BEB. Indeed, the star exhibits an exceptionally high RUWE value of 17.79. 
The case of TOI-1638.01 is more ambiguous: we performed three independent analyses using three available contrast curves and found that in two cases the SEB and TP scenarios have indistinguishable probabilities (SEB at $43.0 \pm 2.5$\% versus TP at $42.9 \pm 3.4$\%, and SEB at $49.3 \pm 2.9$\% versus TP at $34.6 \pm 4.0$\%), while in the third analysis the TP scenario is marginally favored over SEB ($43.6 \pm 2.7$\% versus $43.4 \pm 2.3$\%). Given that the results are overall consistent with an EB scenario in two out of three analyses, and that in one case the SEB probability exceeds the TP probability by approximately 15\%, we conservatively classify TOI-1638.01 as an EB candidate, while acknowledging the ambiguity of its nature.
The latest ExoFOP disposition of TOI-1638.01  remains a planet candidate. Furthermore, even if the star has a RUWE value of 1.53 and $\mathrm{IPD_{FMP}}$ of 6, it is not listed in the {\it Gaia} DR3 NSS catalogue. For TOI-5051.01, the eclipsing binary with twice the reported period (EBx2P) scenario emerged as the most probable.

\subsection{Other candidates}
In addition, we classified 44 objects as Other, with the TP scenario being the most probable. For all candidates the TP scenario probability exceeds 50\%, with the exception of four cases. For TOI-1832.01, the TP scenario has a probability of $46 \pm 23$\%, marginally higher than the second most probable scenario (DTP, $40 \pm 22$\%); the large standard deviations reflect a genuinely uncertain classification, and this target should therefore be treated with caution. For TOI-4766.01, the TP probability is $49.6 \pm 2.5$\%, which remains a factor of 3.6 higher than the second most probable scenario (STP). For TOI-5323.01, the TP probability is $38 \pm 0.5$\%, approximately twice the probability of the second most probable scenario (SEB). For TOI-5955.01, the TP scenario has a probability of $45.4 \pm 0.6$\%, compared to $39.1 \pm 0.5$\% for the second most probable scenario (DTP). We note that, unlike TOI-1832.01, the classifications of TOI-5323.01 and TOI-5955.01 are stable across runs, as indicated by their small standard deviations, suggesting that despite not exceeding the 50\% threshold, the TP scenario remains the most robustly favoured interpretation for these two targets. A subset of these candidates has already been reported in the literature. Two have already been characterised as planets, 13 have been validated and two resulted to be false positive. Notably, four stars exhibits high RUWE or $\mathrm{IPD_{FMP}}$ values. For TOI-240 a RUWE value $>$ 1.4 is reported, but it is not listed in the {\it Gaia} DR3 NSS catalogue and retains a planet candidate disposition on ExoFOP. TOI-486 has an $\mathrm{IPD_{FMP}}$ of 4 and is included in the {\it Gaia} DR3 NSS catalogue, where it is described by a constant-acceleration model. TOI-6077 has a quite high RUWE value (3.28) and it is included in the {\it Gaia} DR3 NSS catalogue with a variable-acceleration model. Nevertheless, both stars are listed on ExoFOP as hosting candidate planets. TOI-6688 exhibits a high RUWE value (1.69) and an $\mathrm{IPD_{FMP}}$ of 29, strongly indicating stellar binarity. Eight targets in the Other category have S/N below 15: TOI-486 (7.38), TOI-727 (7.38), TOI-2274 (10.6), TOI-5747 (8.93), TOI-5955 (6.4), TOI-6077 (3.3), TOI-6086 (6.0), and TOI-6245 (9.4). Thirteen candidates in this category have previously been statistically validated in the literature: TOI-486\,b \citep{2024Tey}, TOI-741\,b \citep[with a mass estimate;][]{2025DiTomasso}, TOI-880,c \citep[with a mass estimate from the][mass–radius relation]{Chen2017,2025Zhang}, TOI-1080\,b \citep[with a mass estimate;][]{toi1080}, TOI-1105\,b \citep{RAVEN}, TOI-1730\,b and TOI-1730\,d \citep{WilsonTOI1730}, TOI-2133\,b \citep{Morello_2026}, TOI-2274\,b \citep[with additional ground-based photometry;][]{2024PelaezTorres}, TOI-6086\,b \citep[with ground-based photometry;][]{2024Barkaoui}, TOI-6276\,b and TOI-6276\,c \citep{2025Kunimoto}, and TOI-6716\,b \citep{scott_6716}. Two targets have been confirmed as planets through mass measurements: TOI-5747\,b \citep{2024Murgas} and TOI-5800\,b \citep{2025jenkins,2025naponiello}. Finally, two objects in this category were classified as false positives by \citet{vatestII}: TOI-4090.01, which in that study was based on only two TESS sectors compared to the five used here, and TOI-5584.01, for which we incorporated available contrast curves to better account for the flux contribution from nearby stars.

\subsection{Ambiguous results}
\label{ambiguous}
The six targets we evaluated are those for which we obtained ambiguous results, i.e. whose classification varies depending on the adopted contrast curves. They are: TOI-880.01, TOI-2079.02, TOI-5319.01, TOI-5507.01, TOI-5720.01, TOI-6255.01. To this subsample we add TOI-233.02, as previously discussed in Section~\ref{validated}, for a total of seven candidates with ambiguous results. For TOI-880.01, our analysis yielded a validated planet with two contrast curves, an Other result with one, and with the two remaining curves we obtained FPP and NFPP values that do not fall in any categories with a DTP scenario. Currently, it is reported as a planet \citep{Doyle_880}. For TOI-2079.02, we obtained three dispositions as validated planet and one likely planet. On ExoFOP is listed as active planet candidate. TOI-5319.01 was classified as Other using two different contrast curves and as a false positive a third one; it has been validated using ground-based photometry by \citet{2024PelaezTorres}. For TOI-5507.01, we obtained two likely planet results and one validated planet, with an active planet candidate status on ExoFOP. The star has a RUWE value of 1.62, but it is not listed in the {\it Gaia} DR3 NSS catalogue. TOI-5720.01 was classified as a validated planet with three different curves and as a likely planet using a forth one, and was confirmed as a planet by \citet{2024Barkaoui}. Finally, TOI-6255.01 yielded one validated planet result and one likely planet result, and \citet{2024Dai} mass-characterised it as a planet. It is also reported in {\it Gaia} NSS with a constant-acceleration model describing the system. TOI-2079.02 and TOI-5507.01 currently lack existing publications and our analysis did not yield homogeneous results across the various contrast curves, yet both fall in the validated/likely planets classification.

\subsection{Candidates with radius $>\,8\,R_{\oplus}$}
The statistical validation framework implemented in \texttt{TRICERATOPS} is designed for targets with radii smaller than $8\,R_{\oplus}$, in accordance with the limits recommended by \citet{2021giacalonevetting}. Consequently, for candidates exceeding this threshold, the classification of validated planet cannot be applied, and we instead adopt the more conservative designation of vetted candidate, including them in the likely planets category. Candidates with $R_{\rm p} > 8\,R_{\oplus}$ that were classified as false positives, eclipsing binary candidates, or assigned to the other or ambiguous results categories have already been discussed in the respective sections above. Here we focus exclusively on the 15 candidates that satisfy the FPP and NFPP criteria for a likely planet disposition.

Three of these vetted candidates show elevated RUWE or $\mathrm{IPD_{FMP}}$ values, warranting additional caution. TOI-3586 has a RUWE of 1.97 and an $\mathrm{IPD_{FMP}}$ of 0; no additional indications of binarity are reported on ExoFOP, which lists it as a planet candidate, and it does not appear in the \textit{Gaia} DR3 NSS catalogue. TOI-6806 has a RUWE below 1.4 but an $\mathrm{IPD_{FMP}}$ of 5; this star appears in the \textit{Gaia} DR3 NSS catalogue with a spectroscopic binary solution and is listed on ExoFOP as an ambiguous planet candidate. TOI-7464 presents a more subtle case: although neither its RUWE nor its $\mathrm{IPD_{FMP}}$ exceed the adopted thresholds, the system appears in the \textit{Gaia} DR3 NSS catalogue with a constant-acceleration model, indicating that the stellar components are unresolved and the photocentre is offset from the barycentre. As the orbital period significantly exceeds the observational baseline, this acceleration remains approximately constant in both magnitude and direction \citep{halbwachs2023}. ExoFOP lists this system as hosting a planet candidate.

Among the vetted candidates in this subsample, TOI-330\,b and TOI-934\,b were vetted by \citet{RAVEN}.
Three have published mass measurements, which illustrate the diversity of objects populating this radius regime. TOI-1823\,b is a confirmed sub-Saturn with a radius of $7.54_{-0.25}^{+0.33}\,R_{\oplus}$ and a mass of $67.4 \pm 8.2\,M_{\oplus}$. TOI-5375\,b has been studied by two independent groups: \citet{Lambert_5375} report a mass of $83.81 \pm 2.10\,M_{\mathrm{Jup}}$ and a radius of $0.1114_{-0.0050}^{+0.0048}\,R_{\mathrm{Jup}}$, classifying the object as a low-mass star, while \citet{Maldonado_5375} derive a mass of $77 \pm 8\,M_{\mathrm{Jup}}$ and a radius of $0.99 \pm 0.16\,R_{\mathrm{Jup}}$, consistent with a brown dwarf. Finally, TOI-6110\,b (also known as NGTS-28Ab) is confirmed as a brown dwarf by \citet{henderson_toi6110}, with a mass of $69.0_{-4.8}^{+5.3}\,M_{\mathrm{Jup}}$ and a radius of $0.95 \pm 0.05\,R_{\mathrm{Jup}}$.

\subsection{Targets with derived host-star parameters}
\label{sec:derived_param}
Among the 153 objects that we could not analyse because of the missing key parameters (at least one among star mass and radius) in TESS Objects of Interest Catalog \citep{Guerrero_2021}, we were able to derive the parameters of 15 of them, as we reported in Table~\ref{info_parameter_derived}, in which we reported which stellar parameter has been derive and its value. Stellar parameters were derived using the parallax reported in {\it Gaia} DR3 catalogue and photometric data \citep[J, H, K, G, and V magnitudes, Bp and Rp colours,][]{GDR2} by interpolating the empirical tables of \citet{PecautMamajek2013}, which are calibrated for main-sequence stars. We note that for stars with radii exceeding 2.5\,$R_{\odot}$, a regime where the main-sequence assumption becomes unreliable and evolved stars may be present. In these cases, we assigned a BEB or NEB classification as the most probable scenario. The results for these targets should therefore be interpreted with caution and we reported them in Table~\ref{info_vetted_derived}.
In particular we recover seven false positives (TOI-885.01, TOI-1215.01, TOI-1976.01, TOI-3583.01, TOI-3606.01, TOI-4276.01, and TOI-4801.01), four likely planets (TOI-579.01, TOI-4625.01, TOI-5809.01, and TOI-6827.01), and three Other with TP scenario being the most probable (TOI-1490.01, TOI-4839.01, and TOI-6645.01). Finally, we identify one eclipsing binary candidates: TOI-4893.01 (EBx2P). Among them two have a S/N below 15: TOI-885 (13.7) and TOI-1215 (4.9).
Several targets in this subsample show elevated RUWE and $\mathrm{IPD_{FMP}}$ values, suggestive of unresolved companions or binarity, although none of them, with the exception of TOI-4893, are reported in the \textit{Gaia} DR3 NSS catalogue. The most extreme cases are TOI-4839 (RUWE\,=\,26.43, $\mathrm{IPD_{FMP}}$\,=\,73), TOI-4276 (RUWE\,=\,18.30, $\mathrm{IPD_{FMP}}$\,=\,5), TOI-3606 (RUWE\,=\,9.98), TOI-3583 (RUWE\,=\,9.39, $\mathrm{IPD_{FMP}}$\,=\,54), TOI-4801 (RUWE\,=\,6.61, $\mathrm{IPD_{FMP}}$\,=\,48), TOI-1976 (RUWE\,=\,5.23, $\mathrm{IPD_{FMP}}$\,=\,85), TOI-579 (RUWE\,=\,6.13, $\mathrm{IPD_{FMP}}$\,=\,45), TOI-4625 (RUWE\,=\,2.73, $\mathrm{IPD_{FMP}}$\,=\,66), and  TOI-4893 (RUWE\,=\,31.25, $\mathrm{IPD_{FMP}}$\,=\,22) is also listed in the \textit{Gaia} DR3 NSS catalogue with a solution compatible with an orbital model. TOI-1215 has a RUWE of 1.74 and $\mathrm{IPD_{FMP}}$ of 8, while TOI-6645 has a RUWE of 9.15 and $\mathrm{IPD_{FMP}}$ of 36; the latter is also flagged as a likely EB on the ExoFOP website. Finally, TOI-6827 shows an elevated $\mathrm{IPD_{FMP}}$ of 12.

\begin{table}
\fontsize{9}{10}\selectfont
\centering

\caption{Derived stellar parameters.}
\label{info_parameter_derived}
\newcolumntype{M}[1]{>{\centering\arraybackslash}m{#1}}

\begin{tabular}{cccc}

\hline
\hline
TOI & Derived Parameters & $R_{\star}$ [$\mathrm{R_{\odot}}$] & $M_{\star}$ [$\mathrm{M_{\odot}}$] \\
\hline
579 & $R_{\star}$, $M_{\star}$ & 0.79 & 0.79 \\
885 & $M_{\star}$ & 8.02 & 1.0 \\
1215 & $R_{\star}$, $M_{\star}$ & 0.72 & 0.64 \\
1490 & $M_{\star}$ & 5.11 & 1.0 \\
1976 & $R_{\star}$, $M_{\star}$ & 3.53 & 1.0 \\
3583 & $R_{\star}$, $M_{\star}$ & 0.55 & 0.55 \\
3606 & $R_{\star}$, $M_{\star}$ & 0.61 & 0.60 \\
4276 & $R_{\star}$, $M_{\star}$ & 0.77 & 0.80 \\
4625 & $M_{\star}$ & 0.72 & 0.70 \\
4801 & $R_{\star}$, $M_{\star}$ & 0.76 & 0.79 \\
4839 & $R_{\star}$, $M_{\star}$ & 0.76 & 0.79 \\
4893 & $R_{\star}$, $M_{\star}$ & 0.76 & 0.79 \\
5809 & $R_{\star}$, $M_{\star}$ & 0.88 & 0.68 \\
6645 & $R_{\star}$, $M_{\star}$ & 0.70 & 0.71 \\
6827 & $M_{\star}$ & 2.53 & 1.0 \\

\hline
\end{tabular}

\end{table}

\section{Discussion}
\label{sec:discussion}

\textit{False positives and follow-up potential}. The vetting analysis presented in this work provides a first systematic prioritisation of TPC candidates for follow-up observations in support of the Ariel mission. However, it is important to stress that a false positive classification from \texttt{TRICERATOPS} does not necessarily imply that a given candidate is not a planet. In particular, candidates classified as false positives for which no clear indicators of stellar binarity are present, such as elevated RUWE or $\mathrm{IPD_{FMP}}$ values, or a listing in the \textit{Gaia} DR3 NSS catalogue, may still harbour genuine planetary signals that could be confirmed through dedicated follow-up observations. The probabilistic nature of our approach establishes a priority ranking rather than a definitive determination of the nature of each signal, and some fraction of the current false positive sample may be reclassified as planet candidates once additional data become available.

\textit{The role of follow-up observations}. Two observational strategies would significantly improve the robustness of the vetting analysis. First, the inclusion of ground-based photometric follow-up, such as that enabled by \texttt{TRICERATOPS+} \citep{Gomez_5706}, would allow the rejection of additional false positive scenarios by constraining the transit chromaticity and ruling out nearby eclipsing binaries at angular separations not accessible to TESS. Second, high-resolution imaging contrast curves play a crucial role in the statistical validation process. As demonstrated by \citet{Collier2026}, 72\% of validated planets in their \texttt{TRICERATOPS}-based analysis would have failed validation without the adopted contrast curves, with the fraction being highest for small planets ($R_p < 1.7\,R_\oplus$). Expanding the availability of contrast curves for our sample would therefore directly increase the number of candidates that can be formally validated.

\textit{Completeness of the analysis.} A significant fraction of the initial sample (138 out of 318 targets) could not be analysed due to missing stellar parameters or unavailable 2-minute cadence TESS light curves. We acknowledge that the priority ranking derived from our analysis is necessarily limited to the subset of targets for which sufficient data were available at the time of writing. However, we argue that this limitation does not undermine the scientific relevance of our results for several reasons. First, the statistical validation of a sample of planet candidates is a valuable result in itself, as it identifies a set of high-priority objects for mass measurements and atmospheric characterisation, directly supporting the preparation of the Ariel mission. Second, the excluded targets were removed solely on the basis of data availability, not on any intrinsic property that would make them less interesting as Ariel candidates. As new TESS sectors are observed and ground-based follow-up campaigns provide improved stellar characterisation, many of these targets will become accessible to future vetting analyses. Third, we stress that this work is explicitly conceived as a preliminary, iterative study. The Ariel launch is currently scheduled for 2031, and both the TPC catalogue and the {\it Known} catalogue will continue to evolve substantially in the next years. Our results should therefore be understood as a first screening of the most accessible candidates rather than a definitive target selection. Fourth, for 15 of the excluded targets we were able to recover stellar parameters from \textit{Gaia} DR3 parallaxes, demonstrating that the incompleteness can be progressively reduced as auxiliary data become available. The extension of this approach to a larger fraction of the excluded sample, combined with the analysis of targets orbiting stars with $T_{\rm eff} > 5000$\,K\, is planned as a natural follow-up of the present work.

\begin{figure}
    \centering
    \includegraphics[width=\hsize]{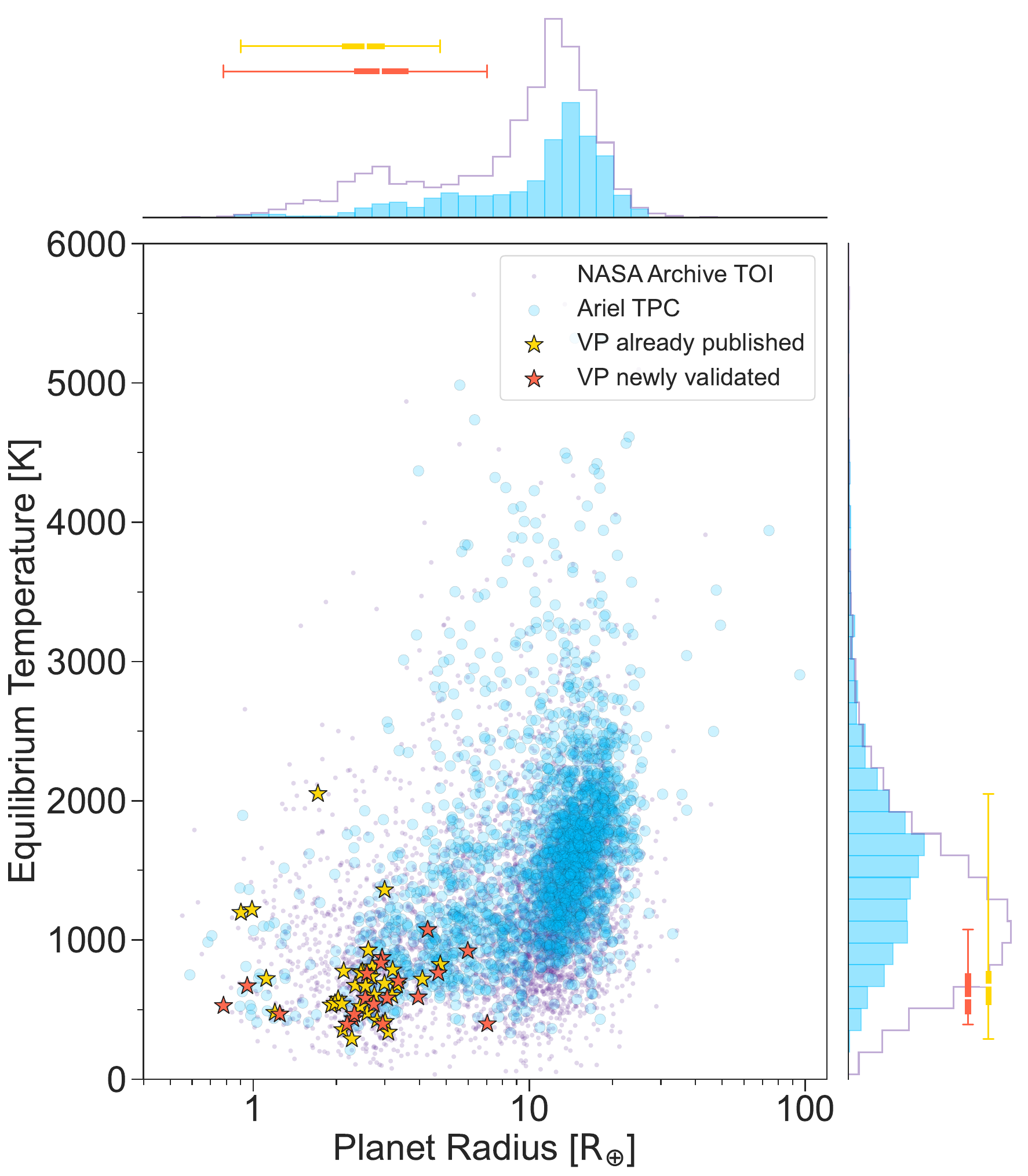}
    \caption{Comparison between TOIs reported in the NASA Exoplanet Archive (October 2026) with the last three Ariel TPCs catalogues (09-07-2024, 18-07-2025, and 18-08-2025) and our validated planets sample.}
    \label{fig:compar_TPC_TOI}
\end{figure}

\begin{figure}
    \centering
    \includegraphics[width=\hsize]{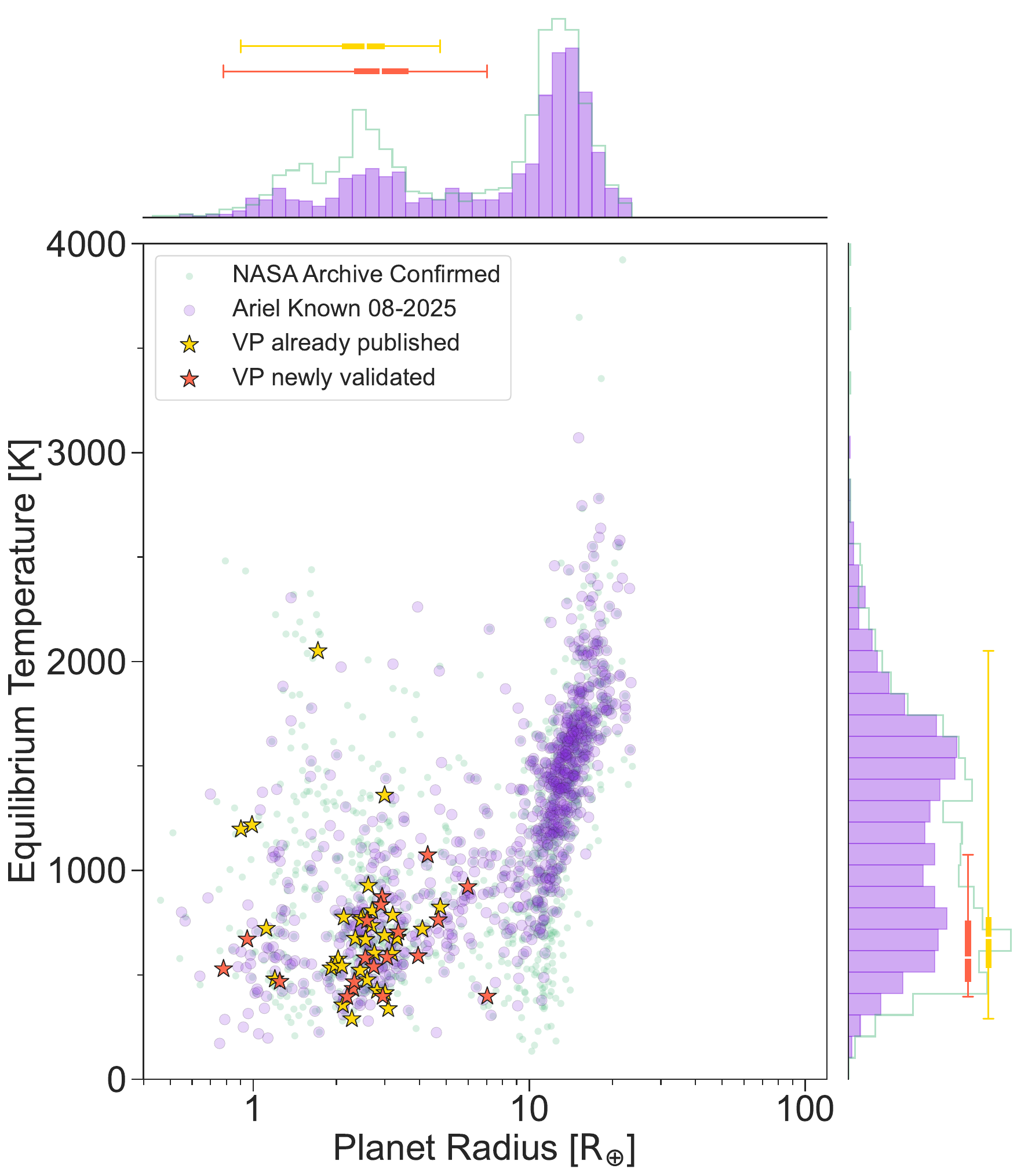}
    \caption{Comparison between confirmed planets with a transit detection reported in the NASA Exoplanet Archive (October 2026) with the Ariel {\it Known} catalogue and our validated planets sample.}
    \label{fig:compar_Known_Conf}
\end{figure}

\textit{Comparison between the NASA Exoplanet Archive and the Ariel catalogues.} To evaluate how well the Ariel TPC catalogues represent the broader known exoplanet population, we compared them with the TESS Objects of Interest reported in the NASA Exoplanet Archive\footnote{\raggedright \url{https://exoplanetarchive.ipac.caltech.edu/}, October 2026}. As shown in Figure~\ref{fig:compar_TPC_TOI}, the three TPC catalogues, combined into a single sample for visual clarity, provide a representative coverage of the Archive population in both planetary radius and equilibrium temperature. Our validated planets predominantly populate the low-temperature regime at radii below $4\,R_{\oplus}$, a region that is currently sparsely sampled, as highlight with the box plot. In Figure~\ref{fig:compar_Known_Conf}, we compare the confirmed transiting planets reported in the NASA Exoplanet Archive with the Ariel {\it Known} catalogue (18-08-2025) and our validated planet sample. The {\it Known} catalogue shows a relatively uniform distribution below 2000\,K and a concentration of objects at radii above $10\,R_{\oplus}$. Our statistically validated planets contribute to filling an underpopulated region of this parameter space, complementing the existing {\it Known} catalogue and expanding its coverage toward smaller, cooler planets.

\textit{Atmospheric characterisation prospects with Ariel.} To assess the suitability of our validated planets for atmospheric characterisation with Ariel, we computed the Transmission Spectroscopy Metric \citep{Kempton_2018} for each target, using the stellar magnitude in the $J$ band. This choice is particularly appropriate in the context of Ariel, whose instruments cover the near- and mid-infrared range (0.5--7.8\,$\mu$m), making the $J$-band magnitude a relevant proxy for the stellar brightness in the mission's primary working range. Following the thresholds defined by \citet{Kempton_2018}, we considered a target suitable for atmospheric characterisation if TSM\,$>$\,10 for terrestrial planets ($R_p < 1.5\,R_\oplus$), TSM\,$>$\,90 for small sub-Neptunes ($1.5 < R_p < 2.75\,R_\oplus$), and TSM\,$>$\,90 for large sub-Neptunes, sub-Jovians, and Jovian-class planets ($R_p > 2.75\,R_\oplus$). Applying these criteria, 11 of our validated targets exceed the relevant threshold and are therefore strong candidates for future atmospheric characterisation with Ariel. Once the masses of these validated planets are measured through radial velocity follow-up, it will be possible to explore a broader region of parameter space compared to that currently covered by the {\it Known} catalogue (18-08-2025), as illustrated in Figures~\ref{fig:KM_KP} and~\ref{fig:M_KP}. In particular, our validated sample extends the coverage toward lower planetary temperatures and longer orbital periods, two regimes that are currently under-represented in the {\it Known} catalogue and that are of key scientific interest for Ariel's investigation of atmospheric diversity.

\section{Summary}
\label{sec:summary}

In this work, we applied a vetting procedure to a subsample of candidates from the last three Ariel TESS Planet Candidates catalogues. To support the development of the consolidated Ariel Mission Candidate Sample, from which the Ariel Mission Reference Sample of up to 1000 confirmed exoplanets will be selected to provide comprehensive coverage of the relevant parameter space, our work primarily focuses on targets suitable for atmospheric characterisation.
The aim of this study is to assign higher observational priority to statistically validated candidates, while assigning lower priority to those identified as potential false positives. It should be emphasised that our results do not provide definitive confirmation of a planet’s nature, nor do they conclusively rule out false positives. Rather, they offer a probabilistic ranking to guide future observations. Since mass characterisation and confirmation of a planet candidate require a substantial number of observations and significant time, allocating more attention to already validated objects can be particularly advantageous.
We used the \texttt{TRICERATOPS} tool, specifically designed for the validation of TESS planet candidates. We focused on targets orbiting stars up to mid K-type ($T _{\rm eff} <$\,5000\,K) and identified 56 statistically validated candidates and 22 false positives. In addition these two groups, we define two intermediate categories: likely planets, comprising 16 targets, and Other, composed of 44 objects for which the FPP and NFPP values lie outside the recommended thresholds of \citet{2021giacalonevetting}, yet a transiting planet scenario remains the most likely. In addition, we identified five eclipsing binary candidates and seven targets with ambiguous classifications. For 15 targets that were initially excluded due to missing stellar parameters, we derived the necessary properties from \textit{Gaia} DR3 parallaxes, recovering additional candidates across all classification categories. Specifically, we found seven false positives, four likely planets, three Other, and one eclipsing binary candidate.
Furthermore, we expanded our parameter space to include giant planets and brown dwarfs, thus considering candidates with $R_{\rm p} >$\,8\,$\mathrm{R_{\oplus}}$, excluded by \cite{2021giacalonevetting}. We vetted 15 candidates with $R_{\rm p}\,>\,8\,\mathrm{R_{\oplus}}$ and added them to the likely planets category, for a total of 35 vetted targets. In this way our total final analysed sample is composed of 56 statistically validated planets, 35 likely planets, 29 false positives, 47 other with TP scenario as most probable one, and six eclipsing binaries candidates.
In Figures \ref{fig:Sample_KM} and \ref{fig:Sample_M} we report the vetted sample compared to the last TPCs catalogue (released on 18-08-2025). Indeed, our analysis has validated objects occupying previously underpopulated regions of the parameter space, including high-temperature planets ($T_{\rm eq} >$\,2000\,K) and sparsely populated regimes at low temperatures ($T_{\rm eq} <$\,500\,K) orbiting early-M and middle-K stars. The majority of the validated and likely planets exhibit intermediate temperatures, thereby increasing the number of objects of interest in these regions. 
We encourage follow-up observations of 34 targets, reported in Table \ref{tab:validated_new}. Among them, 19 candidates have been newly validated within this work and passed our checks for stellar binarity, like the RUWE and $\mathrm{IPD_{FMP}}$ values, and we cross-matched the corresponding sources with the {\it Gaia} DR3 Non-Single Stars catalogue. To these, we add 15 targets validated both in this work and in previous publications (see Section \ref{sec:Results}).
Our next step is to apply the vetting procedure described in this work to the full TPCs catalogue, focusing on planet candidates orbiting stars with $T_{\rm eff} >$\,5000\,K. This will enable a more comprehensive characterisation of the TPC sample and the compilation of a prioritised list of targets for radial velocity follow-up and mass determination, ultimately supporting their inclusion in the final Ariel Mission Candidate Sample.

\section{Data availability}
Tables \ref{results_vetted} and \ref{info_vetted} are only available in electronic form at the CDS via anonymous ftp to \url{cdsarc.u-strasbg.fr (130.79.128.5)} or via \url{http://cdsweb.u-strasbg.fr/cgi-bin/qcat?J/A+A/}.


\begin{acknowledgements}
The Authors thank the anonymous referee for their comments, which greatly improved this work.
The Authors acknowledge Charles Beichman, Courtney Dressing, Mark Everett, Steven Giacalone, Erica Gonzales, Steve Howell, Claire Lamman, and Jennifer Winters for the imaging observations reported in ExoFOP. 
G.P. acknowledges the INAF Mini-Grant “Planetary Systems Architecture with Jovian planets - JASP” according to the INAF Fundamental Astrophysics funding scheme. G.P., C.D.M., and G.M. acknowledge support from the ASI-INAF agreement 2021-5-HH.2-2024 and the European Union - Next Generation EU through the grant n. 2022J7ZFRA - Exo-planetary Cloudy Atmospheres and Stellar High energy (Exo-CASH) funded by MUR - PRIN 2022. This research has made use of the Exoplanet Follow-up Observation Program (ExoFOP; DOI: 10.26134/ExoFOP5) website, which is operated by the California Institute of Technology, under contract with the National Aeronautics and Space Administration under the Exoplanet Exploration Program. This work has made use of data from the European Space Agency (ESA) mission {\it Gaia} (\url{https://www.cosmos.esa.int/gaia}), processed by the {\it Gaia} Data Processing and Analysis Consortium (DPAC, \url{https://www.cosmos.esa.int/web/gaia/dpac/consortium}). Funding for the DPAC has been provided by national institutions, in particular the institutions participating in the {\it Gaia} Multilateral Agreement. Based on observations obtained at the Hale Telescope, Palomar Observatory, as part of a collaborative agreement between the Caltech Optical Observatories and the Jet Propulsion Laboratory operated by Caltech for NASA. Some of the data presented herein were obtained at Keck Observatory, which is a private 501(c)3 non-profit organization operated as a scientific partnership among the California Institute of Technology, the University of California, and the National Aeronautics and Space Administration. The Observatory was made possible by the generous financial support of the W. M. Keck Foundation. The authors wish to recognize and acknowledge the very significant cultural role and reverence that the summit of Maunakea has always had within the Native Hawaiian community. We are most fortunate to have the opportunity to conduct observations from this mountain. This work made use of \texttt{matplotlib} \citep{matplotlib}, \texttt{numpy} \citep{numpy}, and \texttt{pandas} \citep{2022Reback} libraries.

\end{acknowledgements}

\bibliographystyle{aa}
\bibliography{bibliography}

@ARTICLE{roboao,
       author = {{Baranec}, Christoph and {Riddle}, Reed and {Law}, Nicholas M. and {Ramaprakash}, A.~N. and {Tendulkar}, Shriharsh and {Hogstrom}, Kristina and {Bui}, Khanh and {Burse}, Mahesh and {Chordia}, Pravin and {Das}, Hillol and et al.},
        title = "{High-efficiency Autonomous Laser Adaptive Optics}",
      journal = {\apjl},
         year = 2014,
        month = jul,
       volume = {790},
       number = {1},
          eid = {L8},
        pages = {L8},
          doi = {10.1088/2041-8205/790/1/L8},
archivePrefix = {arXiv},
       eprint = {1407.8179},
 primaryClass = {astro-ph.IM},
       adsurl = {https://ui.adsabs.harvard.edu/abs/2014ApJ...790L...8B}
}

@ARTICLE{2025Barkaoui,
       author = {{Barkaoui}, K. and {Korth}, J. and {Gaidos}, E. and {Agol}, E. and {Parviainen}, H. and {Pozuelos}, F.~J. and {Palle}, E. and {Narita}, N. and {Grimm}, S. and {Brady}, M. and {Bean}, J.~L. and {Morello}, G. and {Rackham}, B.~V. and {Burgasser}, A.~J. and {Van Grootel}, V. and {Rojas-Ayala}, B. and {Seifahrt}, A. and {Marfil}, E. and {Passegger}, V.~M. and {Stalport}, M. and {Gillon}, M. and {Collins}, K.~A. and {Shporer}, A. and {Giacalone}, S. and {Yal{\c{c}}{\i}nkaya}, S. and {Ducrot}, E. and {Timmermans}, M. and {Triaud}, A.~H.~M.~J. and {de Wit}, J. and {Soubkiou}, A. and {Watkins}, C.~N. and {Aganze}, C. and {Alonso}, R. and {Amado}, P.~J. and {Basant}, R. and {Ba{\c{s}}t{\"u}rk}, {\"O}. and {Benkhaldoun}, Z. and {Burdanov}, A. and {Calatayud-Borras}, Y. and {Chouqar}, J. and {Conti}, D.~M. and {Collins}, K.~I. and {Davoudi}, F. and {Delrez}, L. and {Dressing}, C.~D. and {de Leon}, J. and {D{\'e}vora-Pajares}, M. and {Demory}, B.~O. and {Dransfield}, G. and {Esparza-Borges}, E. and {Fern{\'a}ndez-Rodriguez}, G. and {Fukuda}, I. and {Fukui}, A. and {Gallardo}, P.~P.~M. and {Garcia}, L. and {Garcia}, N.~A. and {Ghachoui}, M. and {Gerald{\'\i}a-Gonz{\'a}lez}, S. and {G{\'o}mez Maqueo Chew}, Y. and {Gonz{\'a}lez-Rodr{\'\i}guez}, J. and {G{\"u}nther}, M.~N. and {Hayashi}, Y. and {Horne}, K. and {Hooton}, M.~J. and {Hsu}, C.~C. and {Ikuta}, K. and {Isogai}, K. and {Jehin}, E. and {Jenkins}, J.~M. and {Kawauchi}, K. and {Kagetani}, T. and {Kawai}, Y. and {Kasper}, D. and {Kielkopf}, J.~F. and {Klagyivik}, P. and {Lacedelli}, G. and {Latham}, D.~W. and {Libotte}, F. and {Luque}, R. and {Livingston}, J.~H. and {Mancini}, L. and {Massey}, B. and {Mori}, M. and {Mu{\~n}oz Torres}, S. and {Murgas}, F. and {Niraula}, P. and {Orell-Miquel}, J. and {Rapetti}, David and {Rebolo-L{\'o}pez}, R. and {Ricker}, G. and {Papini}, R. and {Pedersen}, P.~P. and {Pel{\'a}ez-Torres}, A. and {P{\'e}rez-Prieto}, J.~A. and {Poultourtzidis}, E. and {Rodriguez}, P.~M. and {Queloz}, D. and {Savel}, A.~B. and {Schanche}, N. and {S{\'a}nchez-Benavente}, M. and {Sibbald}, L. and {Sefako}, R. and {Sohy}, S. and {Sota}, A. and {Schwarz}, R.~P. and {Seager}, S. and {Sebastian}, D. and {Southworth}, J. and {Stangret}, M. and {Stef{\'a}nsson}, G. and {St{\"u}rmer}, J. and {Srdoc}, G. and {Thompson}, S.~J. and {Terada}, Y. and {Vanderspek}, R. and {Wang}, G. and {Watanabe}, N. and {Wilkin}, F.~P. and {Winn}, J. and {Wells}, R.~D. and {Ziegler}, C. and {Z{\'u}{\~n}iga-Fern{\'a}ndez}, S.},
        title = "{TOI-2015 b: A sub-Neptune in strong gravitational interaction with an outer non-transiting planet}",
      journal = {\aap},
         year = 2025,
        month = mar,
       volume = {695},
          eid = {A281},
        pages = {A281},
          doi = {10.1051/0004-6361/202452916},
archivePrefix = {arXiv},
       eprint = {2502.07074},
 primaryClass = {astro-ph.EP},
       adsurl = {https://ui.adsabs.harvard.edu/abs/2025A&A...695A.281B}
}

@ARTICLE{2024Barkaoui,
       author = {{Barkaoui}, K. and {Schwarz}, R.~P. and {Narita}, N. and {Mistry}, P. and {Magliano}, C. and {Hirano}, T. and {Maity}, M. and {Burgasser}, A.~J. and {Rackham}, B.~V. and {Murgas}, F. and {Pozuelos}, F.~J. and {Stassun}, K.~G. and {Everett}, M.~E. and {Ciardi}, D.~R. and {Lamman}, C. and {Pass}, E.~K. and {Bieryla}, A. and {Aganze}, C. and {Esparza-Borges}, E. and {Collins}, K.~A. and {Covone}, G. and {de Leon}, J. and {D{\'e}vora-Pajares}, M. and {de Wit}, J. and {Fukuda}, Izuru and {Fukui}, A. and {Gerasimov}, R. and {Gillon}, M. and {Hayashi}, Y. and {Howell}, S.~B. and {Ikoma}, M. and {Ikuta}, K. and {Jenkins}, J.~M. and {Karpoor}, P.~R. and {Kawai}, Y. and {Kimura}, T. and {Kotani}, T. and {Latham}, D.~W. and {Mori}, M. and {Pall{\'e}}, E. and {Parviainen}, H. and {Patel}, Y.~G. and {Ricker}, G. and {Relles}, H.~M. and {Shporer}, A. and {Seager}, S. and {Softich}, E. and {Srdoc}, G. and {Tamura}, M. and {Theissen}, C.~A. and {Twicken}, J.~D. and {Vanderspek}, R. and {Watanabe}, N. and {Watkins}, C.~N. and {Winn}, J.~N. and {Wohler}, B.},
        title = "{Three short-period Earth-sized planets around M dwarfs discovered by TESS: TOI-5720 b, TOI-6008 b, and TOI-6086 b}",
      journal = {\aap},
         year = 2024,
        month = jul,
       volume = {687},
          eid = {A264},
        pages = {A264},
          doi = {10.1051/0004-6361/202349127},
archivePrefix = {arXiv},
       eprint = {2405.06350},
 primaryClass = {astro-ph.EP},
       adsurl = {https://ui.adsabs.harvard.edu/abs/2024A&A...687A.264B}
}

@ARTICLE{barkaoui_7166,
       author = {{Barkaoui}, Khalid and {Pozuelos}, Francisco J. and {Rackham}, Benjamin V. and {Burgasser}, Adam J. and {Triaud}, Amaury H.~M.~J. and {Serra-Ricart}, Miquel and {Timmermans}, Mathilde and {Yal{\c{c}}{\i}nkaya}, Sel{\c{c}}uk and {Soubkiou}, Abderahmane and {Stassun}, Keivan G. and et al.},
        title = "{TOI-7166 b: a habitable zone mini-Neptune planet around a nearby low-mass star}",
      journal = {\mnras},
         year = 2025,
        month = dec,
       volume = {544},
       number = {2},
        pages = {2637-2652},
          doi = {10.1093/mnras/staf1807},
archivePrefix = {arXiv},
       eprint = {2512.06841},
 primaryClass = {astro-ph.EP},
       adsurl = {https://ui.adsabs.harvard.edu/abs/2025MNRAS.544.2637B}
}

@ARTICLE{Batalha2019,
       author = {{Batalha}, Natasha E. and {Lewis}, Taylor and {Fortney}, Jonathan J. and {Batalha}, Natalie M. and {Kempton}, Eliza and {Lewis}, Nikole K. and {Line}, Michael R.},
        title = "{The Precision of Mass Measurements Required for Robust Atmospheric Characterization of Transiting Exoplanets}",
      journal = {\apjl},
         year = 2019,
        month = nov,
       volume = {885},
       number = {1},
          eid = {L25},
        pages = {L25},
          doi = {10.3847/2041-8213/ab4909},
archivePrefix = {arXiv},
       eprint = {1910.00076},
 primaryClass = {astro-ph.EP},
       adsurl = {https://ui.adsabs.harvard.edu/abs/2019ApJ...885L..25B}
}

@ARTICLE{belokurov_binarity,
       author = {{Belokurov}, Vasily and {Penoyre}, Zephyr and {Oh}, Semyeong and {Iorio}, Giuliano and {Hodgkin}, Simon and {Evans}, N. Wyn and {Everall}, Andrew and {Koposov}, Sergey E. and {Tout}, Christopher A. and {Izzard}, Robert and et al.},
        title = "{Unresolved stellar companions with Gaia DR2 astrometry}",
      journal = {\mnras},
         year = 2020,
        month = aug,
       volume = {496},
       number = {2},
        pages = {1922-1940},
          doi = {10.1093/mnras/staa1522},
archivePrefix = {arXiv},
       eprint = {2003.05467},
 primaryClass = {astro-ph.SR},
       adsurl = {https://ui.adsabs.harvard.edu/abs/2020MNRAS.496.1922B}
}

@ARTICLE{2024brady1450,
       author = {{Brady}, Madison and {Bean}, Jacob L. and {Seifahrt}, Andreas and {Kasper}, David and {Luque}, Rafael and {Stef{\'a}nsson}, Gu{\dj}mundur and {St{\"u}rmer}, Julian and {Charbonneau}, David and {Collins}, Karen A. and {Doty}, John P. and {Essack}, Zahra and {Fukui}, Akihiko and {Grau Horta}, Ferran and {Hedges}, Christina and {Hellier}, Coel and {Jenkins}, Jon M. and {Narita}, Norio and {Quinn}, Samuel N. and {Shporer}, Avi and {Schwarz}, Richard P. and {Seager}, Sara and {Stassun}, Keivan G. and {Striegel}, Stephanie and {Watkins}, Cristilyn N. and {Winn}, Joshua N. and {Zambelli}, Roberto},
        title = "{Early Results from the HUMDRUM Survey: A Small, Earth-mass Planet Orbits TOI-1450A}",
      journal = {\aj},
         year = 2024,
        month = aug,
       volume = {168},
       number = {2},
          eid = {67},
        pages = {67},
          doi = {10.3847/1538-3881/ad500a},
archivePrefix = {arXiv},
       eprint = {2405.10400},
 primaryClass = {astro-ph.EP},
       adsurl = {https://ui.adsabs.harvard.edu/abs/2024AJ....168...67B}
}

@ARTICLE{browndwarf,
       author = {{Carmichael}, Theron W. and {Giacalone}, Steven and {Vowell}, Noah and {Huber}, Daniel and {Wang}, Xian-Yu and {Bossett}, Malik and {Handley}, Luke and {Householder}, Aaron and {Li}, Yaguang and {Fulton}, Benjamin J. and et al.},
        title = "{The OATMEAL Survey. II. The 3D Spin─Orbit Obliquity of an Eccentric Transiting Brown Dwarf in the Ruprecht 147 Open Cluster}",
      journal = {\aj},
         year = 2026,
        month = jan,
       volume = {171},
       number = {1},
          eid = {12},
        pages = {12},
          doi = {10.3847/1538-3881/ae19e8},
archivePrefix = {arXiv},
       eprint = {2506.18971},
 primaryClass = {astro-ph.EP},
       adsurl = {https://ui.adsabs.harvard.edu/abs/2026AJ....171...12C}
}

@ARTICLE{Chabrier1997,
       author = {{Chabrier}, Gilles and {Baraffe}, Isabelle},
        title = "{Structure and evolution of low-mass stars}",
      journal = {\aap},
         year = 1997,
        month = nov,
       volume = {327},
        pages = {1039-1053},
          doi = {10.48550/arXiv.astro-ph/9704118},
archivePrefix = {arXiv},
       eprint = {astro-ph/9704118},
 primaryClass = {astro-ph},
       adsurl = {https://ui.adsabs.harvard.edu/abs/1997A&A...327.1039C}
}

@ARTICLE{2020Changeat,
       author = {{Changeat}, Q. and {Keyte}, L. and {Waldmann}, I.~P. and {Tinetti}, G.},
        title = "{Impact of Planetary Mass Uncertainties on Exoplanet Atmospheric Retrievals}",
      journal = {\apj},
         year = 2020,
        month = jun,
       volume = {896},
       number = {2},
          eid = {107},
        pages = {107},
          doi = {10.3847/1538-4357/ab8f8b},
archivePrefix = {arXiv},
       eprint = {1908.06305},
 primaryClass = {astro-ph.EP},
       adsurl = {https://ui.adsabs.harvard.edu/abs/2020ApJ...896..107C}
}

@ARTICLE{Charnay22,
       author = {{Charnay}, Benjamin and {Mendon{\c{c}}a}, Jo{\~a}o M. and {Kreidberg}, Laura and {Cowan}, Nicolas B. and {Taylor}, Jake and {Bell}, Taylor J. and {Demangeon}, Olivier and {Edwards}, Billy and {Haswell}, Carole A. and {Morello}, Giuseppe and et al.},
        title = "{A survey of exoplanet phase curves with Ariel}",
      journal = {Experimental Astronomy},
         year = 2022,
        month = apr,
       volume = {53},
       number = {2},
        pages = {417-446},
          doi = {10.1007/s10686-021-09715-x},
archivePrefix = {arXiv},
       eprint = {2102.06523},
 primaryClass = {astro-ph.EP},
       adsurl = {https://ui.adsabs.harvard.edu/abs/2022ExA....53..417C}
}

@ARTICLE{Chen2017,
       author = {{Chen}, Jingjing and {Kipping}, David},
        title = "{Probabilistic Forecasting of the Masses and Radii of Other Worlds}",
      journal = {\apj},
         year = 2017,
        month = jan,
       volume = {834},
       number = {1},
          eid = {17},
        pages = {17},
          doi = {10.3847/1538-4357/834/1/17},
archivePrefix = {arXiv},
       eprint = {1603.08614},
 primaryClass = {astro-ph.EP},
       adsurl = {https://ui.adsabs.harvard.edu/abs/2017ApJ...834...17C}
}

@ARTICLE{Chontos2022,
       author = {{Chontos}, Ashley and {Murphy}, Joseph M. Akana and {MacDougall}, Mason G. and {Fetherolf}, Tara and {Van Zandt}, Judah and {Rubenzahl}, Ryan A. and {Beard}, Corey and {Huber}, Daniel and {Batalha}, Natalie M. and {Crossfield}, Ian J.~M. and et al.},
        title = "{The TESS-Keck Survey: Science Goals and Target Selection}",
      journal = {\aj},
         year = 2022,
        month = jun,
       volume = {163},
       number = {6},
          eid = {297},
        pages = {297},
          doi = {10.3847/1538-3881/ac6266},
archivePrefix = {arXiv},
       eprint = {2106.06156},
 primaryClass = {astro-ph.EP},
       adsurl = {https://ui.adsabs.harvard.edu/abs/2022AJ....163..297C}
}

@ARTICLE{Clark26,
       author = {{Clark}, Catherine A. and {van Belle}, Gerard T. and {Ciardi}, David R. and {Lund}, Michael B. and {Howell}, Steve B. and {Everett}, Mark E. and {Beichman}, Charles A. and {Winters}, Jennifer G.},
        title = "{A Dearth of Close-in Stellar Companions to M-dwarf TESS Objects of Interest}",
      journal = {\aj},
         year = 2022,
        month = may,
       volume = {163},
       number = {5},
          eid = {232},
        pages = {232},
          doi = {10.3847/1538-3881/ac6101},
archivePrefix = {arXiv},
       eprint = {2203.12795},
 primaryClass = {astro-ph.SR},
       adsurl = {https://ui.adsabs.harvard.edu/abs/2022AJ....163..232C}
}

@ARTICLE{Cloutier2019,
       author = {{Cloutier}, Ryan},
        title = "{The Independent Discovery of Planet Candidates around Low-mass Stars and Astrophysical False Positives from the First Two TESS Sectors}",
      journal = {\aj},
         year = 2019,
        month = aug,
       volume = {158},
       number = {2},
          eid = {81},
        pages = {81},
          doi = {10.3847/1538-3881/ab27be},
archivePrefix = {arXiv},
       eprint = {1812.08145},
 primaryClass = {astro-ph.EP},
       adsurl = {https://ui.adsabs.harvard.edu/abs/2019AJ....158...81C}
}

@ARTICLE{2024cointepas663,
       author = {{Cointepas}, M. and {Bouchy}, F. and {Almenara}, J.~M. and {Bonfils}, X. and {Astudillo-Defru}, N. and {Knierim}, H. and {Stalport}, M. and {Mignon}, L. and {Grieves}, N. and {Bean}, J. and {Brady}, M. and {Burt}, J. and {Canto Martins}, B.~L. and {Collins}, K.~A. and {Collins}, K.~I. and {Delfosse}, X. and {de Medeiros}, J.~R. and {Demory}, B. -O. and {Dorn}, C. and {Forveille}, T. and {Fukui}, A. and {Gan}, T. and {G{\'o}mez Maqueo Chew}, Y. and {Halverson}, S. and {Helled}, R. and {Helm}, I. and {Hirano}, T. and {Horne}, K. and {Howell}, S.~B. and {Isogai}, K. and {Kasper}, D. and {Kawauchi}, K. and {Livingston}, J.~H. and {Massey}, B. and {Matson}, R.~A. and {Murgas}, F. and {Narita}, N. and {Palle}, E. and {Relles}, H.~M. and {Sabin}, L. and {Schanche}, N. and {Schwarz}, R.~P. and {Seifahrt}, A. and {Shporer}, A. and {Stefansson}, G. and {Sturmer}, J. and {Tamura}, M. and {Tan}, T. -G. and {Twicken}, J.~D. and {Watanabe}, N. and {Wells}, R.~D. and {Wilkin}, F.~P. and {Ricker}, G.~R. and {Seager}, S. and {Winn}, J.~N. and {Jenkins}, J.~M.},
        title = "{TOI-663: A newly discovered multi-planet system with three transiting mini-Neptunes orbiting an early M star}",
      journal = {\aap},
         year = 2024,
        month = may,
       volume = {685},
          eid = {A19},
        pages = {A19},
          doi = {10.1051/0004-6361/202346899},
       adsurl = {https://ui.adsabs.harvard.edu/abs/2024A&A...685A..19C}
}

@ARTICLE{Collier2026,
       author = {{Collier}, Michael and {Giacalone}, Steven and {Derfer}, Brian and {Ciardi}, David R. and {Clark}, Catherine A. and {Crossfield}, Ian J.~M. and {Deveny}, Sarah and {Fajardo-Acosta}, Sergio B. and {Furlan}, Elise and {Gilbert}, Emily A. and et al.},
        title = "{Assessing the Impact of High-Resolution Imaging on Statistical Validation of TESS Planet Candidates}",
      journal = {arXiv e-prints},
         year = 2026,
        month = aug,
          eid = {arXiv:2608.21560},
        pages = {arXiv:2608.21560},
          doi = {10.48550/arXiv.2608.21560},
archivePrefix = {arXiv},
       eprint = {2608.21560},
 primaryClass = {astro-ph.EP},
       adsurl = {https://ui.adsabs.harvard.edu/abs/2026arXiv260821560C}
}

@INPROCEEDINGS{HARPSN,
       author = {{Cosentino}, Rosario and {Lovis}, Christophe and {Pepe}, Francesco and {Collier Cameron}, Andrew and {Latham}, David W. and {Molinari}, Emilio and {Udry}, Stephane and {Bezawada}, Naidu and {Black}, Martin and {Born}, Andy and {Buchschacher}, Nicolas and {Charbonneau}, Dave and {Figueira}, Pedro and {Fleury}, Michel and {Galli}, Alberto and {Gallie}, Angus and {Gao}, Xiaofeng and {Ghedina}, Adriano and {Gonzalez}, Carlos and {Gonzalez}, Manuel and {Guerra}, Jose and {Henry}, David and {Horne}, Keith and {Hughes}, Ian and {Kelly}, Dennis and {Lodi}, Marcello and {Lunney}, David and {Maire}, Charles and {Mayor}, Michel and {Micela}, Giusi and {Ordway}, Mark P. and {Peacock}, John and {Phillips}, David and {Piotto}, Giampaolo and {Pollacco}, Don and {Queloz}, Didier and {Rice}, Ken and {Riverol}, Carlos and {Riverol}, Luis and {San Juan}, Jose and {Sasselov}, Dimitar and {Segransan}, Damien and {Sozzetti}, Alessandro and {Sosnowska}, Danuta and {Stobie}, Brian and {Szentgyorgyi}, Andrew and {Vick}, Andy and {Weber}, Luc},
        title = "{Harps-N: the new planet hunter at TNG}",
    booktitle = {Ground-based and Airborne Instrumentation for Astronomy IV},
         year = 2012,
       editor = {{McLean}, Ian S. and {Ramsay}, Suzanne K. and {Takami}, Hideki},
       series = {Society of Photo-Optical Instrumentation Engineers (SPIE) Conference Series},
       volume = {8446},
        month = sep,
          eid = {84461V},
        pages = {84461V},
          doi = {10.1117/12.925738},
       adsurl = {https://ui.adsabs.harvard.edu/abs/2012SPIE.8446E..1VC}
}

@ARTICLE{2025Crossfield,
       author = {{Crossfield}, Ian J.~M. and {Polanski}, Alex S. and {Robertson}, Paul and {Murphy}, Joseph Akana and {Turtelboom}, Emma V. and {Luque}, Rafael and {Beatty}, Thomas and {Daylan}, Tansu and {Isaacson}, Howard and {Brande}, Jonathan and {Kreidberg}, Laura and {Batalha}, Natalie M. and {Huber}, Daniel and {Rhem}, Maleah and {Dressing}, Courtney and {Kane}, Stephen R. and {Bossett}, Malik and {Gagnebin}, Anna and {Kroft}, Maxwell A. and {Premnath}, Pranav H. and {Rogers}, Claire J. and {Collins}, Karen A. and {Latham}, David W. and {Watkins}, Cristilyn N. and {Ciardi}, David R. and {Howell}, Steve B. and {Savel}, Arjun B. and {Berlind}, Perry and {Calkins}, Michael L. and {Esquerdo}, Gilbert A. and {Mink}, Jessica and {Clark}, Catherine A. and {Lund}, Michael B. and {Matson}, Rachel A. and {Everett}, Mark E. and {Schlieder}, Joshua E. and {Matthews}, Elisabeth C. and {Giacalone}, Steven and {Barclay}, Thomas and {Zambelli}, Roberto and {Plavchan}, Peter and {Ellingson}, Taylor and {Bowen}, Michael and {Srdoc}, Gregor and {McLeod}, Kim K. and {Schwarz}, Richard P. and {Barkaoui}, Khalid and {Kamler}, Jacob and {Murgas}, Felipe and {Palle}, Enric and {Narita}, Norio and {Fukui}, Akihiko and {Relles}, Howard M. and {Bieryla}, Allyson and {Girardin}, Eric and {Massey}, Bob and {Stockdale}, Chris and {Lewin}, Pablo and {Papini}, Riccardo and {Guerra}, Pere and {Conti}, Dennis M. and {Yal{\c{c}}inkaya}, Sel{\c{c}}uk and {Ba{\c{s}}t{\"u}rk}, {\"O}zg{\"u}r and {Mourad}, Ghachoui},
        title = "{OrCAS: Origins, Compositions, and Atmospheres of Sub-Neptunes. I. Survey Definition}",
      journal = {\aj},
         year = 2025,
        month = feb,
       volume = {169},
       number = {2},
          eid = {89},
        pages = {89},
          doi = {10.3847/1538-3881/ad9aa6},
archivePrefix = {arXiv},
       eprint = {2411.16836},
 primaryClass = {astro-ph.EP},
       adsurl = {https://ui.adsabs.harvard.edu/abs/2025AJ....169...89C}
}

@ARTICLE{2024Dai,
       author = {{Dai}, Fei and {Howard}, Andrew W. and {Halverson}, Samuel and {Orell-Miquel}, Jaume and {Pall{\'e}}, Enric and {Isaacson}, Howard and {Fulton}, Benjamin and {Price}, Ellen M. and {Plotnykov}, Mykhaylo and {Rogers}, Leslie A. and {Valencia}, Diana and {Paragas}, Kimberly and {Greklek-McKeon}, Michael and {Gomez Barrientos}, Jonathan and {Knutson}, Heather A. and {Petigura}, Erik A. and {Weiss}, Lauren M. and {Lee}, Rena and {Brinkman}, Casey L. and {Huber}, Daniel and {Stef{\'a}nsson}, Gumundur and {Masuda}, Kento and {Giacalone}, Steven and {Lu}, Cicero X. and {Kite}, Edwin S. and {Hu}, Renyu and {Gaidos}, Eric and {Zhang}, Michael and {Rubenzahl}, Ryan A. and {Winn}, Joshua N. and {Han}, Te and {Beard}, Corey and {Holcomb}, Rae and {Householder}, Aaron and {Gilbert}, Gregory J. and {Lubin}, Jack and {Ong}, J.~M. Joel and {Polanski}, Alex S. and {Saunders}, Nicholas and {Van Zandt}, Judah and {Yee}, Samuel W. and {Zhang}, Jingwen and {Zink}, Jon and {Holden}, Bradford and {Baker}, Ashley and {Brodheim}, Max and {Crossfield}, Ian J.~M. and {Deich}, William and {Edelstein}, Jerry and {Gibson}, Steven R. and {Hill}, Grant M. and {Jelinsky}, Sharon R. and {Kassis}, Marc and {Laher}, Russ R. and {Lanclos}, Kyle and {Lilley}, Scott and {Payne}, Joel N. and {Rider}, Kodi and {Robertson}, Paul and {Roy}, Arpita and {Schwab}, Christian and {Shaum}, Abby P. and {Sirk}, Martin M. and {Smith}, Chris and {Vandenberg}, Adam and {Walawender}, Josh and {Wang}, Sharon X. and {Wang}, Shin-Ywan (Cindy) and {Wishnow}, Edward and {Wright}, Jason T. and {Yeh}, Sherry and {Caballero}, Jos{\'e} A. and {Morales}, Juan C. and {Murgas}, Felipe and {Nagel}, Evangelos and {Reiners}, Ansgar and {Schweitzer}, Andreas and {Tabernero}, Hugo M. and {Zechmeister}, Mathias and {Spencer}, Alton and {Ciardi}, David R. and {Clark}, Catherine A. and {Lund}, Michael B. and {Caldwell}, Douglas A. and {Collins}, Karen A. and {Schwarz}, Richard P. and {Barkaoui}, Khalid and {Watkins}, Cristilyn and {Shporer}, Avi and {Narita}, Norio and {Fukui}, Akihiko and {Srdoc}, Gregor and {Latham}, David W. and {Jenkins}, Jon M. and {Ricker}, George R. and {Seager}, Sara and {Vanderspek}, Roland},
        title = "{An Earth-sized Planet on the Verge of Tidal Disruption}",
      journal = {\aj},
         year = 2024,
        month = sep,
       volume = {168},
       number = {3},
          eid = {101},
        pages = {101},
          doi = {10.3847/1538-3881/ad5a7d},
archivePrefix = {arXiv},
       eprint = {2407.21167},
 primaryClass = {astro-ph.EP},
       adsurl = {https://ui.adsabs.harvard.edu/abs/2024AJ....168..101D}
}

@ARTICLE{DiMaio2023,
       author = {{Di Maio}, C. and {Changeat}, Q. and {Benatti}, S. and {Micela}, G.},
        title = "{Analysis of the planetary mass uncertainties on the accuracy of atmospherical retrieval}",
      journal = {\aap},
         year = 2023,
        month = jan,
       volume = {669},
          eid = {A150},
        pages = {A150},
          doi = {10.1051/0004-6361/202244881},
archivePrefix = {arXiv},
       eprint = {2211.02897},
 primaryClass = {astro-ph.EP},
       adsurl = {https://ui.adsabs.harvard.edu/abs/2023A&A...669A.150D}
}

@ARTICLE{dimaio2026,
       author = {{Di Maio}, C. and {Benatti}, S. and {Locci}, D. and {Spinelli}, R. and {Baratella}, M. and {Biazzo}, K. and {Maldonado}, J. and {Lanza}, A.~F. and {Dorn}, C. and {Cubillos}, P.~E. and et al.},
        title = "{The GAPS Programme at TNG: LXXIII. Confirmation of the hot sub-Neptune TOI-4602 b (HD 25295 b), a key target for future atmospheric characterization}",
      journal = {\aap},
         year = 2026,
        month = may,
       volume = {709},
          eid = {A270},
        pages = {A270},
          doi = {10.1051/0004-6361/202659402},
archivePrefix = {arXiv},
       eprint = {2604.11728},
 primaryClass = {astro-ph.EP},
       adsurl = {https://ui.adsabs.harvard.edu/abs/2026A&A...709A.270D}
}

@ARTICLE{2025DiTomasso,
       author = {{DiTomasso}, Victoria and {L{\'o}pez-Morales}, Mercedes and {Peacock}, Sarah and {Malavolta}, Luca and {Kirk}, James and {Stevenson}, Kevin B. and {Fu}, Guangwei and {Lustig-Yaeger}, Jacob},
        title = "{The Magnetically Induced Radial Velocity Variation of Gliese 341 and an Upper Limit to the Mass of Its Transiting Earth-sized Planet}",
      journal = {\apj},
         year = 2025,
        month = feb,
       volume = {979},
       number = {2},
          eid = {214},
        pages = {214},
          doi = {10.3847/1538-4357/ad9dd3},
archivePrefix = {arXiv},
       eprint = {2502.07867},
 primaryClass = {astro-ph.EP},
       adsurl = {https://ui.adsabs.harvard.edu/abs/2025ApJ...979..214D}
}

@ARTICLE{Doyle_880,
       author = {{Doyle}, Lauren and {Armstrong}, David J. and {Acu{\~n}a}, Lorena and {Osborn}, Ares and {Sousa}, S{\'e}rgio A.~G. and {Castro-Gonz{\'a}lez}, Amadeo and {Bourrier}, Vincent and {Alves}, Douglas and {Barrado}, David and {Barros}, Susana C.~C. and et al.},
        title = "{Exploring the Neptunian desert: insights from a homogeneous planetary sample}",
      journal = {\mnras},
         year = 2025,
        month = jun,
       volume = {539},
       number = {4},
        pages = {3138-3156},
          doi = {10.1093/mnras/staf670},
archivePrefix = {arXiv},
       eprint = {2504.16164},
 primaryClass = {astro-ph.EP},
       adsurl = {https://ui.adsabs.harvard.edu/abs/2025MNRAS.539.3138D}
}

@ARTICLE{Edwards2019,
       author = {{Edwards}, Billy and {Mugnai}, Lorenzo and {Tinetti}, Giovanna and {Pascale}, Enzo and {Sarkar}, Subhajit},
        title = "{An Updated Study of Potential Targets for Ariel}",
      journal = {\aj},
         year = 2019,
        month = jun,
       volume = {157},
       number = {6},
          eid = {242},
        pages = {242},
          doi = {10.3847/1538-3881/ab1cb9},
archivePrefix = {arXiv},
       eprint = {1905.04959},
 primaryClass = {astro-ph.EP},
       adsurl = {https://ui.adsabs.harvard.edu/abs/2019AJ....157..242E}
}

@ARTICLE{targetlist,
       author = {{Edwards}, Billy and {Tinetti}, Giovanna},
        title = "{The Ariel Target List: The Impact of TESS and the Potential for Characterizing Multiple Planets within a System}",
      journal = {\aj},
         year = 2022,
        month = jul,
       volume = {164},
       number = {1},
          eid = {15},
        pages = {15},
          doi = {10.3847/1538-3881/ac6bf9},
archivePrefix = {arXiv},
       eprint = {2205.05073},
 primaryClass = {astro-ph.EP},
       adsurl = {https://ui.adsabs.harvard.edu/abs/2022AJ....164...15E}
}

@ARTICLE{latte,
       author = {{Eisner}, Nora and {Lintott}, Chris and {Aigrain}, Suzanne},
        title = "{LATTE: Lightcurve Analysis Tool for Transiting Exoplanets}",
      journal = {The Journal of Open Source Software},
         year = 2020,
        month = may,
       volume = {5},
       number = {49},
          eid = {2101},
        pages = {2101},
          doi = {10.21105/joss.02101},
       adsurl = {https://ui.adsabs.harvard.edu/abs/2020JOSS....5.2101E}
}

@ARTICLE{Filomeno_5734,
       author = {{Filomeno}, S. and {Trifonov}, T. and {Damasso}, M. and {Baratella}, M. and {Benatti}, S. and {Biazzo}, K. and {Collins}, K.~A. and {Cosentino}, R. and {Desidera}, S. and {Di Maio}, C. and et al.},
        title = "{The GAPS Programme at the TNG: LXXII. TOI-5734b: A hot sub-Neptune orbiting a relatively young K dwarf with an Earth-like density}",
      journal = {\aap},
         year = 2026,
        month = mar,
       volume = {708},
          eid = {A90},
        pages = {A90},
          doi = {10.1051/0004-6361/202555931},
archivePrefix = {arXiv},
       eprint = {2602.18108},
 primaryClass = {astro-ph.EP},
       adsurl = {https://ui.adsabs.harvard.edu/abs/2026A&A...708A..90F}
}

@ARTICLE{Fressin2013,
       author = {{Fressin}, Fran{\c{c}}ois and {Torres}, Guillermo and {Charbonneau}, David and {Bryson}, Stephen T. and {Christiansen}, Jessie and {Dressing}, Courtney D. and {Jenkins}, Jon M. and {Walkowicz}, Lucianne M. and {Batalha}, Natalie M.},
        title = "{The False Positive Rate of Kepler and the Occurrence of Planets}",
      journal = {\apj},
         year = 2013,
        month = apr,
       volume = {766},
       number = {2},
          eid = {81},
        pages = {81},
          doi = {10.1088/0004-637X/766/2/81},
archivePrefix = {arXiv},
       eprint = {1301.0842},
 primaryClass = {astro-ph.EP},
       adsurl = {https://ui.adsabs.harvard.edu/abs/2013ApJ...766...81F}
}

@ARTICLE{Gaia2016,
       author = {{Gaia Collaboration} and {Prusti}, T. and {de Bruijne}, J.~H.~J. and {Brown}, A.~G.~A. and {Vallenari}, A. and {Babusiaux}, C. and {Bailer-Jones}, C.~A.~L. and {Bastian}, U. and {Biermann}, M. and {Evans}, D.~W. and {Eyer}, L. and {Jansen}, F. and {Jordi}, C. and {Klioner}, S.~A. and {Lammers}, U. and {Lindegren}, L. and {Luri}, X. and {Mignard}, F. and {Milligan}, D.~J. and {Panem}, C. and {Poinsignon}, V. and {Pourbaix}, D. and {Randich}, S. and {Sarri}, G. and {Sartoretti}, P. and {Siddiqui}, H.~I. and {Soubiran}, C. and {Valette}, V. and {van Leeuwen}, F. and {Walton}, N.~A. and {Aerts}, C. and {Arenou}, F. and {Cropper}, M. and {Drimmel}, R. and {H{\o}g}, E. and {Katz}, D. and {Lattanzi}, M.~G. and {O'Mullane}, W. and {Grebel}, E.~K. and {Holland}, A.~D. and {Huc}, C. and {Passot}, X. and {Bramante}, L. and {Cacciari}, C. and {Casta{\~n}eda}, J. and {Chaoul}, L. and {Cheek}, N. and {De Angeli}, F. and {Fabricius}, C. and {Guerra}, R. and {Hern{\'a}ndez}, J. and {Jean-Antoine-Piccolo}, A. and {Masana}, E. and {Messineo}, R. and {Mowlavi}, N. and {Nienartowicz}, K. and {Ord{\'o}{\~n}ez-Blanco}, D. and {Panuzzo}, P. and {Portell}, J. and {Richards}, P.~J. and {Riello}, M. and {Seabroke}, G.~M. and {Tanga}, P. and {Th{\'e}venin}, F. and {Torra}, J. and {Els}, S.~G. and {Gracia-Abril}, G. and {Comoretto}, G. and {Garcia-Reinaldos}, M. and {Lock}, T. and {Mercier}, E. and {Altmann}, M. and {Andrae}, R. and {Astraatmadja}, T.~L. and {Bellas-Velidis}, I. and {Benson}, K. and {Berthier}, J. and {Blomme}, R. and {Busso}, G. and {Carry}, B. and {Cellino}, A. and {Clementini}, G. and {Cowell}, S. and {Creevey}, O. and {Cuypers}, J. and {Davidson}, M. and {De Ridder}, J. and {de Torres}, A. and {Delchambre}, L. and {Dell'Oro}, A. and {Ducourant}, C. and {Fr{\'e}mat}, Y. and {Garc{\'\i}a-Torres}, M. and {Gosset}, E. and {Halbwachs}, J.-L. and {Hambly}, N.~C. and {Harrison}, D.~L. and {Hauser}, M. and {Hestroffer}, D. and {Hodgkin}, S.~T. and {Huckle}, H.~E. and {Hutton}, A. and {Jasniewicz}, G. and {Jordan}, S. and {Kontizas}, M. and {Korn}, A.~J. and {Lanzafame}, A.~C. and {Manteiga}, M. and {Moitinho}, A. and {Muinonen}, K. and {Osinde}, J. and {Pancino}, E. and {Pauwels}, T. and {Petit}, J.-M. and {Recio-Blanco}, A. and {Robin}, A.~C. and {Sarro}, L.~M. and {Siopis}, C. and {Smith}, M. and {Smith}, K.~W. and {Sozzetti}, A. and {Thuillot}, W. and {van Reeven}, W. and {Viala}, Y. and {Abbas}, U. and {Abreu Aramburu}, A. and {Accart}, S. and {Aguado}, J.~J. and {Allan}, P.~M. and {Allasia}, W. and {Altavilla}, G. and {{\'A}lvarez}, M.~A. and {Alves}, J. and {Anderson}, R.~I. and {Andrei}, A.~H. and {Anglada Varela}, E. and {Antiche}, E. and {Antoja}, T. and {Ant{\'o}n}, S. and {Arcay}, B. and {Atzei}, A. and {Ayache}, L. and {Bach}, N. and {Baker}, S.~G. and {Balaguer-N{\'u}{\~n}ez}, L. and {Barache}, C. and {Barata}, C. and {Barbier}, A. and {Barblan}, F. and {Baroni}, M. and {Barrado y Navascu{\'e}s}, D. and {Barros}, M. and {Barstow}, M.~A. and {Becciani}, U. and {Bellazzini}, M. and {Bellei}, G. and {Bello Garc{\'\i}a}, A. and {Belokurov}, V. and {Bendjoya}, P. and {Berihuete}, A. and {Bianchi}, L. and {Bienaym{\'e}}, O. and {Billebaud}, F. and {Blagorodnova}, N. and {Blanco-Cuaresma}, S. and {Boch}, T. and {Bombrun}, A. and {Borrachero}, R. and {Bouquillon}, S. and {Bourda}, G. and {Bouy}, H. and {Bragaglia}, A. and {Breddels}, M.~A. and {Brouillet}, N. and {Br{\"u}semeister}, T. and {Bucciarelli}, B. and {Budnik}, F. and {Burgess}, P. and {Burgon}, R. and {Burlacu}, A. and {Busonero}, D. and {Buzzi}, R. and {Caffau}, E. and {Cambras}, J. and {Campbell}, H. and {Cancelliere}, R. and {Cantat-Gaudin}, T. and {Carlucci}, T. and {Carrasco}, J.~M. and {Castellani}, M. and {Charlot}, P. and {Charnas}, J. and {Charvet}, P. and {Chassat}, F. and {Chiavassa}, A. and {Clotet}, M. and {Cocozza}, G. and {Collins}, R.~S. and {Collins}, P. and {Costigan}, G.},
        title = "{The Gaia mission}",
      journal = {\aap},
         year = 2016,
        month = nov,
       volume = {595},
          eid = {A1},
        pages = {A1},
          doi = {10.1051/0004-6361/201629272},
archivePrefix = {arXiv},
       eprint = {1609.04153},
 primaryClass = {astro-ph.IM},
       adsurl = {https://ui.adsabs.harvard.edu/abs/2016A&A...595A...1G}
}

@ARTICLE{GDR2,
       author = {{Gaia Collaboration} and {Brown}, A.~G.~A. and {Vallenari}, A. and {Prusti}, T. and {de Bruijne}, J.~H.~J. and {Babusiaux}, C. and {Bailer-Jones}, C.~A.~L. and {Biermann}, M. and {Evans}, D.~W. and {Eyer}, L. and et al.},
        title = "{Gaia Data Release 2. Summary of the contents and survey properties}",
      journal = {\aap},
         year = 2018,
        month = aug,
       volume = {616},
          eid = {A1},
        pages = {A1},
          doi = {10.1051/0004-6361/201833051},
archivePrefix = {arXiv},
       eprint = {1804.09365},
 primaryClass = {astro-ph.GA},
       adsurl = {https://ui.adsabs.harvard.edu/abs/2018A&A...616A...1G}
}

@ARTICLE{gaia_IPD,
       author = {{Gaia Collaboration} and {Arenou}, F. and {Babusiaux}, C. and {Barstow}, M.~A. and {Faigler}, S. and {Jorissen}, A. and {Kervella}, P. and {Mazeh}, T. and {Mowlavi}, N. and {Panuzzo}, P. and et al.},
        title = "{Gaia Data Release 3. Stellar multiplicity, a teaser for the hidden treasure}",
      journal = {\aap},
         year = 2023,
        month = jun,
       volume = {674},
          eid = {A34},
        pages = {A34},
          doi = {10.1051/0004-6361/202243782},
archivePrefix = {arXiv},
       eprint = {2206.05595},
 primaryClass = {astro-ph.SR},
       adsurl = {https://ui.adsabs.harvard.edu/abs/2023A&A...674A..34G}
}

@ARTICLE{Gaia2023,
       author = {{Gaia Collaboration} and {Vallenari}, A. and {Brown}, A.~G.~A. and {Prusti}, T. and {de Bruijne}, J.~H.~J. and {Arenou}, F. and {Babusiaux}, C. and {Biermann}, M. and {Creevey}, O.~L. and {Ducourant}, C. and {Evans}, D.~W. and {Eyer}, L. and {Guerra}, R. and {Hutton}, A. and {Jordi}, C. and {Klioner}, S.~A. and {Lammers}, U.~L. and {Lindegren}, L. and {Luri}, X. and {Mignard}, F. and {Panem}, C. and {Pourbaix}, D. and {Randich}, S. and {Sartoretti}, P. and {Soubiran}, C. and {Tanga}, P. and {Walton}, N.~A. and {Bailer-Jones}, C.~A.~L. and {Bastian}, U. and {Drimmel}, R. and {Jansen}, F. and {Katz}, D. and {Lattanzi}, M.~G. and {van Leeuwen}, F. and {Bakker}, J. and {Cacciari}, C. and {Casta{\~n}eda}, J. and {De Angeli}, F. and {Fabricius}, C. and {Fouesneau}, M. and {Fr{\'e}mat}, Y. and {Galluccio}, L. and {Guerrier}, A. and {Heiter}, U. and {Masana}, E. and {Messineo}, R. and {Mowlavi}, N. and {Nicolas}, C. and {Nienartowicz}, K. and {Pailler}, F. and {Panuzzo}, P. and {Riclet}, F. and {Roux}, W. and {Seabroke}, G.~M. and {Sordo}, R. and {Th{\'e}venin}, F. and {Gracia-Abril}, G. and {Portell}, J. and {Teyssier}, D. and {Altmann}, M. and {Andrae}, R. and {Audard}, M. and {Bellas-Velidis}, I. and {Benson}, K. and {Berthier}, J. and {Blomme}, R. and {Burgess}, P.~W. and {Busonero}, D. and {Busso}, G. and {C{\'a}novas}, H. and {Carry}, B. and {Cellino}, A. and {Cheek}, N. and {Clementini}, G. and {Damerdji}, Y. and {Davidson}, M. and {de Teodoro}, P. and {Nu{\~n}ez Campos}, M. and {Delchambre}, L. and {Dell'Oro}, A. and {Esquej}, P. and {Fern{\'a}ndez-Hern{\'a}ndez}, J. and {Fraile}, E. and {Garabato}, D. and {Garc{\'\i}a-Lario}, P. and {Gosset}, E. and {Haigron}, R. and {Halbwachs}, J.-L. and {Hambly}, N.~C. and {Harrison}, D.~L. and {Hern{\'a}ndez}, J. and {Hestroffer}, D. and {Hodgkin}, S.~T. and {Holl}, B. and {Jan{\ss}en}, K. and {Jevardat de Fombelle}, G. and {Jordan}, S. and {Krone-Martins}, A. and {Lanzafame}, A.~C. and {L{\"o}ffler}, W. and {Marchal}, O. and {Marrese}, P.~M. and {Moitinho}, A. and {Muinonen}, K. and {Osborne}, P. and {Pancino}, E. and {Pauwels}, T. and {Recio-Blanco}, A. and {Reyl{\'e}}, C. and {Riello}, M. and {Rimoldini}, L. and {Roegiers}, T. and {Rybizki}, J. and {Sarro}, L.~M. and {Siopis}, C. and {Smith}, M. and {Sozzetti}, A. and {Utrilla}, E. and {van Leeuwen}, M. and {Abbas}, U. and {{\'A}brah{\'a}m}, P. and {Abreu Aramburu}, A. and {Aerts}, C. and {Aguado}, J.~J. and {Ajaj}, M. and {Aldea-Montero}, F. and {Altavilla}, G. and {{\'A}lvarez}, M.~A. and {Alves}, J. and {Anders}, F. and {Anderson}, R.~I. and {Anglada Varela}, E. and {Antoja}, T. and {Baines}, D. and {Baker}, S.~G. and {Balaguer-N{\'u}{\~n}ez}, L. and {Balbinot}, E. and {Balog}, Z. and {Barache}, C. and {Barbato}, D. and {Barros}, M. and {Barstow}, M.~A. and {Bartolom{\'e}}, S. and {Bassilana}, J.-L. and {Bauchet}, N. and {Becciani}, U. and {Bellazzini}, M. and {Berihuete}, A. and {Bernet}, M. and {Bertone}, S. and {Bianchi}, L. and {Binnenfeld}, A. and {Blanco-Cuaresma}, S. and {Blazere}, A. and {Boch}, T. and {Bombrun}, A. and {Bossini}, D. and {Bouquillon}, S. and {Bragaglia}, A. and {Bramante}, L. and {Breedt}, E. and {Bressan}, A. and {Brouillet}, N. and {Brugaletta}, E. and {Bucciarelli}, B. and {Burlacu}, A. and {Butkevich}, A.~G. and {Buzzi}, R. and {Caffau}, E. and {Cancelliere}, R. and {Cantat-Gaudin}, T. and {Carballo}, R. and {Carlucci}, T. and {Carnerero}, M.~I. and {Carrasco}, J.~M. and {Casamiquela}, L. and {Castellani}, M. and {Castro-Ginard}, A. and {Chaoul}, L. and {Charlot}, P. and {Chemin}, L. and {Chiaramida}, V. and {Chiavassa}, A. and {Chornay}, N. and {Comoretto}, G. and {Contursi}, G. and {Cooper}, W.~J. and {Cornez}, T. and {Cowell}, S. and {Crifo}, F. and {Cropper}, M. and {Crosta}, M. and {Crowley}, C. and {Dafonte}, C. and {Dapergolas}, A. and {David}, M. and {David}, P. and {de Laverny}, P. and {De Luise}, F. and {De March}, R.},
        title = "{Gaia Data Release 3. Summary of the content and survey properties}",
      journal = {\aap},
         year = 2023,
        month = jun,
       volume = {674},
          eid = {A1},
        pages = {A1},
          doi = {10.1051/0004-6361/202243940},
archivePrefix = {arXiv},
       eprint = {2208.00211},
 primaryClass = {astro-ph.GA},
       adsurl = {https://ui.adsabs.harvard.edu/abs/2023A&A...674A...1G}
}

@ARTICLE{JWST,
       author = {{Gardner}, Jonathan P. and {Mather}, John C. and {Clampin}, Mark and {Doyon}, Rene and {Greenhouse}, Matthew A. and {Hammel}, Heidi B. and {Hutchings}, John B. and {Jakobsen}, Peter and {Lilly}, Simon J. and {Long}, Knox S. and et al.},
        title = "{The James Webb Space Telescope}",
      journal = {\ssr},
         year = 2006,
        month = apr,
       volume = {123},
       number = {4},
        pages = {485-606},
          doi = {10.1007/s11214-006-8315-7},
archivePrefix = {arXiv},
       eprint = {astro-ph/0606175},
 primaryClass = {astro-ph},
       adsurl = {https://ui.adsabs.harvard.edu/abs/2006SSRv..123..485G}
}

@article{2025gerald4491,
       author = {{Gerald{\'\i}a-Gonz{\'a}lez}, S. and {Orell-Miquel}, J. and {Pall{\'e}}, E. and {Murgas}, F. and {Lacedelli}, G. and {B{\'e}jar}, V.~J.~S. and {Caballero}, J.~A. and {Duque-Arribas}, C. and {Lillo-Box}, J. and {Montes}, D. and et al.},
        title = "{Discovery of a transiting hot water-world candidate orbiting Ross 176 with TESS and CARMENES}",
      journal = {\aap},
         year = 2025,
        month = aug,
       volume = {700},
          eid = {A216},
        pages = {A216},
          doi = {10.1051/0004-6361/202553719},
archivePrefix = {arXiv},
       eprint = {2507.15763},
 primaryClass = {astro-ph.EP},
       adsurl = {https://ui.adsabs.harvard.edu/abs/2025A&A...700A.216G}
}

@software{triceratopscode,
       author = {{Giacalone}, Steven and {Dressing}, Courtney D.},
        title = "{triceratops: Candidate exoplanet rating tool}",
 howpublished = {Astrophysics Source Code Library, record ascl:2002.004},
         year = 2020,
        month = feb,
          eid = {ascl:2002.004},
archivePrefix = {ascl},
       eprint = {2002.004},
       adsurl = {https://ui.adsabs.harvard.edu/abs/2020ascl.soft02004G}
}

@ARTICLE{2021giacalonevetting,
       author = {{Giacalone}, Steven and {Dressing}, Courtney D. and {Jensen}, Eric L.~N. and {Collins}, Karen A. and {Ricker}, George R. and {Vanderspek}, Roland and {Seager}, S. and {Winn}, Joshua N. and {Jenkins}, Jon M. and {Barclay}, Thomas and {Barkaoui}, Khalid and {Cadieux}, Charles and {Charbonneau}, David and {Collins}, Kevin I. and {Conti}, Dennis M. and {Doyon}, Ren{\'e} and {Evans}, Phil and {Ghachoui}, Mourad and {Gillon}, Micha{\"e}l and {Guerrero}, Natalia M. and {Hart}, Rhodes and {Jehin}, Emmanu{\"e}l and {Kielkopf}, John F. and {McLean}, Brian and {Murgas}, Felipe and {Palle}, Enric and {Parviainen}, Hannu and {Pozuelos}, Francisco J. and {Relles}, Howard M. and {Shporer}, Avi and {Socia}, Quentin and {Stockdale}, Chris and {Tan}, Thiam-Guan and {Torres}, Guillermo and {Twicken}, Joseph D. and {Waalkes}, William C. and {Waite}, Ian A.},
        title = "{Vetting of 384 TESS Objects of Interest with TRICERATOPS and Statistical Validation of 12 Planet Candidates}",
      journal = {\aj},
         year = 2021,
        month = jan,
       volume = {161},
       number = {1},
          eid = {24},
        pages = {24},
          doi = {10.3847/1538-3881/abc6af},
archivePrefix = {arXiv},
       eprint = {2002.00691},
 primaryClass = {astro-ph.EP},
       adsurl = {https://ui.adsabs.harvard.edu/abs/2021AJ....161...24G}
}

@ARTICLE{2022Giacalone,
       author = {{Giacalone}, Steven and {Dressing}, Courtney D. and {Hedges}, Christina and {Kostov}, Veselin B. and {Collins}, Karen A. and {Jensen}, Eric L.~N. and {Yahalomi}, Daniel A. and {Bieryla}, Allyson and {Ciardi}, David R. and {Howell}, Steve B. and {Lillo-Box}, Jorge and {Barkaoui}, Khalid and {Winters}, Jennifer G. and {Matthews}, Elisabeth and {Livingston}, John H. and {Quinn}, Samuel N. and {Safonov}, Boris S. and {Cadieux}, Charles and {Furlan}, E. and {Crossfield}, Ian J.~M. and {Mandell}, Avi M. and {Gilbert}, Emily A. and {Kruse}, Ethan and {Quintana}, Elisa V. and {Ricker}, George R. and {Seager}, S. and {Winn}, Joshua N. and {Jenkins}, Jon M. and {Duffy Adkins}, Britt and {Baker}, David and {Barclay}, Thomas and {Barrado}, David and {Batalha}, Natalie M. and {Belinski}, Alexander A. and {Benkhaldoun}, Zouhair and {Buchhave}, Lars A. and {Cacciapuoti}, Luca and {Charbonneau}, David and {Chontos}, Ashley and {Christiansen}, Jessie L. and {Cloutier}, Ryan and {Collins}, Kevin I. and {Conti}, Dennis M. and {Cutting}, Neil and {Dixon}, Scott and {Doyon}, Ren{\'e} and {Mufti}, Mohammed El and {Esparza-Borges}, Emma and {Essack}, Zahra and {Fukui}, Akihiko and {Gan}, Tianjun and {Gary}, Kaz and {Ghachoui}, Mourad and {Gillon}, Micha{\"e}l and {Girardin}, Eric and {Glidden}, Ana and {Gonzales}, Erica J. and {Guerra}, Pere and {Horch}, Elliott P. and {He{\l}miniak}, Krzysztof G. and {Howard}, Andrew W. and {Huber}, Daniel and {Irwin}, Jonathan M. and {Isopi}, Giovanni and {Jehin}, Emmanu{\"e}l and {Kagetani}, Taiki and {Kane}, Stephen R. and {Kawauchi}, Kiyoe and {Kielkopf}, John F. and {Lewin}, Pablo and {Luker}, Lindy and {Lund}, Michael B. and {Mallia}, Franco and {Mao}, Shude and {Massey}, Bob and {Matson}, Rachel A. and {Mireles}, Ismael and {Mori}, Mayuko and {Murgas}, Felipe and {Narita}, Norio and {O'Dwyer}, Tanner and {Petigura}, Erik A. and {Polanski}, Alex S. and {Pozuelos}, Francisco J. and {Palle}, Enric and {Parviainen}, Hannu and {Plavchan}, Peter P. and {Relles}, Howard M. and {Robertson}, Paul and {Rose}, Mark E. and {Rowden}, Pamela and {Roy}, Arpita and {Savel}, Arjun B. and {Schlieder}, Joshua E. and {Schnaible}, Chloe and {Schwarz}, Richard P. and {Sefako}, Ramatholo and {Selezneva}, Aleksandra and {Skinner}, Brett and {Stockdale}, Chris and {Strakhov}, Ivan A. and {Tan}, Thiam-Guan and {Torres}, Guillermo and {Tronsgaard}, Ren{\'e} and {Twicken}, Joseph D. and {Vermilion}, David and {Waite}, Ian A. and {Walter}, Bradley and {Wang}, Gavin and {Ziegler}, Carl and {Zou}, Yujie},
        title = "{Validation of 13 Hot and Potentially Terrestrial TESS Planets}",
      journal = {\aj},
         year = 2022,
        month = feb,
       volume = {163},
       number = {2},
          eid = {99},
        pages = {99},
          doi = {10.3847/1538-3881/ac4334},
archivePrefix = {arXiv},
       eprint = {2201.12661},
 primaryClass = {astro-ph.EP},
       adsurl = {https://ui.adsabs.harvard.edu/abs/2022AJ....163...99G}
}

@ARTICLE{Gomez_5706,
       author = {{Gomez Barrientos}, Jonathan and {Greklek-McKeon}, Michael and {Knutson}, Heather A. and {Giacalone}, Steven and {Levine}, W. Garrett and {Saidel}, Morgan and {Vissapragada}, Shreyas and {Ciardi}, David R. and {Collins}, Karen A. and {Latham}, David W. and et al.},
        title = "{Validation of TESS Planet Candidates with Multicolor Transit Photometry and TRICERATOPS+}",
      journal = {\aj},
         year = 2025,
        month = sep,
       volume = {170},
       number = {3},
          eid = {148},
        pages = {148},
          doi = {10.3847/1538-3881/ade68b},
archivePrefix = {arXiv},
       eprint = {2508.02782},
 primaryClass = {astro-ph.EP},
       adsurl = {https://ui.adsabs.harvard.edu/abs/2025AJ....170..148G}
}

@ARTICLE{toi1080,
       author = {{G{\'o}mez Maqueo Chew}, Y. and {Dransfield}, G. and {Barkaoui}, K. and {Cadieux}, C. and {Ducrot}, E. and {Rackham}, B.~V. and {Timmermans}, M. and {Burgasser}, A.~J. and {Segura}, A. and {Stassun}, K.~G. and et al.},
        title = "{TOI-1080 b: a temperate, rocky planet orbiting a quiet M4V host}",
      journal = {\mnras},
         year = 2026,
        month = may,
       volume = {548},
       number = {1},
          eid = {stag438},
        pages = {stag438},
          doi = {10.1093/mnras/stag438},
archivePrefix = {arXiv},
       eprint = {2603.00385},
 primaryClass = {astro-ph.EP},
       adsurl = {https://ui.adsabs.harvard.edu/abs/2026MNRAS.548ag438G}
}

@article{Guerrero_2021,
doi = {10.3847/1538-4365/abefe1},
url = {https://doi.org/10.3847/1538-4365/abefe1},
year = {2021},
month = {jun},
publisher = {The American Astronomical Society},
volume = {254},
number = {2},
pages = {39},
author = {Guerrero, Natalia M. and Seager, S. and Huang, Chelsea X. and Vanderburg, Andrew and Soto, Aylin Garcia and Mireles, Ismael and Hesse, Katharine and Fong, William and Glidden, Ana and Shporer, Avi and Latham, David W. and Collins, Karen A. and Quinn, Samuel N. and Burt, Jennifer and Dragomir, Diana and Crossfield, Ian and Vanderspek, Roland and Fausnaugh, Michael and Burke, Christopher J. and Ricker, George and Daylan, Tansu and Essack, Zahra and Günther, Maximilian N. and Osborn, Hugh P. and Pepper, Joshua and Rowden, Pamela and Sha, Lizhou and Villanueva Jr., Steven and Yahalomi, Daniel A. and Yu, Liang and Ballard, Sarah and Batalha, Natalie M. and Berardo, David and Chontos, Ashley and Dittmann, Jason A. and Esquerdo, Gilbert A. and Mikal-Evans, Thomas and Jayaraman, Rahul and Krishnamurthy, Akshata and Louie, Dana R. and Mehrle, Nicholas and Niraula, Prajwal and Rackham, Benjamin V. and Rodriguez, Joseph E. and Rowden, Stephen J. L. and Sousa-Silva, Clara and Watanabe, David and Wong, Ian and Zhan, Zhuchang and Zivanovic, Goran and Christiansen, Jessie L. and Ciardi, David R. and Swain, Melanie A. and Lund, Michael B. and Mullally, Susan E. and Fleming, Scott W. and Rodriguez, David R. and Boyd, Patricia T. and Quintana, Elisa V. and Barclay, Thomas and Colón, Knicole D. and Rinehart, S. A. and Schlieder, Joshua E. and Clampin, Mark and Jenkins, Jon M. and Twicken, Joseph D. and Caldwell, Douglas A. and Coughlin, Jeffrey L. and Henze, Chris and Lissauer, Jack J. and Morris, Robert L. and Rose, Mark E. and Smith, Jeffrey C. and Tenenbaum, Peter and Ting, Eric B. and Wohler, Bill and Bakos, G. Á. and Bean, Jacob L. and Berta-Thompson, Zachory K. and Bieryla, Allyson and Bouma, Luke G. and Buchhave, Lars A. and Butler, Nathaniel and Charbonneau, David and Doty, John P. and Ge, Jian and Holman, Matthew J. and Howard, Andrew W. and Kaltenegger, Lisa and Kane, Stephen R. and Kjeldsen, Hans and Kreidberg, Laura and Lin, Douglas N. C. and Minsky, Charlotte and Narita, Norio and Paegert, Martin and Pál, András and Palle, Enric and Sasselov, Dimitar D. and Spencer, Alton and Sozzetti, Alessandro and Stassun, Keivan G. and Torres, Guillermo and Udry, Stephane and Winn, Joshua N.},
title = {The TESS Objects of Interest Catalog from the TESS Prime Mission},
journal = {The Astrophysical Journal Supplement Series}
}

@ARTICLE{PHARO,
       author = {{Hayward}, T.~L. and {Brandl}, B. and {Pirger}, B. and {Blacken}, C. and {Gull}, G.~E. and {Schoenwald}, J. and {Houck}, J.~R.},
        title = "{PHARO: A Near-Infrared Camera for the Palomar Adaptive Optics System}",
      journal = {\pasp},
         year = 2001,
        month = jan,
       volume = {113},
       number = {779},
        pages = {105-118},
          doi = {10.1086/317969},
       adsurl = {https://ui.adsabs.harvard.edu/abs/2001PASP..113..105H}
}

@ARTICLE{halbwachs2023,
       author = {{Halbwachs}, Jean-Louis and {Pourbaix}, Dimitri and {Arenou}, Fr{\'e}d{\'e}ric and {Galluccio}, Laurent and {Guillout}, Patrick and {Bauchet}, Nathalie and {Marchal}, Olivier and {Sadowski}, Gilles and {Teyssier}, David},
        title = "{Gaia Data Release 3. Astrometric binary star processing}",
      journal = {\aap},
         year = 2023,
        month = jun,
       volume = {674},
          eid = {A9},
        pages = {A9},
          doi = {10.1051/0004-6361/202243969},
archivePrefix = {arXiv},
       eprint = {2206.05726},
 primaryClass = {astro-ph.SR},
       adsurl = {https://ui.adsabs.harvard.edu/abs/2023A&A...674A...9H}
}

@ARTICLE{numpy,
       author = {{Harris}, Charles R. and {Millman}, K. Jarrod and {van der Walt}, St{\'e}fan J. and {Gommers}, Ralf and {Virtanen}, Pauli and {Cournapeau}, David and {Wieser}, Eric and {Taylor}, Julian and {Berg}, Sebastian and {Smith}, Nathaniel J. and {Kern}, Robert and {Picus}, Matti and {Hoyer}, Stephan and {van Kerkwijk}, Marten H. and {Brett}, Matthew and {Haldane}, Allan and {del R{\'\i}o}, Jaime Fern{\'a}ndez and {Wiebe}, Mark and {Peterson}, Pearu and {G{\'e}rard-Marchant}, Pierre and {Sheppard}, Kevin and {Reddy}, Tyler and {Weckesser}, Warren and {Abbasi}, Hameer and {Gohlke}, Christoph and {Oliphant}, Travis E.},
        title = "{Array programming with NumPy}",
      journal = {\nat},
         year = 2020,
        month = sep,
       volume = {585},
       number = {7825},
        pages = {357-362},
          doi = {10.1038/s41586-020-2649-2},
archivePrefix = {arXiv},
       eprint = {2006.10256},
 primaryClass = {cs.MS},
       adsurl = {https://ui.adsabs.harvard.edu/abs/2020Natur.585..357H}
}

@ARTICLE{henderson_toi6110,
       author = {{Henderson}, Beth A. and {Casewell}, Sarah L. and {Goad}, Michael R. and {Acton}, Jack S. and {G{\"u}nther}, Maximilian N. and {Nielsen}, Louise D. and {Burleigh}, Matthew R. and {Belardi}, Claudia and {Tilbrook}, Rosanna H. and {Turner}, Oliver and et al.},
        title = "{NGTS-28Ab: a short period transiting brown dwarf}",
      journal = {\mnras},
         year = 2024,
        month = may,
       volume = {530},
       number = {1},
        pages = {318-339},
          doi = {10.1093/mnras/stae508},
archivePrefix = {arXiv},
       eprint = {2402.09943},
 primaryClass = {astro-ph.EP},
       adsurl = {https://ui.adsabs.harvard.edu/abs/2024MNRAS.530..318H}
}

@ARTICLE{GeminiNIRI,
       author = {{Hodapp}, Klaus W. and {Jensen}, Joseph B. and {Irwin}, Everett M. and {Yamada}, Hubert and {Chung}, Randolph and {Fletcher}, Kent and {Robertson}, Louis and {Hora}, Joseph L. and {Simons}, Douglas A. and {Mays}, Wendy and et al.},
        title = "{The Gemini Near-Infrared Imager (NIRI)}",
      journal = {\pasp},
         year = 2003,
        month = dec,
       volume = {115},
       number = {814},
        pages = {1388-1406},
          doi = {10.1086/379669},
       adsurl = {https://ui.adsabs.harvard.edu/abs/2003PASP..115.1388H}
}

@ARTICLE{Holl2023,
       author = {{Holl}, B. and {Sozzetti}, A. and {Sahlmann}, J. and {Giacobbe}, P. and {S{\'e}gransan}, D. and {Unger}, N. and {Delisle}, J. -B. and {Barbato}, D. and {Lattanzi}, M.~G. and {Morbidelli}, R. and et al.},
        title = "{Gaia Data Release 3. Astrometric orbit determination with Markov chain Monte Carlo and genetic algorithms: Systems with stellar, sub-stellar, and planetary mass companions}",
      journal = {\aap},
         year = 2023,
        month = jun,
       volume = {674},
          eid = {A10},
        pages = {A10},
          doi = {10.1051/0004-6361/202244161},
archivePrefix = {arXiv},
       eprint = {2206.05439},
 primaryClass = {astro-ph.EP},
       adsurl = {https://ui.adsabs.harvard.edu/abs/2023A&A...674A..10H}
}

@ARTICLE{Horch_DSSI,
       author = {{Horch}, Elliott P. and {Veillette}, Daniel R. and {Baena Gall{\'e}}, Roberto and {Shah}, Sagar C. and {O'Rielly}, Grant V. and {van Altena}, William F.},
        title = "{Observations of Binary Stars with the Differential Speckle Survey Instrument. I. Instrument Description and First Results}",
      journal = {\aj},
         year = 2009,
        month = jun,
       volume = {137},
       number = {6},
        pages = {5057-5067},
          doi = {10.1088/0004-6256/137/6/5057},
       adsurl = {https://ui.adsabs.harvard.edu/abs/2009AJ....137.5057H}
}

@ARTICLE{Horch_DSSI2,
       author = {{Horch}, Elliott P. and {van Belle}, Gerard T. and {Davidson}, Jr., James W. and {Ciastko}, Lindsay A. and {Everett}, Mark E. and {Bjorkman}, Karen S.},
        title = "{Observations of Binary Stars with the Differential Speckle Survey Instrument. VI. Measures during 2014 at the Discovery Channel Telescope}",
      journal = {\aj},
         year = 2015,
        month = nov,
       volume = {150},
       number = {5},
          eid = {151},
        pages = {151},
          doi = {10.1088/0004-6256/150/5/151},
archivePrefix = {arXiv},
       eprint = {1509.03498},
 primaryClass = {astro-ph.SR},
       adsurl = {https://ui.adsabs.harvard.edu/abs/2015AJ....150..151H}
}

@ARTICLE{Hord2024,
       author = {{Hord}, Benjamin J. and {Kempton}, Eliza M. -R. and {Evans-Soma}, Thomas M. and {Latham}, David W. and {Ciardi}, David R. and {Dragomir}, Diana and {Col{\'o}n}, Knicole D. and {Ross}, Gabrielle and {Vanderburg}, Andrew and {de Beurs}, Zoe L. and {Collins}, Karen A. and {Watkins}, Cristilyn N. and {Bean}, Jacob and {Cowan}, Nicolas B. and {Daylan}, Tansu and {Morley}, Caroline V. and {Ih}, Jegug and {Baker}, David and {Barkaoui}, Khalid and {Batalha}, Natalie M. and {Behmard}, Aida and {Belinski}, Alexander and {Benkhaldoun}, Zouhair and {Benni}, Paul and {Bernacki}, Krzysztof and {Bieryla}, Allyson and {Binnenfeld}, Avraham and {Bosch-Cabot}, Pau and {Bouchy}, Fran{\c{c}}ois and {Bozza}, Valerio and {Brahm}, Rafael and {Buchhave}, Lars A. and {Calkins}, Michael and {Chontos}, Ashley and {Clark}, Catherine A. and {Cloutier}, Ryan and {Cointepas}, Marion and {Collins}, Kevin I. and {Conti}, Dennis M. and {Crossfield}, Ian J.~M. and {Dai}, Fei and {de Leon}, Jerome P. and {Dransfield}, Georgina and {Dressing}, Courtney and {Dustor}, Adam and {Esquerdo}, Gilbert and {Evans}, Phil and {Fajardo-Acosta}, Sergio B. and {Fio{\l}ka}, Jerzy and {For{\'e}s-Toribio}, Raquel and {Frasca}, Antonio and {Fukui}, Akihiko and {Fulton}, Benjamin and {Furlan}, Elise and {Gan}, Tianjun and {Gandolfi}, Davide and {Ghachoui}, Mourad and {Giacalone}, Steven and {Gilbert}, Emily A. and {Gillon}, Micha{\"e}l and {Girardin}, Eric and {Gonzales}, Erica and {Grau Horta}, Ferran and {Gregorio}, Joao and {Greklek-McKeon}, Michael and {Guerra}, Pere and {Hartman}, J.~D. and {Hellier}, Coel and {Helm}, Ian and {He{\l}miniak}, Krzysztof G. and {Henning}, Thomas and {Hill}, Michelle L. and {Horne}, Keith and {Howard}, Andrew W. and {Howell}, Steve B. and {Huber}, Daniel and {Isopi}, Giovanni and {Jehin}, Emmanuel and {Jenkins}, Jon M. and {Jensen}, Eric L.~N. and {Johnson}, Marshall C. and {Jord{\'a}n}, Andr{\'e}s and {Kane}, Stephen R. and {Kielkopf}, John F. and {Krushinsky}, Vadim and {Lasota}, S{\l}awomir and {Lee}, Elena and {Lewin}, Pablo and {Livingston}, John H. and {Lubin}, Jack and {Lund}, Michael B. and {Mallia}, Franco and {Mann}, Christopher R. and {Marino}, Giuseppi and {Maslennikova}, Nataliia and {Massey}, Bob and {Matson}, Rachel and {Matthews}, Elisabeth and {Mayo}, Andrew W. and {Mazeh}, Tsevi and {McLeod}, Kim K. and {Michaels}, Edward J. and {Mo{\v{c}}nik}, Teo and {Mori}, Mayuko and {Mraz}, Georgia and {Mu{\~n}oz}, Jose A. and {Narita}, Norio and {Natarajan}, Krupa and {Dyregaard Nielsen}, Louise and {Osborn}, Hugh and {Palle}, Enric and {Panahi}, Aviad and {Papini}, Riccardo and {Plavchan}, Peter and {Polanski}, Alex S. and {Popowicz}, Adam and {Pozuelos}, Francisco J. and {Quinn}, Samuel N. and {Radford}, Don J. and {Reed}, Phillip A. and {Relles}, Howard M. and {Rice}, Malena and {Robertson}, Paul and {Rodriguez}, Joseph E. and {Rosenthal}, Lee J. and {Rubenzahl}, Ryan A. and {Schanche}, Nicole and {Schlieder}, Joshua and {Schwarz}, Richard P. and {Sefako}, Ramotholo and {Shporer}, Avi and {Sozzetti}, Alessandro and {Srdoc}, Gregor and {Stockdale}, Chris and {Tarasenkov}, Alexander and {Tan}, Thiam-Guan and {Timmermans}, Mathilde and {Ting}, Eric B. and {Van Zandt}, Judah and {Vignes}, JP and {Waite}, Ian and {Watanabe}, Noriharu and {Weiss}, Lauren M. and {Wittrock}, Justin and {Zhou}, George and {Ziegler}, Carl and {Zucker}, Shay},
        title = "{Identification of the Top TESS Objects of Interest for Atmospheric Characterization of Transiting Exoplanets with JWST}",
      journal = {\aj},
         year = 2024,
        month = may,
       volume = {167},
       number = {5},
          eid = {233},
        pages = {233},
          doi = {10.3847/1538-3881/ad3068},
archivePrefix = {arXiv},
       eprint = {2308.09617},
 primaryClass = {astro-ph.EP},
       adsurl = {https://ui.adsabs.harvard.edu/abs/2024AJ....167..233H}
}

@ARTICLE{2024Hori,
       author = {{Hori}, Yasunori and {Fukui}, Akihiko and {Hirano}, Teruyuki and {Narita}, Norio and {de Leon}, Jerome P. and {Ishikawa}, Hiroyuki Tako and {Hartman}, Joel D. and {Morello}, Giuseppe and {Garc{\'\i}a}, Nestor Abreu and {{\'A}lvarez Hern{\'a}ndez}, Leticia and {B{\'e}jar}, V{\'\i}ctor J.~S. and {Calatayud-Borras}, Y{\'e}ssica and {Carleo}, Ilaria and {Enoc}, Gareb and {Esparza-Borges}, Emma and {Fukuda}, Izuru and {Gal{\'a}n}, Daniel and {Gerald{\'\i}a-Gonz{\'a}lez}, Samuel and {Hayashi}, Yuya and {Ikoma}, Masahiro and {Ikuta}, Kai and {Isogai}, Keisuke and {Kagetani}, Taiki and {Kawai}, Yugo and {Kawauchi}, Kiyoe and {Kimura}, Tadahiro and {Kodama}, Takanori and {Korth}, Judith and {Kusakabe}, Nobuhiko and {Laza-Ramos}, Andr{\'e}s and {Livingston}, John H. and {Luque}, Rafael and {Miyakawa}, Kohei and {Mori}, Mayuko and {Murgas}, Felipe and {Orell-Miquel}, Jaume and {Palle}, Enric and {Parviainen}, Hannu and {Pel{\'a}ez-Torres}, Alberto and {Puig-Subir{\`a}}, Marta and {S{\'a}nchez-Benavente}, Manuel and {Sosa-Guill{\'e}n}, Paula and {Stangret}, Monika and {Terada}, Yuka and {Mu{\~n}oz Torres}, Sara and {Watanabe}, Noriharu and {Bakos}, Gaspar {\'A}. and {Barkaoui}, Khalid and {Beichman}, Charles and {Benkhaldoun}, Zouhair and {Boyle}, Andrew W. and {Ciardi}, David R. and {Clark}, Catherine A. and {Collins}, Karen A. and {Collins}, Kevin I. and {Conti}, Dennis M. and {Crossfield}, Ian J.~M. and {Everett}, Mark E. and {Furlan}, Elise and {Ghachoui}, Mourad and {Gillon}, Micha{\"e}l and {Gonzales}, Erica J. and {Higuera}, Jesus and {Horne}, Keith and {Howell}, Steve B. and {Jehin}, Emmanu{\"e}l and {Lester}, Kathryn V. and {Lund}, Michael B. and {Matson}, Rachel and {Matthews}, Elisabeth C. and {Pozuelos}, Francisco J. and {Safonov}, Boris S. and {Schlieder}, Joshua E. and {Schwarz}, Richard P. and {Sefako}, Ramotholo and {Srdoc}, Gregor and {Strakhov}, Ivan A. and {Timmermans}, Mathilde and {Waalkes}, William C. and {Ziegler}, Carl and {Charbonneau}, David and {Essack}, Zahra and {Guerrero}, Natalia M. and {Harakawa}, Hiroki and {Hedges}, Christina and {Ishizuka}, Masato and {Jenkins}, Jon M. and {Konishi}, Mihoko and {Kotani}, Takayuki and {Kudo}, Tomoyuki and {Kurokawa}, Takashi and {Kuzuhara}, Masayuki and {Nishikawa}, Jun and {Omiya}, Masashi and {Ricker}, George R. and {Seager}, Sara and {Serizawa}, Takuma and {Striegel}, Stephanie and {Tamura}, Motohide and {Ueda}, Akitoshi and {Vanderspek}, Roland and {Vievard}, S{\'e}bastien and {Winn}, Joshua N.},
        title = "{The Discovery and Follow-up of Four Transiting Short-period Sub-Neptunes Orbiting M Dwarfs}",
      journal = {\aj},
         year = 2024,
        month = jun,
       volume = {167},
       number = {6},
          eid = {289},
        pages = {289},
          doi = {10.3847/1538-3881/ad4115},
archivePrefix = {arXiv},
       eprint = {2405.12637},
 primaryClass = {astro-ph.EP},
       adsurl = {https://ui.adsabs.harvard.edu/abs/2024AJ....167..289H}
}

@ARTICLE{Howell_nessi,
       author = {{Howell}, Steve B. and {Matson}, Rachel A. and {Ciardi}, David R. and {Everett}, Mark E. and {Livingston}, John H. and {Scott}, Nicholas J. and {Horch}, Elliott P. and {Winn}, Joshua N.},
        title = "{Speckle Observations of TESS Exoplanet Host Stars: Understanding the Binary Exoplanet Host Star Orbital Period Distribution}",
      journal = {\aj},
         year = 2021,
        month = apr,
       volume = {161},
       number = {4},
          eid = {164},
        pages = {164},
          doi = {10.3847/1538-3881/abdec6},
archivePrefix = {arXiv},
       eprint = {2101.08671},
 primaryClass = {astro-ph.EP},
       adsurl = {https://ui.adsabs.harvard.edu/abs/2021AJ....161..164H}
}

@ARTICLE{speckle_Howell,
       author = {{Howell}, Steve B. and {Everett}, Mark E. and {Sherry}, William and {Horch}, Elliott and {Ciardi}, David R.},
        title = "{Speckle Camera Observations for the NASA Kepler Mission Follow-up Program}",
      journal = {\aj},
         year = 2011,
        month = jul,
       volume = {142},
       number = {1},
          eid = {19},
        pages = {19},
          doi = {10.1088/0004-6256/142/1/19},
       adsurl = {https://ui.adsabs.harvard.edu/abs/2011AJ....142...19H}
}

@Article{matplotlib,
  Author    = {Hunter, J. D.},
  Title     = {Matplotlib: A 2D graphics environment},
  Journal   = {Computing in Science \& Engineering},
  Volume    = {9},
  Number    = {3},
  Pages     = {90--95},
  publisher = {IEEE COMPUTER SOC},
  doi       = {10.1109/MCSE.2007.55},
  year      = 2007
}

@ARTICLE{JanoMunoz2025,
       author = {{Jan{\'o} Mu{\~n}oz}, C. and {Hooton}, M.~J. and {Pedersen}, P.~P. and {Barkaoui}, K. and {Rackham}, B.~V. and {Burgasser}, A.~J. and {Pozuelos}, F.~J. and {Stassun}, K.~G. and {Queloz}, D. and {Triaud}, A.~H.~M.~J. and {Ziegler}, C. and {Almenara}, J.~M. and {Timmermans}, M. and {Bonfils}, X. and {Collins}, K.~A. and {Demory}, B.~O. and {Dransfield}, G. and {Ghachoui}, M. and {Gillon}, M. and {Jehin}, E. and {Mann}, A.~W. and {Sebastian}, D. and {Thompson}, S. and {Twicken}, J.~D. and {de Wit}, J. and {Z{\'u}{\~n}iga-Fern{\'a}ndez}, S.},
        title = "{TOI-2407 b: a warm Neptune in the desert}",
      journal = {\mnras},
         year = 2025,
        month = jul,
       volume = {541},
       number = {1},
        pages = {630-641},
          doi = {10.1093/mnras/staf946},
archivePrefix = {arXiv},
       eprint = {2506.06195},
 primaryClass = {astro-ph.EP},
       adsurl = {https://ui.adsabs.harvard.edu/abs/2025MNRAS.541..630J}
}

@ARTICLE{2025jenkins,
       author = {{Jenkins}, Sydney A. and {Vanderburg}, Andrew and {Sethi}, Ritika and {Millholland}, Sarah and {Rodriguez}, Joseph E. and {Fossati}, Luca and {Krenn}, Andreas and {Pass}, Emily and {Venner}, Alexander and {Butler}, R. Paul and {Osborn}, Hugh and {Householder}, Aaron and {Ziegler}, Carl and {Becker}, Juliette and {Berlind}, Perry and {Bieryla}, Allyson and {Broeg}, Christopher and {Calkins}, Michael L. and {Crane}, Jeffrey D. and {Daylan}, Tansu and {de Wit}, Julien and {Eastman}, Jason D. and {Ehrenreich}, David and {Esquerdo}, Gilbert A. and {Fausnaugh}, Michael and {Gaibor}, Yadira and {Latham}, David W. and {Lendl}, Monika and {Mayo}, Andrew W. and {Scandariato}, Gaetano and {Shectman}, Steve and {Striegel}, Stephanie and {Teske}, Johanna and {Wilson}, Thomas G.},
        title = "{An Eccentric Sub-Neptune Moving Into the Evaporation Desert}",
      journal = {\apjl},
         year = 2025,
        month = aug,
       volume = {989},
       number = {2},
          eid = {L20},
        pages = {L20},
          doi = {10.3847/2041-8213/adea46},
archivePrefix = {arXiv},
       eprint = {2505.10324},
 primaryClass = {astro-ph.EP},
       adsurl = {https://ui.adsabs.harvard.edu/abs/2025ApJ...989L..20J}
}

@ARTICLE{2026Jiang7166,
       author = {{Jiang}, Chengzi and {Selezneva}, Aleksandra and {Parviainen}, Hannu and {Murgas}, Felipe and {Pall{\'e}}, Enric and {Fern{\'a}ndez-Rodr{\'\i}guez}, Gareb and {Gerald{\'\i}a-Gonz{\'a}lez}, Samuel and {Orell-Miquel}, Jaume and {Narita}, Norio and {Fukui}, Akihiko and et al.},
        title = "{Multicolour validation of two temperate mini-Neptunes around M-dwarf habitable zones}",
      journal = {\mnras},
         year = 2026,
        month = apr,
       volume = {547},
       number = {2},
          eid = {stag367},
        pages = {stag367},
          doi = {10.1093/mnras/stag367},
archivePrefix = {arXiv},
       eprint = {2512.06817},
 primaryClass = {astro-ph.EP},
       adsurl = {https://ui.adsabs.harvard.edu/abs/2026MNRAS.547ag367J}
}

@ARTICLE{2024Jones,
       author = {{Jones}, Sinclaire E. and {Stef{\'a}nsson}, Gu{\dj}mundur and {Masuda}, Kento and {Libby-Roberts}, Jessica E. and {Gardner}, Cristilyn N. and {Holcomb}, Rae and {Beard}, Corey and {Robertson}, Paul and {Ca{\~n}as}, Caleb I. and {Mahadevan}, Suvrath and {Kanodia}, Shubham and {Lin}, Andrea S.~J. and {Kobulnicky}, Henry A. and {Parker}, Brock A. and {Bender}, Chad F. and {Cochran}, William D. and {Diddams}, Scott A. and {Fernandes}, Rachel B. and {Gupta}, Arvind F. and {Halverson}, Samuel and {Hawley}, Suzanne L. and {Hearty}, Fred R. and {Hebb}, Leslie and {Kowalski}, Adam and {Lubin}, Jack and {Monson}, Andrew and {Ninan}, Joe P. and {Ramsey}, Lawrence and {Roy}, Arpita and {Schwab}, Christian and {Terrien}, Ryan C. and {Wisniewski}, John},
        title = "{TOI-2015 b: A Warm Neptune with Transit Timing Variations Orbiting an Active Mid-type M Dwarf}",
      journal = {\aj},
         year = 2024,
        month = aug,
       volume = {168},
       number = {2},
          eid = {93},
        pages = {93},
          doi = {10.3847/1538-3881/ad55ea},
archivePrefix = {arXiv},
       eprint = {2310.11775},
 primaryClass = {astro-ph.EP},
       adsurl = {https://ui.adsabs.harvard.edu/abs/2024AJ....168...93J}
}

@ARTICLE{Kempton_2018,
       author = {{Kempton}, Eliza M.-R. and {Bean}, Jacob L. and {Louie}, Dana R. and {Deming}, Drake and {Koll}, Daniel D.~B. and {Mansfield}, Megan and {Christiansen}, Jessie L. and {L{\'o}pez-Morales}, Mercedes and {Swain}, Mark R. and {Zellem}, Robert T. and et al.},
        title = "{A Framework for Prioritizing the TESS Planetary Candidates Most Amenable to Atmospheric Characterization}",
      journal = {\pasp},
         year = 2018,
        month = nov,
       volume = {130},
       number = {993},
        pages = {114401},
          doi = {10.1088/1538-3873/aadf6f},
archivePrefix = {arXiv},
       eprint = {1805.03671},
 primaryClass = {astro-ph.EP},
       adsurl = {https://ui.adsabs.harvard.edu/abs/2018PASP..130k4401K}
}

@ARTICLE{kostov,
       author = {{Kostov}, Veselin B. and {Powell}, Brian P. and {Fornear}, Aline U. and {Di Fraia}, Marco Z. and {Gagliano}, Robert and {Jacobs}, Thomas L. and {de Lambilly}, Julien S. and {Durantini Luca}, Hugo A. and {Majewski}, Steven R. and {Omohundro}, Mark and et al.},
        title = "{The TESS Ten Thousand Catalog: 10,001 Uniformly Vetted and Validated Eclipsing Binary Stars Detected in Full-frame Image Data by Machine Learning and Analyzed by Citizen Scientists}",
      journal = {\apjs},
         year = 2025,
        month = aug,
       volume = {279},
       number = {2},
          eid = {50},
        pages = {50},
          doi = {10.3847/1538-4365/ade2d8},
archivePrefix = {arXiv},
       eprint = {2506.05631},
 primaryClass = {astro-ph.SR},
       adsurl = {https://ui.adsabs.harvard.edu/abs/2025ApJS..279...50K}
}

@ARTICLE{LEOvetter,
       author = {{Kunimoto}, Michelle and {Bryson}, Steve and {Jaffee}, Drayson and {Rowe}, Jason F. and {Daylan}, Tansu and {Giacalone}, Steven and {Lissauer}, Jack J. and {Matesic}, Michael R.~B. and {Mullally}, Susan E. and {Eschen}, Yoshi Nike Emilia},
        title = "{LEO-Vetter: Fully Automated Flux- and Pixel-level Vetting of TESS Planet Candidates to Support Occurrence Rates}",
      journal = {\aj},
         year = 2025,
        month = nov,
       volume = {170},
       number = {5},
          eid = {280},
        pages = {280},
          doi = {10.3847/1538-3881/ae070a},
archivePrefix = {arXiv},
       eprint = {2509.10619},
 primaryClass = {astro-ph.EP},
       adsurl = {https://ui.adsabs.harvard.edu/abs/2025AJ....170..280K}
}

@ARTICLE{2025Kunimoto,
       author = {{Kunimoto}, Michelle and {Lin}, Zifan and {Millholland}, Sarah and {Venner}, Alexander and {Hinkel}, Natalie R. and {Shporer}, Avi and {Vanderburg}, Andrew and {Bailey}, Jeremy and {Brahm}, Rafael and {Burt}, Jennifer A. and {Butler}, R. Paul and {Carter}, Brad and {Ciardi}, David R. and {Collins}, Karen A. and {Collins}, Kevin I. and {Col{\'o}n}, Knicole D. and {Crane}, Jeffrey D. and {Daylan}, Tansu and {D{\'\i}az}, Mat{\'\i}as R. and {Doty}, John P. and {Feng}, Fabo and {Guenther}, Eike W. and {Horner}, Jonathan and {Howell}, Steve B. and {Janik}, Jan and {Jones}, Hugh R.~A. and {Kab{\'a}th}, Petr and {Kanodia}, Shubham and {Littlefield}, Colin and {Osborn}, Hugh P. and {O'Toole}, Simon and {Paegert}, Martin and {Pintr}, Pavel and {Schwarz}, Richard P. and {Shectman}, Steve and {Srdoc}, Gregor and {Stassun}, Keivan G. and {Teske}, Johanna K. and {Twicken}, Joseph D. and {Vanzi}, Leonardo and {Wang}, Sharon X. and {Wittenmyer}, Robert A. and {Jenkins}, Jon M. and {Ricker}, George R. and {Seager}, Sara and {Winn}, Joshua},
        title = "{Two Earth-size Planets and an Earth-size Candidate Transiting the nearby Star HD 101581}",
      journal = {\aj},
         year = 2025,
        month = jan,
       volume = {169},
       number = {1},
          eid = {47},
        pages = {47},
          doi = {10.3847/1538-3881/ad9266},
archivePrefix = {arXiv},
       eprint = {2412.08863},
 primaryClass = {astro-ph.EP},
       adsurl = {https://ui.adsabs.harvard.edu/abs/2025AJ....169...47K}
}

@ARTICLE{Ikuta_toi654,
       author = {{Ikuta}, Kai and {Narita}, Norio and {Takarada}, Takuya and {Hirano}, Teruyuki and {Fukui}, Akihiko and {Ishikawa}, Hiroyuki Tako and {Hori}, Yasunori and {Kimura}, Tadahiro and {Kodama}, Takanori and {Ikoma}, Masahiro and et al.},
        title = "{The mass of TOI-654 b: A short-period sub-Neptune transiting a mid-M dwarf}",
      journal = {\pasj},
         year = 2025,
        month = oct,
       volume = {77},
       number = {5},
        pages = {1101-1112},
          doi = {10.1093/pasj/psaf090},
archivePrefix = {arXiv},
       eprint = {2507.16222},
 primaryClass = {astro-ph.EP},
       adsurl = {https://ui.adsabs.harvard.edu/abs/2025PASJ...77.1101I}
}

@ARTICLE{2025Lacedelli,
       author = {{Lacedelli}, G. and {Pall{\'e}}, E. and {Davis}, Y.~T. and {Luque}, R. and {Morello}, G. and {Tabernero}, H.~M. and {Zapatero Osorio}, M.~R. and {Pozuelos}, F.~J. and {Jankowski}, D. and {Nowak}, G. and {Murgas}, F. and {Orell-Miquel}, J. and {Akana Murphy}, J.~M. and {Barkaoui}, K. and {Charbonneau}, D. and {Dransfield}, G. and {Ducrot}, E. and {Gerald{\'\i}a-Gonz{\'a}lez}, S. and {Irwin}, J. and {Jehin}, E. and {Osborne}, H.~L.~M. and {Pedersen}, P.~P. and {Rackham}, B.~V. and {Scott}, M.~G. and {Timmermans}, M. and {Triaud}, A. and {Van Eylen}, V.},
        title = "{A transiting rocky super-Earth and a non-transiting sub-Neptune orbiting the M dwarf TOI-771}",
      journal = {\aap},
         year = 2025,
        month = jun,
       volume = {698},
          eid = {A223},
        pages = {A223},
          doi = {10.1051/0004-6361/202554161},
archivePrefix = {arXiv},
       eprint = {2504.18223},
 primaryClass = {astro-ph.EP},
       adsurl = {https://ui.adsabs.harvard.edu/abs/2025A&A...698A.223L}
}

@ARTICLE{2026Lacedelli,
       author = {{Lacedelli}, G. and {Pall{\'e}}, E. and {Luque}, R. and {Ikuta}, K. and {Tabernero}, H.~M. and {Zapatero Osorio}, M.~R. and {Almenara}, J.~M. and {Pozuelos}, F.~J. and {Jankowski}, D. and {Narita}, N. and et al.},
        title = "{Sibling sub-Neptunes around sibling M dwarfs: TOI-521 and TOI-912}",
      journal = {\aap},
         year = 2026,
        month = jan,
       volume = {705},
          eid = {A260},
        pages = {A260},
          doi = {10.1051/0004-6361/202557304},
archivePrefix = {arXiv},
       eprint = {2509.14782},
 primaryClass = {astro-ph.EP},
       adsurl = {https://ui.adsabs.harvard.edu/abs/2026A&A...705A.260L}
}

@ARTICLE{RAVEN,
       author = {{Lafarga}, M. and {Armstrong}, D.~J. and {Cui}, K. and {Hadjigeorghiou}, A. and {Kunovac}, V. and {Doyle}, L. and {Bryant}, E.~M. and {D{\'\i}az}, R.~F. and {Nieto}, L.~A. and {Osborn}, A.},
        title = "{Automatic search for transiting planets in TESS─SPOC FFIs with RAVEN: over 100 newly validated planets and over 2000 vetted candidates}",
      journal = {\mnras},
         year = 2026,
        month = may,
       volume = {548},
       number = {3},
          eid = {stag512},
        pages = {stag512},
          doi = {10.1093/mnras/stag512},
archivePrefix = {arXiv},
       eprint = {2603.22597},
 primaryClass = {astro-ph.EP},
       adsurl = {https://ui.adsabs.harvard.edu/abs/2026MNRAS.548ag512L}
}

@ARTICLE{Lambert_5375,
       author = {{Lambert}, Mika and {Bender}, Chad F. and {Kanodia}, Shubham and {Ca{\~n}as}, Caleb I. and {Monson}, Andrew and {Stef{\'a}nsson}, Gudmundur and {Cochran}, William D. and {Everett}, Mark E. and {Gupta}, Arvind F. and {Hearty}, Fred and et al.},
        title = "{TOI-5375 B: A Very Low Mass Star at the Hydrogen-burning Limit Orbiting an Early M-type Star}",
      journal = {\aj},
         year = 2023,
        month = may,
       volume = {165},
       number = {5},
          eid = {218},
        pages = {218},
          doi = {10.3847/1538-3881/acc651},
archivePrefix = {arXiv},
       eprint = {2303.16193},
 primaryClass = {astro-ph.SR},
       adsurl = {https://ui.adsabs.harvard.edu/abs/2023AJ....165..218L}
}

@ARTICLE{Lamman2020,
       author = {{Lamman}, Claire and {Baranec}, Christoph and {Berta-Thompson}, Zachory K. and {Law}, Nicholas M. and {Schonhut-Stasik}, Jessica and {Ziegler}, Carl and {Salama}, Ma{\"\i}ssa and {Jensen-Clem}, Rebecca and {Duev}, Dmitry A. and {Riddle}, Reed and et al.},
        title = "{Robo-AO M-dwarf Multiplicity Survey: Catalog}",
      journal = {\aj},
         year = 2020,
        month = apr,
       volume = {159},
       number = {4},
          eid = {139},
        pages = {139},
          doi = {10.3847/1538-3881/ab6ef1},
archivePrefix = {arXiv},
       eprint = {2001.05988},
 primaryClass = {astro-ph.SR},
       adsurl = {https://ui.adsabs.harvard.edu/abs/2020AJ....159..139L}
}

@ARTICLE{2025ALee,
       author = {{Lee}, Rena A. and {Dai}, Fei and {Howard}, Andrew W. and {Halverson}, Samuel and {Gomez Barrientos}, Jonathan and {Greklek-McKeon}, Michael and {Knutson}, Heather A. and {Fulton}, Benjamin J. and {Stef{\'a}nsson}, Gu{\dj}mundur and {Lubin}, Jack and {Isaacson}, Howard and {Brinkman}, Casey L. and {Saunders}, Nicholas and {Hey}, Daniel and {Huber}, Daniel and {Weiss}, Lauren M. and {Rogers}, Leslie A. and {Valencia}, Diana and {Plotnykov}, Mykhaylo and {Paragas}, Kimberly and {Hu}, Renyu and {Han}, Te and {Petigura}, Erik A. and {Rubenzahl}, Ryan and {Ciardi}, David R. and {Householder}, Aaron and {Gilbert}, Gregory J. and {Ong}, J.~M. Joel and {Zhang}, Jingwen and {Luhn}, Jacob and {Handley}, Luke and {Beard}, Corey and {Giacalone}, Steven and {Holcomb}, Rae and {Van Zandt}, Judah and {Baker}, Ashley D. and {Brodheim}, Max and {Brown}, Matt and {Charbonneau}, David and {Collins}, Karen A. and {Crossfield}, Ian J.~M. and {Deich}, William and {Dumusque}, Xavier and {Gibson}, Steven R. and {Gilbert}, Emily and {Hill}, Grant M. and {Holden}, Bradford and {Jenkins}, Jon M. and {Kaye}, Stephen and {Laher}, Russ R. and {Lanclos}, Kyle and {Levine}, W. Garett and {Payne}, Joel and {Polanski}, Alex S. and {O'Meara}, John and {Ricker}, George R. and {Rider}, Kodi and {Robertson}, Paul and {Roy}, Arpita and {Schlieder}, Joshua E. and {Schwab}, Christian and {Seager}, Sara and {Shaum}, Abby P. and {Sirk}, Martin M. and {Striegel}, Stephanie and {Teske}, Johanna and {Valliant}, John and {Vanderspek}, Roland and {Vasisht}, Gautam and {Walawender}, Josh and {Wang}, Sharon Xuesong and {Winn}, Joshua N. and {Wishnow}, Edward and {Yeh}, Sherry},
        title = "{TOI-6324 b: An Earth-mass Ultra-short-period Planet Transiting a Nearby M Dwarf}",
      journal = {\apjl},
         year = 2025,
        month = apr,
       volume = {983},
       number = {2},
          eid = {L36},
        pages = {L36},
          doi = {10.3847/2041-8213/adadd7},
archivePrefix = {arXiv},
       eprint = {2502.16087},
 primaryClass = {astro-ph.EP},
       adsurl = {https://ui.adsabs.harvard.edu/abs/2025ApJ...983L..36L}
}

@INPROCEEDINGS{NaCo,
       author = {{Lenzen}, Rainer and {Hartung}, Markus and {Brandner}, Wolfgang and {Finger}, Gert and {Hubin}, Norbert N. and {Lacombe}, Francois and {Lagrange}, Anne-Marie and {Lehnert}, Matthew D. and {Moorwood}, Alan F.~M. and {Mouillet}, David},
        title = "{NAOS-CONICA first on sky results in a variety of observing modes}",
    booktitle = {Instrument Design and Performance for Optical/Infrared Ground-based Telescopes},
         year = 2003,
       editor = {{Iye}, Masanori and {Moorwood}, Alan F.~M.},
       series = {Society of Photo-Optical Instrumentation Engineers (SPIE) Conference Series},
       volume = {4841},
        month = mar,
        pages = {944-952},
          doi = {10.1117/12.460044},
       adsurl = {https://ui.adsabs.harvard.edu/abs/2003SPIE.4841..944L}
}

@ARTICLE{Lindegren2021,
       author = {{Lindegren}, L. and {Klioner}, S.~A. and {Hern{\'a}ndez}, J. and {Bombrun}, A. and {Ramos-Lerate}, M. and {Steidelm{\"u}ller}, H. and {Bastian}, U. and {Biermann}, M. and {de Torres}, A. and {Gerlach}, E. and et al.},
        title = "{Gaia Early Data Release 3. The astrometric solution}",
      journal = {\aap},
         year = 2021,
        month = may,
       volume = {649},
          eid = {A2},
        pages = {A2},
          doi = {10.1051/0004-6361/202039709},
archivePrefix = {arXiv},
       eprint = {2012.03380},
 primaryClass = {astro-ph.IM},
       adsurl = {https://ui.adsabs.harvard.edu/abs/2021A&A...649A...2L}
}

@ARTICLE{Maldonado_5375,
       author = {{Maldonado}, J. and {Petralia}, A. and {Mantovan}, G. and {Rainer}, M. and {Lanza}, A.~F. and {Di Maio}, C. and {Colombo}, S. and {Nardiello}, D. and {Benatti}, S. and {Borsato}, L. and et al.},
        title = "{The GAPS programme at TNG. XLIII. A massive brown dwarf orbiting the active M dwarf TOI-5375}",
      journal = {\aap},
         year = 2023,
        month = jun,
       volume = {674},
          eid = {A132},
        pages = {A132},
          doi = {10.1051/0004-6361/202346096},
archivePrefix = {arXiv},
       eprint = {2304.04477},
 primaryClass = {astro-ph.SR},
       adsurl = {https://ui.adsabs.harvard.edu/abs/2023A&A...674A.132M}
}

@ARTICLE{2024Martioli,
       author = {{Martioli}, E. and {Petrucci}, R.~P. and {Jofr{\'e}}, E. and {H{\'e}brard}, G. and {Ghezzi}, L. and {G{\'o}mez Maqueo Chew}, Y. and {D{\'\i}az}, R.~F. and {Perottoni}, H.~D. and {Garcia}, L.~H. and {Rapetti}, D. and {Lecavelier des Etangs}, A. and {de Almeida}, L. and {Arnold}, L. and {Artigau}, {\'E}. and {Basant}, R. and {Bean}, J.~L. and {Bieryla}, A. and {Boisse}, I. and {Bonfils}, X. and {Brady}, M. and {Cadieux}, C. and {Carmona}, A. and {Cook}, N.~J. and {Delfosse}, X. and {Donati}, J. -F. and {Doyon}, R. and {Furlan}, E. and {Howell}, S.~B. and {Jenkins}, J.~M. and {Kasper}, D. and {Kiefer}, F. and {Latham}, D.~W. and {Levine}, A.~M. and {Lorenzo-Oliveira}, D. and {Luque}, R. and {McLeod}, K.~K. and {Melendez}, J. and {Moutou}, C. and {Netto}, Y. and {Pritchard}, T.~A. and {Rowden}, P. and {Seifahrt}, A. and {Stef{\'a}nsson}, G. and {St{\"u}rmer}, J. and {Twicken}, J.~D.},
        title = "{TOI-3568 b: A super-Neptune in the sub-Jovian desert}",
      journal = {\aap},
         year = 2024,
        month = oct,
       volume = {690},
          eid = {A312},
        pages = {A312},
          doi = {10.1051/0004-6361/202450334},
archivePrefix = {arXiv},
       eprint = {2409.03704},
 primaryClass = {astro-ph.EP},
       adsurl = {https://ui.adsabs.harvard.edu/abs/2024A&A...690A.312M}
}

@ARTICLE{vatestI,
       author = {{Mistry}, Priyashkumar and {Pathak}, Kamlesh and {Lekkas}, Georgios and {Prasad}, Aniket and {Bhattarai}, Surendra and {Maity}, Mousam and {Beichman}, Charles A. and {Ciardi}, David R. and {Evans}, Phil and {Bieryla}, Allyson and {Eastman}, Jason D. and {Esquerdo}, Gilbert A. and {Lucero}, Jennifer P.},
        title = "{VaTEST I: validation of sub-Saturn exoplanet TOI-181b in narrow orbit from its host star}",
      journal = {\mnras},
         year = 2023,
        month = may,
       volume = {521},
       number = {1},
        pages = {1066-1078},
          doi = {10.1093/mnras/stad543},
archivePrefix = {arXiv},
       eprint = {2211.07957},
 primaryClass = {astro-ph.EP},
       adsurl = {https://ui.adsabs.harvard.edu/abs/2023MNRAS.521.1066M}
}

@ARTICLE{vatestII,
       author = {{Mistry}, Priyashkumar and {Pathak}, Kamlesh and {Prasad}, Aniket and {Lekkas}, Georgios and {Bhattarai}, Surendra and {Gharat}, Sarvesh and {Maity}, Mousam and {Kumar}, Dhruv and {Collins}, Karen A. and {Schwarz}, Richard P. and {Mann}, Christopher R. and {Furlan}, Elise and {Howell}, Steve B. and {Ciardi}, David and {Bieryla}, Allyson and {Matthews}, Elisabeth C. and {Gonzales}, Erica and {Ziegler}, Carl and {Crossfield}, Ian and {Giacalone}, Steven and {Tan}, Thiam-Guan and {Evans}, Phil and {He{\l}miniak}, Krzysztof G. and {Collins}, Kevin I. and {Narita}, Norio and {Fukui}, Akihiko and {Pozuelos}, Francisco J. and {Dressing}, Courtney and {Soubkiou}, Abderahmane and {Benkhaldoun}, Zouhair and {Schlieder}, Joshua E. and {Suarez}, Olga and {Barkaoui}, Khalid and {Palle}, Enric and {Murgas}, Felipe and {Srdoc}, Gregor and {Goliguzova}, Maria V. and {Strakhov}, Ivan A. and {Gnilka}, Crystal and {Lester}, Kathryn and {Littlefield}, Colin and {Scott}, Nic and {Matson}, Rachel and {Gillon}, Micha{\"e}l and {Jehin}, Emmanuel and {Timmermans}, Mathilde and {Ghachoui}, Mourad and {Abe}, Lyu and {Bendjoya}, Philippe and {Guillot}, Tristan and {Triaud}, Amaury H.~M.~J.},
        title = "{VaTEST. II. Statistical Validation of 11 TESS-detected Exoplanets Orbiting K-type Stars}",
      journal = {\aj},
         year = 2023,
        month = jul,
       volume = {166},
       number = {1},
          eid = {9},
        pages = {9},
          doi = {10.3847/1538-3881/acd548},
archivePrefix = {arXiv},
       eprint = {2301.09865},
 primaryClass = {astro-ph.EP},
       adsurl = {https://ui.adsabs.harvard.edu/abs/2023AJ....166....9M}
}

@ARTICLE{vatestIII,
       author = {{Mistry}, Priyashkumar and {Prasad}, Aniket and {Maity}, Mousam and {Pathak}, Kamlesh and {Gharat}, Sarvesh and {Lekkas}, Georgios and {Bhattarai}, Surendra and {Kumar}, Dhruv and {Lissauer}, Jack J. and {Twicken}, Joseph D. and {Soubkiou}, Abderahmane and {Pozuelos}, Francisco J. and {Jenkins}, Jon and {Horne}, Keith and {Giacalone}, Steven and {Barkaoui}, Khalid and {Timmermans}, Mathilde and {Watkins}, Cristilyn N. and {Sefako}, Ramotholo and {Collins}, Karen A. and {Ciardi}, David R. and {Clark}, Catherine A. and {Safonov}, Boris S. and {Shporer}, Avi and {Schlieder}, Joshua E. and {Benkhaldoun}, Zouhair and {Stockdale}, Chris and {Ziegler}, Carl and {Gilbert}, Emily A. and {Emmanu{\"e}l}, Jehin and {Murgas}, Felipe and {Crossfield}, Ian J.~M. and {Paegert}, Martin and {Lund}, Michael B. and {Narita}, Norio and {Schwarz}, Richard P. and {Goeke}, Robert F. and {Fajardo-Acosta}, Sergio B. and {Howell}, Steve B. and {Tan}, Thiam-Guan and {Barclay}, Thomas and {Kawai}, Yugo},
        title = "{VaTEST III: Validation of eight potential super-earths from TESS data}",
      journal = {\pasa},
         year = 2024,
        month = apr,
       volume = {41},
          eid = {e030},
        pages = {e030},
          doi = {10.1017/pasa.2024.29},
archivePrefix = {arXiv},
       eprint = {2311.00688},
 primaryClass = {astro-ph.EP},
       adsurl = {https://ui.adsabs.harvard.edu/abs/2024PASA...41...30M}
}

@ARTICLE{Morello_2026,
       author = {{Morello}, G. and {Pel{\'a}ez-Torres}, A. and {Pozuelos}, F.~J. and {D{\'e}vora-Pajares}, M. and {Murgas}, F. and {de Leon}, J.~P. and {Rojas-Ayala}, B. and {Bieryla}, A. and {Barkaoui}, K. and {Yal{\c{c}}inkaya}, S. and et al.},
        title = "{The K-dwarfs Survey I. Four Validated Planets from the Radius Valley to the Neptune Desert}",
      journal = {\mnras},
         year = 2026,
        month = jan,
          doi = {10.1093/mnras/stag183},
       adsurl = {https://ui.adsabs.harvard.edu/abs/2026MNRAS.tmp..176M}
}

@software{vespa,
       author = {{Morton}, Timothy D.},
        title = "{VESPA: False positive probabilities calculator}",
 howpublished = {Astrophysics Source Code Library, record ascl:1503.011},
         year = 2015,
        month = mar,
          eid = {ascl:1503.011},
archivePrefix = {ascl},
       eprint = {1503.011},
       adsurl = {https://ui.adsabs.harvard.edu/abs/2015ascl.soft03011M}
}

@ARTICLE{2023Morton,
       author = {{Morton}, Timothy D. and {Giacalone}, Steven and {Bryson}, Steve},
        title = "{A Recommendation to Retire VESPA for Exoplanet Validation}",
      journal = {Research Notes of the American Astronomical Society},
         year = 2023,
        month = may,
       volume = {7},
       number = {5},
          eid = {107},
        pages = {107},
          doi = {10.3847/2515-5172/acd9a6},
       adsurl = {https://ui.adsabs.harvard.edu/abs/2023RNAAS...7..107M}
}

@ARTICLE{arielrad,
       author = {{Mugnai}, Lorenzo V. and {Pascale}, Enzo and {Edwards}, Billy and {Papageorgiou}, Andreas and {Sarkar}, Subhajit},
        title = "{ArielRad: the Ariel radiometric model}",
      journal = {Experimental Astronomy},
         year = 2020,
        month = oct,
       volume = {50},
       number = {2-3},
        pages = {303-328},
          doi = {10.1007/s10686-020-09676-7},
archivePrefix = {arXiv},
       eprint = {2009.07824},
 primaryClass = {astro-ph.IM},
       adsurl = {https://ui.adsabs.harvard.edu/abs/2020ExA....50..303M}
}

@ARTICLE{2024Murgas,
       author = {{Murgas}, F. and {Pall{\'e}}, E. and {Orell-Miquel}, J. and {Carleo}, I. and {Pe{\~n}a-Mo{\~n}ino}, L. and {P{\'e}rez-Torres}, M. and {Watkins}, C.~N. and {Jeffers}, S.~V. and {Azzaro}, M. and {Barkaoui}, K. and {Belinski}, A.~A. and {Caballero}, J.~A. and {Charbonneau}, D. and {Cheryasov}, D.~V. and {Ciardi}, D.~R. and {Collins}, K.~A. and {Cort{\'e}s-Contreras}, M. and {de Leon}, J. and {Duque-Arribas}, C. and {Enoc}, G. and {Esparza-Borges}, E. and {Fukui}, A. and {Gerald{\'\i}a-Gonz{\'a}lez}, S. and {Gilbert}, E.~A. and {Hatzes}, A.~P. and {Hayashi}, Y. and {Henning}, Th. and {Herrero}, E. and {Jenkins}, J.~M. and {Lillo-Box}, J. and {Lodieu}, N. and {Lund}, M.~B. and {Luque}, R. and {Montes}, D. and {Nagel}, E. and {Narita}, N. and {Parviainen}, H. and {Polanski}, A.~S. and {Reffert}, S. and {Schlecker}, M. and {Sch{\"o}fer}, P. and {Schwarz}, R.~P. and {Schweitzer}, A. and {Seager}, S. and {Stassun}, K.~G. and {Tabernero}, H.~M. and {Terada}, Y. and {Twicken}, J.~D. and {Vanaverbeke}, S. and {Winn}, J.~N. and {Zambelli}, R. and {Amado}, P.~J. and {Quirrenbach}, A. and {Reiners}, A. and {Ribas}, I.},
        title = "{Wolf 327b: A new member of the pack of ultra-short-period super-Earths around M dwarfs}",
      journal = {\aap},
         year = 2024,
        month = apr,
       volume = {684},
          eid = {A83},
        pages = {A83},
          doi = {10.1051/0004-6361/202348813},
archivePrefix = {arXiv},
       eprint = {2401.12150},
 primaryClass = {astro-ph.EP},
       adsurl = {https://ui.adsabs.harvard.edu/abs/2024A&A...684A..83M}
}

@ARTICLE{2025naponiello,
       author = {{Naponiello}, L. and {Vissapragada}, S. and {Bonomo}, A.~S. and {Steinmeyer}, M. -L. and {Filomeno}, S. and {D'Orazi}, V. and {Dorn}, C. and {Sozzetti}, A. and {Mancini}, L. and {Lanza}, A.~F. and {Biazzo}, K. and {Watkins}, C.~N. and {H{\'e}brard}, G. and {Lissauer}, J.~J. and {Howell}, S.~B. and {Ciardi}, D.~R. and {Mantovan}, G. and {Baker}, D. and {Bourrier}, V. and {Buchhave}, L.~A. and {Clark}, C.~A. and {Collins}, K.~A. and {Cosentino}, R. and {Damasso}, M. and {Dumusque}, X. and {Fiorenzano}, A. and {Forveille}, T. and {Heidari}, N. and {Latham}, D.~W. and {Littlefield}, C. and {L{\'o}pez-Morales}, M. and {Lund}, M.~B. and {Malavolta}, L. and {Manni}, F. and {Nardiello}, D. and {Pinamonti}, M. and {Yee}, S.~W. and {Zambelli}, R. and {Ziegler}, C. and {Zingales}, T.},
        title = "{The Hot-Neptune Initiative (HONEI): I. Two hot sub-Neptunes on a close-in eccentric orbit (TOI-5800 b) and a farther-out circular orbit (TOI-5817 b)}",
      journal = {\aap},
         year = 2025,
        month = sep,
       volume = {701},
          eid = {A79},
        pages = {A79},
          doi = {10.1051/0004-6361/202555523},
archivePrefix = {arXiv},
       eprint = {2505.10123},
 primaryClass = {astro-ph.EP},
       adsurl = {https://ui.adsabs.harvard.edu/abs/2025A&A...701A..79N}
}

@software{pandas,
    author       = {The pandas development team},
    title        = {pandas-dev/pandas: Pandas},
    month        = feb,
    year         = 2020,
    publisher    = {Zenodo},
    version      = {latest},
    doi          = {10.5281/zenodo.3509134},
    url          = {https://doi.org/10.5281/zenodo.3509134}
}

@ARTICLE{parc_toi756,
       author = {{Parc}, L{\'e}na and {Bouchy}, Fran{\c{c}}ois and {Cook}, Neil J. and {Grieves}, Nolan and {Artigau}, {\'E}tienne and {L'Heureux}, Alexandrine and {Doyon}, Ren{\'e} and {Messias}, Yuri S. and {Baron}, Fr{\'e}d{\'e}rique and {Barros}, Susana C.~C. and et al.},
        title = "{NIRPS and TESS reveal a peculiar system around the M dwarf TOI-756: A transiting sub-Neptune and a cold eccentric giant}",
      journal = {\aap},
         year = 2025,
        month = oct,
       volume = {702},
          eid = {A138},
        pages = {A138},
          doi = {10.1051/0004-6361/202555684},
archivePrefix = {arXiv},
       eprint = {2510.14927},
 primaryClass = {astro-ph.EP},
       adsurl = {https://ui.adsabs.harvard.edu/abs/2025A&A...702A.138P}
}

@ARTICLE{PecautMamajek2013,
       author = {{Pecaut}, Mark J. and {Mamajek}, Eric E.},
        title = "{Intrinsic Colors, Temperatures, and Bolometric Corrections of Pre-main-sequence Stars}",
      journal = {\apjs},
         year = 2013,
        month = sep,
       volume = {208},
       number = {1},
          eid = {9},
        pages = {9},
          doi = {10.1088/0067-0049/208/1/9},
archivePrefix = {arXiv},
       eprint = {1307.2657},
 primaryClass = {astro-ph.SR},
       adsurl = {https://ui.adsabs.harvard.edu/abs/2013ApJS..208....9P}
}

@ARTICLE{2024PelaezTorres,
       author = {{Pel{\'a}ez-Torres}, A. and {Esparza-Borges}, E. and {Pall{\'e}}, E. and {Parviainen}, H. and {Murgas}, F. and {Morello}, G. and {Zapatero-Osorio}, M.~R. and {Korth}, J. and {Narita}, N. and {Fukui}, A. and {Carleo}, I. and {Luque}, R. and {Garc{\'\i}a}, N. Abreu and {Barkaoui}, K. and {Boyle}, A. and {B{\'e}jar}, V.~J.~S. and {Calatayud-Borras}, Y. and {Cheryasov}, D.~V. and {Christiansen}, J.~L. and {Ciardi}, D.~R. and {Enoc}, G. and {Essack}, Z. and {Fukuda}, I. and {Furesz}, G. and {Gal{\'a}n}, D. and {Gerald{\'\i}a-Gonz{\'a}lez}, S. and {Giacalone}, S. and {Gill}, H. and {Gonzales}, E.~J. and {Hayashi}, Y. and {Ikuta}, K. and {Isogai}, K. and {Kagetani}, T. and {Kawai}, Y. and {Kawauchi}, K. and {Klagyvik}, P. and {Kodama}, T. and {Kusakabe}, N. and {Laza-Ramos}, A. and {de Leon}, J.~P. and {Livingston}, J.~H. and {Lund}, M.~B. and {Madrigal-Aguado}, A. and {Meni}, P. and {Mori}, M. and {Torres}, S. Mu{\~n}oz and {Orell-Miquel}, J. and {Puig}, M. and {Ricker}, G. and {S{\'a}nchez-Benavente}, M. and {Savel}, A.~B. and {Schlieder}, J.~E. and {Schwarz}, R.~P. and {Sefako}, R. and {Sosa-Guill{\'e}n}, P. and {Stangret}, M. and {Stockdale}, C. and {Tamura}, M. and {Terada}, Y. and {Twicken}, J.~D. and {Watanabe}, N. and {Winn}, J. and {Zheltoukhov}, S.~G. and {Ziegler}, C. and {Zou}, Y.},
        title = "{Validation of up to seven TESS planet candidates through multi-colour transit photometry using MuSCAT2 data}",
      journal = {\aap},
         year = 2024,
        month = oct,
       volume = {690},
          eid = {A62},
        pages = {A62},
          doi = {10.1051/0004-6361/202347251},
archivePrefix = {arXiv},
       eprint = {2409.07400},
 primaryClass = {astro-ph.EP},
       adsurl = {https://ui.adsabs.harvard.edu/abs/2024A&A...690A..62P}
}

@ARTICLE{2024polanskitess,
       author = {{Polanski}, Alex S. and {Lubin}, Jack and {Beard}, Corey and {Akana Murphy}, Joseph M. and {Rubenzahl}, Ryan and {Hill}, Michelle L. and {Crossfield}, Ian J.~M. and {Chontos}, Ashley and {Robertson}, Paul and {Isaacson}, Howard and {Kane}, Stephen R. and {Ciardi}, David R. and {Batalha}, Natalie M. and {Dressing}, Courtney and {Fulton}, Benjamin and {Howard}, Andrew W. and {Huber}, Daniel and {Petigura}, Erik A. and {Weiss}, Lauren M. and {Angelo}, Isabel and {Behmard}, Aida and {Blunt}, Sarah and {Brinkman}, Casey L. and {Dai}, Fei and {Dalba}, Paul A. and {Fetherolf}, Tara and {Giacalone}, Steven and {Hirsch}, Lea A. and {Holcomb}, Rae and {Kosiarek}, Molly R. and {Mayo}, Andrew W. and {MacDougall}, Mason G. and {Mo{\v{c}}nik}, Teo and {Pidhorodetska}, Daria and {Rice}, Malena and {Rosenthal}, Lee J. and {Scarsdale}, Nicholas and {Turtelboom}, Emma V. and {Tyler}, Dakotah and {Van Zandt}, Judah and {Yee}, Samuel W. and {Coria}, David R. and {Dulz}, Shannon D. and {Hartman}, Joel D. and {Householder}, Aaron and {Lange}, Sarah and {Langford}, Andrew and {Louden}, Emma M. and {Siegel}, Jared C. and {Gilbert}, Emily A. and {Gonzales}, Erica J. and {Schlieder}, Joshua E. and {Boyle}, Andrew W. and {Christiansen}, Jessie L. and {Clark}, Catherine A. and {Fernandes}, Rachel B. and {Lund}, Michael B. and {Savel}, Arjun B. and {Gill}, Holden and {Beichman}, Charles and {Matson}, Rachel and {Matthews}, Elisabeth C. and {Furlan}, E. and {Howell}, Steve B. and {Scott}, Nicholas J. and {Everett}, Mark E. and {Livingston}, John H. and {Ershova}, Irina O. and {Cheryasov}, Dmitry V. and {Safonov}, Boris and {Lillo-Box}, Jorge and {Barrado}, David and {Morales-Calder{\'o}n}, Mar{\'\i}a},
        title = "{The TESS-Keck Survey. XX. 15 New TESS Planets and a Uniform RV Analysis of All Survey Targets}",
      journal = {\apjs},
         year = 2024,
        month = jun,
       volume = {272},
       number = {2},
          eid = {32},
        pages = {32},
          doi = {10.3847/1538-4365/ad4484},
archivePrefix = {arXiv},
       eprint = {2405.14786},
 primaryClass = {astro-ph.EP},
       adsurl = {https://ui.adsabs.harvard.edu/abs/2024ApJS..272...32P}
}

@ARTICLE{TOI4529,
       author = {{Poultourtzidis}, E. and {Lacedelli}, G. and {Pall{\'e}}, E. and {Carleo}, I. and {Magliano}, C. and {Gerald{\'\i}a-Gonz{\'a}lez}, S. and {Caballero}, J.~A. and {Morello}, G. and {Orell-Miquel}, J. and {Tabernero}, H.~M. and et al.},
        title = "{Characterization of two new transiting sub-Neptunes and a terrestrial planet around M-dwarf hosts}",
      journal = {\aap},
         year = 2026,
        month = mar,
       volume = {707},
          eid = {A106},
        pages = {A106},
          doi = {10.1051/0004-6361/202557155},
archivePrefix = {arXiv},
       eprint = {2601.07414},
 primaryClass = {astro-ph.EP},
       adsurl = {https://ui.adsabs.harvard.edu/abs/2026A&A...707A.106P}
}

@ARTICLE{2024psaridi815,
       author = {{Psaridi}, Angelica and {Osborn}, Hugh and {Bouchy}, Fran{\c{c}}ois and {Lendl}, Monika and {Parc}, L{\'e}na and {Billot}, Nicolas and {Broeg}, Christopher and {Sousa}, S{\'e}rgio G. and {Adibekyan}, Vardan and {Attia}, Mara and {Bonfanti}, Andrea and {Chakraborty}, Hritam and {Collins}, Karen A. and {Davoult}, Jeanne and {Delgado-Mena}, Elisa and {Grieves}, Nolan and {Guillot}, Tristan and {Heitzmann}, Alexis and {Helled}, Ravit and {Hellier}, Coel and {Jenkins}, Jon M. and {Knierim}, Henrik and {Krenn}, Andreas and {Lissauer}, Jack J. and {Luque}, Rafael and {Rapetti}, David and {Santos}, Nuno C. and {Su{\'a}rez}, Olga and {Venturini}, Julia and {Wilkin}, Francis P. and {Wilson}, Thomas G. and {Winn}, Joshua N. and {Ziegler}, Carl and {Zingales}, Tiziano and {Alibert}, Yann and {Brandeker}, Alexis and {Egger}, Jo Ann and {Gandolfi}, Davide and {Hooton}, Matthew J. and {Tuson}, Amy and {Ulmer-Moll}, Sol{\`e}ne and {Abe}, Lyu and {Allart}, Romain and {Alonso}, Roi and {Anderson}, David R. and {Escud{\'e}}, Guillem Anglada and {B{\'a}rczy}, Tamas and {Barrado}, David and {Barros}, Susana C.~C. and {Baumjohann}, Wolfgang and {Beck}, Mathias and {Beck}, Thomas and {Benz}, Willy and {Bonfils}, Xavier and {Borsato}, Luca and {Bourrier}, Vincent and {Ciardi}, David R. and {Cameron}, Andrew Collier and {Charnoz}, S{\'e}bastien and {Cointepas}, Marion and {Csizmadia}, Szil{\'a}rd and {Cubillos}, Patricio and {Lo Curto}, Gaspare and {Davies}, Melvyn B. and {Daylan}, Tansu and {Deleuil}, Magali and {Deline}, Adrien and {Delrez}, Laetitia and {Demangeon}, Olivier D.~S. and {Demory}, Brice-Olivier and {Dorn}, Caroline and {Dumusque}, Xavier and {Ehrenreich}, David and {Erikson}, Anders and {Lecavelier des Etangs}, Alain and {de Miguel}, Diana and {Fortier}, Andrea and {Fossati}, Luca and {Frensch}, Yolanda G.~C. and {Fridlund}, Malcolm and {Gillon}, Micha{\"e}l and {G{\"u}del}, Manuel and {G{\"u}nther}, Maximilian N. and {Hagelberg}, Janis and {Helling}, Christiane and {Hoyer}, Sergio and {Isaak}, Kate G. and {Kiss}, Laszlo L. and {Lam}, Kristine W.~F. and {Laskar}, Jacques and {Lavie}, Baptiste and {Lovis}, Christophe and {Magrin}, Demetrio and {Marafatto}, Luca and {Maxted}, Pierre and {McDermott}, Scott and {M{\'e}karnia}, Djamel and {Mordasini}, Christoph and {Murgas}, Felipe and {Nascimbeni}, Valerio and {Nielsen}, Louise D. and {Olofsson}, G{\"o}ran and {Ottensamer}, Roland and {Pagano}, Isabella and {Pall{\'e}}, Enric and {Peter}, Gisbert and {Piotto}, Giampaolo and {Pollacco}, Don and {Queloz}, Didier and {Ragazzoni}, Roberto and {Ramos}, Devin and {Rando}, Nicola and {Rauer}, Heike and {Reimers}, Christian and {Ribas}, Ignasi and {Seager}, Sara and {S{\'e}gransan}, Damien and {Scandariato}, Gaetano and {Simon}, Attila E. and {Smith}, Alexis M.~S. and {Stalport}, Manu and {Steller}, Manfred and {Szab{\'o}}, Gyula and {Thomas}, Nicolas and {Pritchard}, Tyler A. and {Udry}, St{\'e}phane and {Corral Van Damme}, Carlos and {Van Grootel}, Val{\'e}rie and {Villaver}, Eva and {Walter}, Ingo and {Walton}, Nicholas and {Watkins}, Cristilyn N. and {West}, Richard G.},
        title = "{Discovery of two warm mini-Neptunes with contrasting densities orbiting the young K3V star TOI-815}",
      journal = {\aap},
         year = 2024,
        month = may,
       volume = {685},
          eid = {A5},
        pages = {A5},
          doi = {10.1051/0004-6361/202348678},
archivePrefix = {arXiv},
       eprint = {2401.15709},
 primaryClass = {astro-ph.EP},
       adsurl = {https://ui.adsabs.harvard.edu/abs/2024A&A...685A...5P}
}

@software{2022Reback,
       author = {{Reback}, Jeff and {Jbrockmendel} and {McKinney}, Wes and {Van Den Bossche}, Joris and {Augspurger}, Tom and {Roeschke}, Matthew and {Hawkins}, Simon and {Cloud}, Phillip and {Gfyoung} and {Sinhrks} and {Hoefler}, Patrick and {Klein}, Adam and {Petersen}, Terji and {Tratner}, Jeff and {She}, Chang and {Ayd}, William and {Naveh}, Shahar and {Darbyshire}, JHM and {Garcia}, Marc and {Shadrach}, Richard and {Schendel}, Jeremy and {Hayden}, Andy and {Saxton}, Daniel and {Gorelli}, Marco Edward and {Li}, Fangchen and {Zeitlin}, Matthew and {Jancauskas}, Vytautas and {McMaster}, Ali and {W{\"o}rtwein}, Torsten and {Battiston}, Pietro},
        title = "{pandas-dev/pandas: Pandas 1.4.2}",
         year = 2022,
        month = apr,
          eid = {10.5281/zenodo.6408044},
          doi = {10.5281/zenodo.6408044},
      version = {v1.4.2},
    publisher = {Zenodo},
       adsurl = {https://ui.adsabs.harvard.edu/abs/2022zndo...6408044R}
}

@ARTICLE{NaCo_decom,
       author = {{de Regt}, S. and {Ginski}, C. and {Kenworthy}, M.~A. and {Caceres}, C. and {Garufi}, A. and {Gledhill}, T.~M. and {Hales}, A.~S. and {Huelamo}, N. and {K{\'o}sp{\'a}l}, {\'A}. and {Millar-Blanchaer}, M.~A. and et al.},
        title = "{Polarimetric differential imaging with VLT/NACO. A comprehensive PDI pipeline for NACO data (PIPPIN)}",
      journal = {\aap},
         year = 2024,
        month = apr,
       volume = {684},
          eid = {A73},
        pages = {A73},
          doi = {10.1051/0004-6361/202348736},
archivePrefix = {arXiv},
       eprint = {2404.02222},
 primaryClass = {astro-ph.EP},
       adsurl = {https://ui.adsabs.harvard.edu/abs/2024A&A...684A..73D}
}

@ARTICLE{2024rescigno2134,
       author = {{Rescigno}, F. and {H{\'e}brard}, G. and {Vanderburg}, A. and {Mann}, A.~W. and {Mortier}, A. and {Morrell}, S. and {Buchhave}, L.~A. and {Collins}, K.~A. and {Mann}, C.~R. and {Hellier}, C. and {Haywood}, R.~D. and {West}, R. and {Stalport}, M. and {Heidari}, N. and {Anderson}, D. and {Huang}, C.~X. and {L{\'o}pez-Morales}, M. and {Cort{\'e}s-Zuleta}, P. and {Lewis}, H.~M. and {Dumusque}, X. and {Boisse}, I. and {Rowden}, P. and {Collier Cameron}, A. and {Deleuil}, M. and {Vezie}, M. and {Pepe}, F.~A. and {Delfosse}, X. and {Charbonneau}, D. and {Rice}, K. and {Demangeon}, O. and {Quinn}, S.~N. and {Udry}, S. and {Forveille}, T. and {Winn}, J.~N. and {Sozzetti}, A. and {Hoyer}, S. and {Seager}, S. and {Wilson}, T.~G. and {Dalal}, S. and {Martioli}, E. and {Striegel}, S. and {Boschin}, W. and {Dragomir}, D. and {Mart{\'\i}nez Fiorenzano}, A.~F. and {Cosentino}, R. and {Ghedina}, A. and {Malavolta}, L. and {Affer}, L. and {Lakeland}, B.~S. and {Nicholson}, B.~A. and {Foschino}, S. and {W{\"u}nsche}, A. and {Barkaoui}, K. and {Srdoc}, G. and {Randolph}, J. and {Guillet}, B. and {Conti}, D.~M. and {Ghachoui}, M. and {Gillon}, M. and {Benkhaldoun}, Z. and {Pozuelos}, F.~J. and {Timmermans}, M. and {Girardin}, E. and {Matutano}, S. and {Bosch-Cabot}, P. and {Mu{\~n}oz}, J.~A. and {For{\'e}s-Toribio}, R.},
        title = "{A hot mini-Neptune and a temperate, highly eccentric sub-Saturn around the bright K-dwarf TOI-2134}",
      journal = {\mnras},
         year = 2024,
        month = jan,
       volume = {527},
       number = {3},
        pages = {5385-5407},
          doi = {10.1093/mnras/stad3255},
archivePrefix = {arXiv},
       eprint = {2310.13623},
 primaryClass = {astro-ph.EP},
       adsurl = {https://ui.adsabs.harvard.edu/abs/2024MNRAS.527.5385R}
}

@INPROCEEDINGS{NaCo2,
       author = {{Rousset}, Gerard and {Lacombe}, Francois and {Puget}, Pascal and {Hubin}, Norbert N. and {Gendron}, Eric and {Fusco}, Thierry and {Arsenault}, Robin and {Charton}, Julien and {Feautrier}, Philippe and {Gigan}, Pierre and et al.},
        title = "{NAOS, the first AO system of the VLT: on-sky performance}",
    booktitle = {Adaptive Optical System Technologies II},
         year = 2003,
       editor = {{Wizinowich}, Peter L. and {Bonaccini}, Domenico},
       series = {Society of Photo-Optical Instrumentation Engineers (SPIE) Conference Series},
       volume = {4839},
        month = feb,
        pages = {140-149},
          doi = {10.1117/12.459332},
       adsurl = {https://ui.adsabs.harvard.edu/abs/2003SPIE.4839..140R}
}

@INPROCEEDINGS{roboao_KP,
       author = {{Salama}, Ma{\"\i}ssa and {Baranec}, Christoph and {Jensen-Clem}, Rebecca and {Riddle}, Reed and {Duev}, Dmitry and {Kulkarni}, Shrinivas and {Law}, Nicholas M.},
        title = "{Robo-AO Kitt Peak: status of the system and deployment of a sub-electron readnoise IR camera to detect low-mass companions}",
    booktitle = {Adaptive Optics Systems V},
         year = 2016,
       editor = {{Marchetti}, Enrico and {Close}, Laird M. and {V{\'e}ran}, Jean-Pierre},
       series = {Society of Photo-Optical Instrumentation Engineers (SPIE) Conference Series},
       volume = {9909},
        month = jul,
          eid = {99091A},
        pages = {99091A},
          doi = {10.1117/12.2233741},
archivePrefix = {arXiv},
       eprint = {1608.02985},
 primaryClass = {astro-ph.IM},
       adsurl = {https://ui.adsabs.harvard.edu/abs/2016SPIE.9909E..1AS}
}

@ARTICLE{Savel22,
       author = {{Savel}, Arjun B. and {Hirsch}, Lea A. and {Gill}, Holden and {Dressing}, Courtney D. and {Ciardi}, David R.},
        title = "{SImMER: A Pipeline for Reducing and Analyzing Images of Stars}",
      journal = {\pasp},
         year = 2022,
        month = dec,
       volume = {134},
       number = {1042},
          eid = {124501},
        pages = {124501},
          doi = {10.1088/1538-3873/aca4f9},
archivePrefix = {arXiv},
       eprint = {2212.00641},
 primaryClass = {astro-ph.IM},
       adsurl = {https://ui.adsabs.harvard.edu/abs/2022PASP..134l4501S}
}

@ARTICLE{Schlieder2021,
       author = {{Schlieder}, Joshua E. and {Gonzales}, Erica J. and {Ciardi}, David R. and {Patel}, Rahul I. and {Crossfield}, Ian J.~M. and {Crepp}, Justin R. and {Dressing}, Courtney D. and {Barclay}, Thomas and {Howard}, Andrew W.},
        title = "{Follow-up and Validation of K2 and TESS Planetary Systems with Keck NIRC2 Adaptive Optics Imaging}",
      journal = {Frontiers in Astronomy and Space Sciences},
         year = 2021,
        month = jun,
       volume = {8},
          eid = {63},
        pages = {63},
          doi = {10.3389/fspas.2021.628396},
       adsurl = {https://ui.adsabs.harvard.edu/abs/2021FrASS...8...63S}
}

@ARTICLE{NESSI,
       author = {{Scott}, Nicholas J. and {Howell}, Steve B. and {Horch}, Elliott P. and {Everett}, Mark E.},
        title = "{The NN-explore Exoplanet Stellar Speckle Imager: Instrument Description and Preliminary Results}",
      journal = {\pasp},
         year = 2018,
        month = may,
       volume = {130},
       number = {987},
        pages = {054502},
          doi = {10.1088/1538-3873/aab484},
       adsurl = {https://ui.adsabs.harvard.edu/abs/2018PASP..130e4502S}
}

@ARTICLE{alopekezorro,
       author = {{Scott}, Nicholas J. and {Howell}, Steve B. and {Gnilka}, Crystal L. and {Stephens}, Andrew W. and {Salinas}, Ricardo and {Matson}, Rachel A. and {Furlan}, Elise and {Horch}, Elliott P. and {Everett}, Mark E. and {Ciardi}, David R. and et al.},
        title = "{Twin High-resolution, High-speed Imagers for the Gemini Telescopes: Instrument description and science verification results}",
      journal = {Frontiers in Astronomy and Space Sciences},
         year = 2021,
        month = sep,
       volume = {8},
          eid = {138},
        pages = {138},
          doi = {10.3389/fspas.2021.716560},
       adsurl = {https://ui.adsabs.harvard.edu/abs/2021FrASS...8..138S}
}

@ARTICLE{scott_6716,
       author = {{Scott}, Madison G. and {Dransfield}, Georgina and {Timmermans}, Mathilde and {Triaud}, Amaury H.~M.~J. and {Rackham}, Benjamin V. and {Barkaoui}, Khalid and {Burgasser}, Adam J. and {Collins}, Karen A. and {Gillon}, Micha{\"e}l and {Howell}, Steve B. and et al.},
        title = "{Two temperate Earth- and Neptune-sized planets orbiting fully convective M dwarfs}",
      journal = {\mnras},
         year = 2026,
        month = mar,
       volume = {547},
       number = {1},
          eid = {stag070},
        pages = {stag070},
          doi = {10.1093/mnras/stag070},
archivePrefix = {arXiv},
       eprint = {2601.05799},
 primaryClass = {astro-ph.EP},
       adsurl = {https://ui.adsabs.harvard.edu/abs/2026MNRAS.547ag070S}
}

\begin{appendix}
\nolinenumbers

\section{High-resolution imaging data}
\label{app:app_imaging}
We present a more detailed overview of the observing facilities whose data were used within this work. All the data are publicly available on ExoFOP website. Table \ref{info_vetted} summarises the information on the contrast curves used in our analysis.

\subsection*{A.1: GEMINI-N/\textquotesingle Alopeke, GEMINI-N/NIRI, GEMINI-S/Zorro} 
The Gemini Observatory comprises the Gemini South telescope on Cerro Pachón, Chile, and the Frederick C. Gillett Gemini North telescope on Mauna Kea, Hawai\textquotesingle i. The primary near-infrared imaging instrument at Gemini North is the Gemini Near-Infrared Imager (NIRI) \citep{GeminiNIRI}. The data we used were obtained with the Br$\gamma$ filter (centred at 2.1686\,$\mu$m). \textquotesingle Alopeke and Zorro \citep{alopekezorro} are two identical high-resolution speckle imaging instruments mounted on the 8.1-meter Gemini North and South telescopes, respectively. For this study, we used images acquired with the 562nm and 832nm filters \citep{speckle_Howell}.

\subsection*{A.2: Keck II/NIRC2}
The Keck adaptive optics (AO) facility is installed on the left Nasmyth platform of the Keck II Telescope at the W. M. Keck Observatory, at Mauna Kea, Hawai\textquotesingle i \citep{NIRC2_1996}. NIRC2 is a near-infrared imager designed to operate with the Keck II adaptive optics system \citep{NIRC2_2000,NIRC2_2004}. We used of data taken with the Br$\gamma$, $K$ (centred at 2.196 $\mu$m), $J_{\rm cont}$ (1.2132\,$\mu$m), $H_{\rm cont}$ (1.5804\,$\mu$m), and $K_{\rm cont}$ (2.2706\,$\mu$m) filters \citep{Schlieder2021}.
 
\subsection*{A.3: KPNO/Robo-AO}
The observations were carried out using the Robo-AO laser AO system \citep{roboao,roboao_KP}, mounted on the 2.1-m (KP-2.1m) telescope at Kitt Peak National Observatory (KPNO) in Tucson, Arizona. The images that we used were taken using $i'$ filter (0.75\,$\mu$m) \citep{Lamman2020}.

\subsection*{A.4: LDT/DSSI}
The Differential Speckle Survey Instrument \citep[DSSI;][]{Horch_DSSI,Horch_DSSI2} is mounted on the 4.3-m Lowell Discovery Telescope (LDT) at Lowell Observatory in Flagstaff, Arizona. The data used were obtained with the 692 nm and 880 nm filters \citep{Clark26}.

\subsection*{A.5: Hale/PHARO}
The 5-m Hale Telescope at Palomar Observatory on Palomar Mountain, California, is equipped with an AO system coupled with the near-infrared camera Palomar High Angular Resolution Observer \citep[PHARO;][]{PHARO}. The data used in this work were obtained with the Br$\gamma$ and $H_{\rm cont}$ filters.

\subsection*{A.6: SHANE/ShARCS}   
The 3-m Shane Telescope at Lick Observatory on Mount Hamilton, California, is equipped with the ShaneAO system, which since 2014 operates with the Shane Adaptive Red Camera and Spectrometer \citep[ShARCS;][]{2012SPIE.8447E..3GK, 2014SPIE.9148E..05G, 2014SPIE.9148E..3AM}. The data used in this work were acquired with the $J$ and $Ks$ filter (2.150\,$\mu$m in vacuum) \citep[Dressing et al., in prep]{Savel22}.

\subsection*{A.7: VLT/NaCo}
The Nasmyth Adaptive Optics System (NAOS) combined with the Near-Infrared Imager and Spectrograph (CONICA; \citealt{NaCo,NaCo2}, NaCo) served as the near-infrared imager and spectrometer at the 8-m Very Large Telescope (VLT) in Cerro Paranal, Chile. NaCo was decommissioned, with the final image obtained on 1 October 2019 \citep{NaCo_decom}. The data used in this work were acquired with the $Ks$ and the 2.08316 $\mu$m filters.

\subsection*{A.8: WIYN/NESSI}
The 3.5-m Wisconsin-Indiana-Yale-NOAO (WIYN) Telescope, located at Kitt Peak National Observatory (KPNO in Tucson, Arizona), is equipped with the NN-EXPLORE Exoplanet Stellar Speckle Imager (NESSI; \citealt{NESSI}). The data used in this work were acquired with the 562\,nm, 832\,nm filters, $r$ (620\,nm), and $z$ (943.3\,nm) filters \citep{Howell_nessi}.

\begin{table*}
\centering

\caption{Imaging observations.}
\label{info_vetted}
\newcolumntype{M}[1]{>{\centering\arraybackslash}m{#1}}

\begin{tabular}{ccccccccc}

\hline
\hline
TOI & Observatory & Telescope & Instr. & Filters & Pixel Scale ('') & FWHM ('') & PI & Date \\
\hline

133 & Gemini & Gemini-South & Zorro & 562, 832 & 0.010 & 0.020 & Howell & 02-07-23 \\

213 & Gemini & Gemini-South & Zorro & 562, 832 & 0.010 & 0.020 & Howell & 29-11-20 \\

\makecell[t]{233} &
\makecell[t]{Gemini \\ Palomar} &
\makecell[t]{Gemini-South \\ Hale} &
\makecell[t]{Zorro \\ PHARO} &
\makecell[t]{562, 832 \\ Br$\gamma$} &
\makecell[t]{0.010 \\ 0.025} &
\makecell[t]{0.020 \\ 0.009} &
\makecell[t]{Howell \\ Ciardi} &
\makecell[t]{12-09-19 \\ 14-07-19} \\
\hline

\end{tabular}
\\
\footnotesize\raggedright{{\textbf{Note.} The complete table is available in a machine-readable form.}}

\end{table*}

\section{Details of the analysed sample}
Table~\ref{tab:validated_new} lists the planetary and stellar parameters of the newly validated planets and those previously validated in the literature. Table~\ref{tab:prop_vetted} reports the planetary and stellar parameters along with the classification results for the full analysed sample. Table~\ref{info_vetted_derived} presents the properties of the candidates orbiting stars for which the stellar parameters were independently derived.

\renewcommand\arraystretch{1.2}
\begin{table*}
\centering
\caption{Validated planets from this work.}
\label{tab:validated_new}
\begin{tabular}{lcccccccc}

\hline
\hline

TOI & $R$ [$\mathrm{R_{\oplus}}$] & $P$ [d] & $M_\star$ [$M_{\rm \odot}$] & $V$ (mag) & $J$ (mag) & TSM & RUWE & $\mathrm{IPD_{FMP}}$ \\
\hline

\multicolumn{9}{c}{Newly validated in this work} \\ 
233 b$^a$	&	2.02	$\pm$	0.51	&11.670029  $\pm$	1.5$\times10^{-5}$	&	0.37	$\pm$	0.02	&	13.3	&	9.9	&	63.2	&	1.44	&	0	\\
475 b	&	2.34	$\pm$	0.93	&8.261766   $\pm$	2.3$\times10^{-5}$	&	0.62	$\pm$	0.08	&	12.2	&	9.7	&	46.8	&	0.96	&	0	\\
493 b	&	4.50	$\pm$	0.66	&5.9444410   $\pm$	6.3$\times10^{-6}$	&	0.65	$\pm$	0.08	&	12.6	&	10.4	&	56.6	&	1.11	&	0	\\
533 b	&	3.61	$\pm$	0.54	&19.571586  $\pm$	2.1$\times10^{-5}$	&	0.74	$\pm$	0.09	&	11.8	&	10.1	&	34.5	&	0.91	&	0	\\
1027 b	&	2.88	$\pm$	0.31	&3.2834565   $\pm$	4.6$\times10^{-6}$	&	0.67	$\pm$	0.09	&	11.3	&	9.2	&	96.7$^{*}$	&	1.23	&	1	\\
1027 c	&	2.77	$\pm$	0.27	&11.028750  $\pm$	3.5$\times10^{-5}$	&	0.67	$\pm$	0.09	&	11.3	&	9.2	&	61.4	&	1.23	&	1	\\
1027 d	&	2.52	$\pm$	0.30	&5.011329   $\pm$	1.2$\times10^{-5}$	&	0.67	$\pm$	0.09	&	11.3	&	9.2	&	69.7	&	1.23	&	1	\\
2540 c	&	2.29	$\pm$	0.25	&22.081997  $\pm$	8.2$\times10^{-5}$	&	0.63	$\pm$	0.08	&	9.9	    &	7.1	&	95.4$^{*}$	&	1.05	&	0 \\
4129 b	&	5.68	$\pm$	0.46	&3.2724973   $\pm$	6.5$\times10^{-6}$	&	0.67	$\pm$	0.08	&	12.9	&	10.8	&	87.0	&	1.01	&	0	\\
4642 b	&	1.17	$\pm$	1.04	&3.60224   $\pm$	3.1$\times10^{-4}$	&	0.21	$\pm$	0.02	&	14.8	&	10.3	&	12.4$^{*}$	&	1.35	&	0	\\
5112 b	&	2.18	$\pm$	0.13	&15.53281  $\pm$	1.1$\times10^{-4}$	&	0.59	$\pm$	0.02	&	11.5	&	9.0	&	47.1	&	0.93	&	0	\\
5172 b$^a$	&	3.19	$\pm$	0.31	&4.726929   $\pm$	1.5$\times10^{-5}$	&	0.72	$\pm$	0.08	&	11.6	&	9.7	&	73.5	&	1.54	&	0	\\
5554 b	&	2.68	$\pm$	0.64	&26.6943  $\pm$  3.0$\times10^{-3}$	    &	0.61	$\pm$	0.02	&	11.1	&	8.7	&	58.8	&	1.08	&	0	\\
5722 b	&	6.39	$\pm$	0.71	&61.3191  $\pm$	1.0$\times10^{-3}$	    &	0.80	$\pm$	0.09	&	13.7	&	12.1	&	21.0	&	1.04	&	1	\\
5735 b	&	0.84	$\pm$	0.15	&2.120816   $\pm$	1.2$\times10^{-5}$	&	0.18	$\pm$	0.02	&	15.0	&	10.6	&	19.2$^{*}$	&	1.14	&	0	\\
6276 d$^b$	&	1.03	$\pm$	0.42	&7.87087   $\pm$	7.7$\times10^{-4}$	&	0.74	$\pm$	0.09	&	7.8	    &	5.8	&	22.5$^{*}$	&	0.99	&	0	\\
6715 b	&	3.90	$\pm$	1.82	&2.86188   $\pm$	2.2$\times10^{-4}$	&	0.76	$\pm$	0.09	&	12.8	&	10.6	&	67.3	&	0.91	&	0 \\
7032 b	&	2.45	$\pm$	0.20	&17.745604  $\pm$	6.2$\times10^{-5}$	&	0.71	$\pm$	0.08	&	9.2	    &	7.2	&	107.2$^{*}$	&	0.95	&	0 \\
7062 b	&	3.19	$\pm$	0.23	&10.5138  $\pm$	1.3$\times10^{-3}$	    &	0.81	$\pm$	0.10	&	9.8	    &	8.1	&   116.3$^{*}$	&	0.91	&	0 \\
\hline
\multicolumn{9}{c}{Validated also in previous publications} \\
213 b	&	2.31	$\pm$	0.21	&23.519905  $\pm$	6.2$\times10^{-5}$	&	0.72	$\pm$	0.08	&	10.9	&	8.8	&	47.2	&	1.02	&	0 \\
406 b	&	1.96	$\pm$	0.45	&13.175729  $\pm$	3.1$\times10^{-5}$	&	0.38	$\pm$	0.02	&	13.8	&	9.7	&	57.6	&	1.26	&	0	\\
435 b	&	5.29	$\pm$	0.60	&3.352907   $\pm$	1.4$\times10^{-5}$	&	0.63	$\pm$	0.08	&	14.3	&	11.7	&	57.0	&	0.98	&	0	\\
783 b	&	2.38	$\pm$	1.39	&8.112515   $\pm$	3.1$\times10^{-5}$	&	0.71	$\pm$	0.08	&	11.2	&	9.0	&	64.1	&	0.93	&	0	\\
1301 b	&	2.59	$\pm$	0.30	&6.0964074   $\pm$	5.7$\times10^{-6}$	&	0.77	$\pm$	0.09	&	11.3	&	9.7	&	52.9	&	0.96	&	0	\\
1683 b	&	2.49	$\pm$	0.73	&3.0575250   $\pm$	3.8$\times10^{-6}$	&	0.69	$\pm$	0.09	&	11.0	&	8.8	&	93.7$^{*}$	&	1.12	&	0	\\
1806 b	&	2.84	$\pm$	0.83	&15.15121  $\pm$	1.4$\times10^{-4}$	&	0.39	$\pm$	0.02	&	15.2	&	10.9	&	48.1	&	1.05	&	0	\\
1806 c	&	2.77	$\pm$	1.20	&8.196076   $\pm$	2.9$\times10^{-5}$	&	0.39	$\pm$	0.02	&	15.2	&	10.9	&	57.0	&	1.05	&	0	\\
2407 b	&	3.79	$\pm$	0.53	&2.70292   $\pm$	1.0$\times10^{-4}$	&	0.52	$\pm$	0.02	&	14.7	&	11.1	&	81.4	&	1.09	&	0	\\
2453 b	&	3.00	$\pm$	1.15	&4.4414275   $\pm$	8.8$\times10^{-6}$	&	0.50	$\pm$	0.02	&	14.4	&	11.0	&	55.4	&	1.11	&	0	\\
2540 b	&	2.59	$\pm$	0.30	&12.718725  $\pm$	4.1$\times10^{-5}$	&	0.63	$\pm$	0.08	&	9.9	    &	7.1	&	134.6$^{*}$	&	1.05	&	0	\\
4324 b	&	1.13	$\pm$	1.27	&6.250947   $\pm$	1.2$\times10^{-5}$	&	0.39	$\pm$	0.02	&	12.0	&	8.3	&	12.0$^{*}$	&	1.06	&	0	\\
5799 b	&	1.99	$\pm$	0.18	&4.16455   $\pm$	7.5$\times10^{-4}$	&	0.33	$\pm$	0.02	&	13.3	&	9.7	&	110.5$^{*}$	&	1.18	&	0	\\
6662 b	&	0.94	$\pm$	0.48	&0.2939493   $\pm$	4.0$\times10^{-6}$	&	0.31	$\pm$	0.02	&	13.4	&	9.3	&	30.9$^{*}$	&	1.23	&	1	\\
7166 b	&	2.10	$\pm$	0.32	&12.9228  $\pm$	3.0$\times10^{-3}$	    &	0.19	$\pm$	0.02	&	15.8	&	11.4	&	71.6	&	1.01	&	0	\\
\hline

\end{tabular}
\\ \footnotesize\raggedright{{\textbf{Note.} $^a$ Retained in the validated sample despite elevated RUWE value, as only one binarity 
indicator exceeds the adopted threshold.\\
$^b$ Although the statistical validation criteria are met, \cite{2025Kunimoto} do not consider this target as confirmed due to its low transit S/N. \\ $^{*}$ Candidates exceeding the TSM threshold recommended by \citet{Kempton_2018} for atmospheric characterisation.}}
\end{table*}

\large
\renewcommand\arraystretch{1.2}
\longtab[2]{
\begin{longtable}{lccccccccc}
\caption{Properties and disposition of the vetted systems.} \label{tab:prop_vetted}\\
\hline
\hline
TOI & $R$ [$\mathrm{R_{\oplus}}$] & $P$ [d] & $M_\star$ [$M_{\rm \odot}$] & $V$ (mag) & $J$ (mag) & TSM & RUWE & $\mathrm{IPD_{FMP}}$ & Result \\
\hline
\endfirsthead

\caption{(continued.)} \\
TOI & $R$ [$\mathrm{R_{\oplus}}$] & $P$ [d] & $M_\star$ [$M_{\rm \odot}$] & $V$ (mag) & $J$ (mag) & TSM & RUWE & $\mathrm{IPD_{FMP}}$ & Result \\
\hline
\endhead

\hline
\endfoot

133.01	&	1.92	$\pm$	1.44	&8.199537   $\pm$	1.7$\times10^{-5}$	&	0.67	$\pm$	0.09	&	11.6	&	8.9	&	51.1	&	1.07	&	1	&	LP	\\
210.01	&	2.15	$\pm$	0.20	&9.0105414   $\pm$	9.4$\times10^{-6}$	&	0.31	$\pm$	0.02	&	15.4	&	10.9	&	55.6	&	1.22	&	0	&	FP	\\
233 c	&	1.71	$\pm$	1.28	&7.2011383   $\pm$	1.8$\times10^{-5}$	&	0.37	$\pm$	0.02	&	13.3	&   9.9 &	59.6	&	1.44	&	0	&	AR	\\
240.01	&	3.61	$\pm$	0.40	&19.472302  $\pm$	7.7$\times10^{-5}$	&	0.67	$\pm$	0.08	&	11.8	&	9.6	&	56.3	&	1.92	&	0	&   OT	\\
278.01	&	2.52	$\pm$	0.24	&0.2990390   $\pm$	6.0$\times10^{-6}$	&	0.28	$\pm$	0.02	&	16.5	&   11.2 &	231.5$^{*}$	&	1.28	&	1	&	(BEBx2P)	\\
330 b	&	9.01	$\pm$	0.87	&4.2794669   $\pm$	3.1$\times10^{-6}$	&	0.67	$\pm$	0.08	&	14.0	&	11.8	&    102.6$^{*}$	&	1.06	&	0	&	LP	\\
450.01	&	14.04	$\pm$	5.99	&10.71487  $\pm$	6.0$\times10^{-4}$	&	0.42	$\pm$	0.02	&	15.6	&   10.6 &	391.9	&	1.32	&	0	&	(SEBx2P)	\\
486 b	&	0.58	$\pm$	0.03	&1.7447002   $\pm$	2.1$\times10^{-6}$	&	0.42	$\pm$	0.02	&	11.6	&	7.9	&	29.4$^{*}$	&	1.38	&	4	&	OT	\\
521 b	&	1.97	$\pm$	1.12	&1.5428486   $\pm$	4.1$\times10^{-6}$	&	0.41	$\pm$	0.02	&	14.7	&	10.9	&	62.6	&	1.12	&	0	&	VP	\\
611.01	&	5.30	$\pm$	1.75	&3.147166   $\pm$	1.1$\times10^{-5}$	&	0.79	$\pm$	0.09	&	11.5	&	9.6	&	141.8$^{*}$	&	0.85	&	0	&	FP	\\
654 b	&	2.61	$\pm$	0.13	&1.5275639   $\pm$	3.1$\times10^{-6}$	&	0.42	$\pm$	0.02	&	14.5	&	10.7	&	93.4$^{*}$	&	1.16	&	0	&	VP	\\
663 b	&	2.43	$\pm$	0.40	&2.5989031   $\pm$	6.4$\times10^{-6}$	&	0.51	$\pm$	0.02	&	13.7	&	10.4	&	65.6	&	1.60	&	0	&	VP	\\
663 c	&	2.66	$\pm$	0.20	&4.695599   $\pm$	1.2$\times10^{-5}$	&	0.51	$\pm$	0.02	&	13.7	&	10.4	&	60.6	&	1.60	&	0	&	VP	\\
727.02	&	1.08	$\pm$	1.54	&0.749971   $\pm$	1.3$\times10^{-5}$	&	0.50	$\pm$	0.02	&	12.7	&	9.7	&	9.9	&	1.14	&	0	&	OT	\\
741 b	&	0.87	$\pm$	0.07	&7.576833   $\pm$	3.5$\times10^{-5}$	&	0.52	$\pm$	0.02	&	9.5    	&	6.4	&	21.4$^{*}$	&	1.03	&	0	&	OT	\\
756 b	&	3.18	$\pm$	0.21	&1.2392569   $\pm$	5.5$\times10^{-6}$	&	0.51	$\pm$	0.02	&	14.6	&	11.1	&	85.1	&	1.24	&	0	&	FP	\\
761.01	&	2.78	$\pm$	0.62	&10.555733  $\pm$	2.1$\times10^{-5}$	&	0.63	$\pm$	0.08	&	12.3	&	9.8	&	45.4	&	1.22	&	0	&	OT	\\
771 b	&	1.40	$\pm$	0.44	&2.3260077   $\pm$	4.9$\times10^{-6}$	&	0.21	$\pm$	0.02	&	14.9	&	10.5	&	15.5$^{*}$	&	1.28	&	0	&	FP	\\
782 b	&	2.59	$\pm$	0.92	&8.023981   $\pm$	1.7$\times10^{-5}$	&	0.40	$\pm$	0.02	&	14.8	&	10.7	&	57.2	&	1.25	&	1	&	VP	\\
815 b	&	2.62	$\pm$	0.16	&11.197277  $\pm$	2.6$\times10^{-5}$	&	0.82	$\pm$	0.09	&	10.2	&	8.5	&	71.8	&	1.01	&	0	&	VP	\\
864.01	&	1.00	$\pm$	0.78	&0.5206710   $\pm$	8.1$\times10^{-6}$	&	0.39	$\pm$	0.02	&	13.8	&	10.1	&	13.5$^{*}$	&	1.18	&	0	&	LP	\\
880.01	&	5.04	$\pm$	0.33	&6.3872524   $\pm$	8.0$\times10^{-6}$	&	0.81	$\pm$	0.10	&	10.1	&   8.5 &	169.6$^{*}$	&	1.00	&	0	&	AR	\\
880 c	&	2.78	$\pm$	0.23	&2.57359   $\pm$	5.5$\times10^{-4}$	&	0.81	$\pm$	0.10	&	10.1	&	8.5	&	118.2$^{*}$	&	1.00	&	0	&	OT	\\
900.01	&	2.88	$\pm$	0.69	&4.849524   $\pm$	1.5$\times10^{-5}$	&	0.67	$\pm$	0.08	&	13.7	&	10.8	&	37.8	&	0.98	&	0	&	FP	\\
912 b	&	2.25	$\pm$	0.09	&4.6780590   $\pm$	4.4$\times10^{-6}$	&	0.42	$\pm$	0.02	&	12.7	&	9.1	&    116.0$^{*}$	&	1.02	&	0	&	VP	\\
934 b	&	12.44	$\pm$	0.82	&3.78173833  $\pm$	7.6$\times10^{-7}$	&	0.79	$\pm$	0.10	&	13.6	&	11.8	&	124.5	&	1.15	&	0	&	LP	\\
1080 b	&	1.23	$\pm$	1.06	&3.965248   $\pm$	1.2$\times10^{-5}$	&	0.17	$\pm$	0.02	&	15.4	&	11.0	&	11.4$^{*}$	&	1.19	&	0	&	OT	\\
1082.01	&	3.51	$\pm$	0.38	&16.346905  $\pm$	5.0$\times10^{-5}$	&	0.64	$\pm$	0.08	&	13.2	&	10.7	&	37.0	&	1.23	&	0	&	LP	\\
1083.01	&	2.42	$\pm$	1.68	&12.980295  $\pm$	3.4$\times10^{-5}$	&	0.49	$\pm$	0.02	&	14.2	&	10.7	&	35.3	&	1.21	&	0	&	OT	\\
1105 b	&	2.91	$\pm$	0.25	&10.868052  $\pm$	3.8$\times10^{-5}$	&	0.74	$\pm$	0.09	&	11.3	&	9.4	&	52.7	&	0.99	&	0	&	OT	\\
1180 b	&	3.00	$\pm$	0.22	&9.686752   $\pm$	1.0$\times10^{-5}$	&	0.75	$\pm$	0.09	&	11.0	&	9.2	&	67.3	&	0.94	&	0	&	VP	\\
1184 b	&	2.35	$\pm$	0.19	&5.7484439   $\pm$	3.9$\times10^{-6}$	&	0.71	$\pm$	0.08	&	11.0	& 9.0 &	69.0	&	1.08	&	0	&	LP	\\
1224 b	&	2.10	$\pm$	0.30	&4.1782656   $\pm$	5.2$\times10^{-6}$	&	0.40	$\pm$	0.02	&	14.0	&	10.0	&	78.2	&	1.48	&	0	&	LP	\\
1274 b	&	9.39	$\pm$	0.69	&19.320348  $\pm$	1.3$\times10^{-5}$	&	0.75	$\pm$	0.09	&	12.8	&	11.0	&	82.2	&	0.99	&	0	&	LP	\\
1280 b	&	3.29	$\pm$	0.38	&9.6926392   $\pm$	7.8$\times10^{-6}$	&	0.74	$\pm$	0.09	&	11.6	&	9.7	&	59.2	&	0.72	&	0	&	VP$^{**}$	\\
1410 b	&	3.43	$\pm$	0.31	&1.216873   $\pm$	5.3$\times10^{-5}$	&	0.71	$\pm$	0.09	&	11.1	&	9.3	&	139.5$^{*}$	&	1.14	&	0	&	VP	\\
1412.01	&	6.97	$\pm$	1.38	&4.7319770   $\pm$	5.0$\times10^{-6}$	&	0.66	$\pm$	0.08	&	12.3	&   9.9 &	176.4$^{*}$	&	17.79	&	0	&	(SEB)	\\
1424.01	&	3.52	$\pm$	0.33	&4.896248   $\pm$	1.6$\times10^{-5}$	&	0.71	$\pm$	0.08	&	11.3	&	9.6	&	76.9	&	1.22	&	0	&	FP	\\
1448 b	&	2.64	$\pm$	0.84	&8.0971   $\pm$	2.7$\times10^{-3}$	    &	0.37	$\pm$	0.02	&	15.7	&	11.6	&	44.4	&	1.08	&	0	&	FP	\\
1450 b	&	1.09	$\pm$	0.37	&2.0439268   $\pm$	4.1$\times10^{-6}$	&	0.47	$\pm$	0.02	&	12.3	&	8.5	&	12.6$^{*}$	&	-	&	0	&	VP	\\
1467 b	&	1.87	$\pm$	1.22	&5.971136   $\pm$	1.4$\times10^{-5}$	&	0.49	$\pm$	0.02	&	12.3	&	9.4	&	64.4	&	1.47	&	0	&	VP	\\
1638.01	&	5.39	$\pm$	0.34	&0.9150623   $\pm$	9.0$\times10^{-6}$	&	0.66	$\pm$	0.20	&	13.9	&   10.9 & 	120.1$^{*}$	&	1.53	&	6	&	(TP/SEB)	\\
1730.01	&	2.76	$\pm$	0.96	&6.226232   $\pm$	2.8$\times10^{-5}$	&	0.53	$\pm$	0.02	&	12.2	&	9.1	&	105.4$^{*}$	&	1.03	&	0	&	OT	\\
1730.03	&	2.44	$\pm$	0.11	&12.5740  $\pm$	1.3$\times10^{-3}$	    &	0.53	$\pm$	0.02	&	12.2	&	9.1	&	69.9	&	1.03	&	0	&	OT	\\
1738.01	&	4.38	$\pm$	0.40	&3.698582   $\pm$	1.4$\times10^{-5}$	&	0.64	$\pm$	0.02	&	14.1	&	11.3	&	52.1	&	1.09	&	0	&	OT	\\
1802.01	&	2.51	$\pm$	1.02	&16.7961  $\pm$	5.8$\times10^{-3}$	    &	0.57	$\pm$	0.02	&	12.8	&	9.9	&	39.3	&	1.24	&	0	&	OT	\\
1803 b	&	4.22	$\pm$	0.45	&12.885863  $\pm$	6.3$\times10^{-5}$	&	0.79	$\pm$	0.10	&	11.9	&	10.3	&	57.7	&	0.96	&	0	&	LP	\\
1803 c	&	2.99	$\pm$	0.29	&6.293211   $\pm$	4.0$\times10^{-5}$	&	0.79	$\pm$	0.10	&	11.9	&	10.3	&	52.1	&	0.96	&	0	&	VP	\\
1823 b	&	8.21	$\pm$	0.56	&38.813550  $\pm$	3.5$\times10^{-5}$	&	0.77	$\pm$	0.09	&	10.7	&	8.9	&	139.9$^{*}$	&	0.95	&	1	&	LP	\\
1832.01	&	7.61	$\pm$	0.59	&4.1508569   $\pm$	4.1$\times10^{-6}$	&	0.74	$\pm$	0.08	&	13.9	&	12.0	&	67.5	&	0.99	&	0	&	OT	\\
1858.01	&	18.37	$\pm$	4.65	&2.0912089   $\pm$	2.5$\times10^{-6}$	&	0.66	$\pm$	0.08	&	14.0	&	11.8	&	179.8	&	1.04	&	0	&	LP	\\
2015 b	&	5.38	$\pm$	2.27	&3.349141   $\pm$	1.6$\times10^{-5}$	&	0.30	$\pm$	0.02	&	16.1	&	11.1	&	218.9$^{*}$	&	1.13	&	0	&	FP	\\
2079.02	&	2.06	$\pm$	1.61	&9.317411   $\pm$	2.3$\times10^{-5}$	&	0.48	$\pm$	0.02	&	12.8	&   9.5     &	59.0	&	1.39	&	0	&	AR	\\
2133 b	&	2.45	$\pm$	1.50	&3.826659   $\pm$	1.6$\times10^{-5}$	&	0.65	$\pm$	0.08	&	12.0	&	9.8	&	51.0	&	1.07	&	0	&	OT	\\
2134 b	&	3.22	$\pm$	0.27	&9.2292078   $\pm$	5.6$\times10^{-6}$	&	0.69	$\pm$	0.08	&	8.9	    &	6.8	&	204.9$^{*}$	&	1.02	&	0	&	VP	\\
2142.01	&	1.56	$\pm$	0.34	&9.300435   $\pm$	5.1$\times10^{-5}$	&	0.35	$\pm$	0.02	&	13.3	&	9.6	&	60.2	&	1.14	&	0	&	OT	\\
2274 b	&	1.83	$\pm$	1.87	&2.6796164   $\pm$	7.2$\times10^{-6}$	&	0.61	$\pm$	0.02	&	12.0	&	9.2	&	60.0	&	1.06	&	0	&	OT	\\
2290.01	&	1.20	$\pm$	2.04	&0.3862200   $\pm$	9.1$\times10^{-6}$	&	0.56	$\pm$	0.02	&	12.6	&	9.9	&	8.8	&	1.04	&	0	&	FP	\\
2405.01	&	4.19	$\pm$	0.56	&1.9792573   $\pm$	9.0$\times10^{-6}$	&	0.62	$\pm$	0.08	&	13.9	&	11.3	&	50.1	&	1.06	&	0	&	LP	\\
2408.01	&	5.46	$\pm$	0.45	&9.463615   $\pm$	4.1$\times10^{-5}$	&	0.81	$\pm$	0.10	&	12.8	&	11.0	&	56.7	&	1.01	&	0	&	LP	\\
2431 b	&	5.60	$\pm$	1.96	&0.22419500  $\pm$	4.0$\times10^{-7}$	&	0.64	$\pm$	0.08	&	10.9	&	8.4	&	707.2$^{*}$	&	1.07	&	0	&	VP	\\
2457.01	&	2.76	$\pm$	1.30	&14.1994  $\pm$	4.0$\times10^{-3}$	    &	0.63	$\pm$	0.08	&	12.2	&	9.7	&	46.7	&	1.55	&	0	&	LP	\\
2486.01	&	3.17	$\pm$	0.31	&1.5411764   $\pm$	3.5$\times10^{-6}$	&	0.66	$\pm$	0.08	&	12.3	&	10.1	&	76.3	&	0.98	&	0	&	OT	\\
2552.01	&	3.08	$\pm$	2.73	&2.370112   $\pm$	1.5$\times10^{-5}$	&	0.63	$\pm$	0.08	&	14.5	&	10.7	&	56.7	&	1.11	&	1	&	FP	\\
2626 b	&	10.56	$\pm$	0.89	&3.4562767   $\pm$	6.5$\times10^{-6}$	&	0.69	$\pm$	0.09	&	13.6	&	11.6	&	135.8$^{*}$	&	1.02	&	0	&	LP	\\
2952.01	&	6.26	$\pm$	0.24	&10.784493  $\pm$	1.7$\times10^{-5}$	&	0.51	$\pm$	0.02	&	14.8	&	11.2	&	85.0	&	1.14	&	0	&	LP	\\
3398.01	&	3.57	$\pm$	0.94	&6.2088   $\pm$	1.7$\times10^{-3}$	    &	0.74	$\pm$	0.09	&	13.3	&	11.2	&	45.6	&	1.11	&	0	&	LP	\\
3433.01	&	6.08	$\pm$	0.50	&2.7765784   $\pm$	9.4$\times10^{-6}$	&	0.68	$\pm$	0.08	&	13.8	&	11.7	&	63.6	&	3.55	&	16	&	FP	\\
3494 b	&	2.22	$\pm$	0.80	&7.748612   $\pm$	2.2$\times10^{-5}$	&	0.29	$\pm$	0.02	&	15.0	&	10.5	&	76.9	&	3.73	&	77	&	VP$^{**}$	\\
3498.01	&	4.72	$\pm$	0.44	&4.876878   $\pm$	5.4$\times10^{-5}$	&	0.72	$\pm$	0.09	&	13.0	&	11.0	&	56.3	&	0.91	&	0	&	FP	\\
3568 b	&	6.40	$\pm$	0.73	&4.41780   $\pm$	2.0$\times10^{-4}$	&	0.79	$\pm$	0.09	&	12.9	&	11.2	&	78.3	&	0.98	&	0	&	FP	\\
3585 b	&	4.64	$\pm$	0.35	&3.5410781   $\pm$	6.1$\times10^{-6}$	&	0.74	$\pm$	0.09	&	12.2	&	10.3	&	76.7	&	1.19	&	0	&	VP$^{**}$	\\
3586 b	&	12.76	$\pm$	2.80	&11.320123  $\pm$	1.7$\times10^{-5}$	&	0.82	$\pm$	0.10	&	13.1	&	11.5	&	133.9$^{*}$	&	1.97	&	0	&	LP	\\
3868.01	&	7.83	$\pm$	0.59	&18.565813  $\pm$	4.8$\times10^{-5}$	&	0.74	$\pm$	0.09	&	13.2	&	11.4	&	54.6	&	1.10	&	1	&	LP	\\
3913.01	&	7.98	$\pm$	0.78	&4.0384773   $\pm$	5.4$\times10^{-6}$	&	0.65	$\pm$	0.08	&	13.7	&	11.5	&	76.2	&	1.06	&	0	&	LP	\\
4090.01	&	5.10	$\pm$	0.40	&3.177514   $\pm$	1.8$\times10^{-5}$	&	0.76	$\pm$	0.09	&	13.4	&	11.6	&	49.1	&	1.02	&	0	&	OT	\\
4109.01	&	5.11	$\pm$	0.50	&12.320331  $\pm$	3.1$\times10^{-5}$	&	0.64	$\pm$	0.08	&	13.9	&	11.6	&	40.8	&	1.07	&	0	&	LP	\\
4114.01	&	4.10	$\pm$	0.47	&4.803854   $\pm$	9.0$\times10^{-5}$	&	0.64	$\pm$	0.08	&	13.5	&	11.1	&	46.5	&	1.10	&	0	&	FP	\\
4409 b	&	7.77	$\pm$	0.51	&92.49181  $\pm$	1.7$\times10^{-4}$	&	0.81	$\pm$	0.09	&	12.7	&	11.0	&	44.6	&	1.04	&	0	&	LP	\\
4491 b	&	2.22	$\pm$	0.25	&5.0066220   $\pm$	6.6$\times10^{-6}$	&	0.63	$\pm$	0.08	&	11.5	&	9.2	&	65.7	&	0.86	&	0	&	VP	\\
4527 b	&	0.91	$\pm$	0.95	&0.399444   $\pm$	1.3$\times10^{-5}$	&	0.48	$\pm$	0.02	&	10.9	&	7.8	&	31.9$^{*}$	&	0.98	&	0	&	FP	\\
4529 b	&	1.87	$\pm$	0.56	&5.879575   $\pm$	1.9$\times10^{-5}$	&	0.49	$\pm$	0.02	&	12.2	&	8.8	&	78.5	&	1.23	&	0	&	VP	\\
4552.01	&	1.28	$\pm$	0.24	&0.301151   $\pm$	2.9$\times10^{-5}$	&	0.26	$\pm$	0.02	&	14.4	&	10.3	&	21.8$^{*}$	&	1.05	&	0	&	FP	\\
4556.01	&	1.46	$\pm$	0.22	&3.792456   $\pm$	1.1$\times10^{-5}$	&	0.19	$\pm$	0.02	&	18.3	&	11.5	&	10.1$^{*}$	&	1.27	&	0	&	OT	\\
4576.01	&	3.51	$\pm$	0.38	&10.483406  $\pm$	6.6$\times10^{-5}$	&	0.67	$\pm$	0.08	&	11.5	&	9.1	&	87.6	&	1.18	&	0	&	FP	\\
4646.01	&	12.66	$\pm$	1.12	&8.92995   $\pm$	5.7$\times10^{-4}$	&	0.69	$\pm$	0.08	&	13.1	&	11.2	&	140.4	&	0.90	&	0	&	FP	\\
4766.01	&	10.51	$\pm$	0.82	&1.93266   $\pm$	1.1$\times10^{-4}$	&	0.80	$\pm$	0.10	&	13.4	&	11.8	&	128.8	&	0.99	&	0	&	OT	\\
5051.01	&	14.80	$\pm$	1.11	&3.598072   $\pm$	1.0$\times10^{-5}$	&	0.75	$\pm$	0.09	&	12.7	&   10.7    &	80.9	&	0.91	&	0	&	(EBx2P)	\\
5319.01	&	3.75	$\pm$	0.62	&4.078527   $\pm$	1.7$\times10^{-5}$	&	0.50	$\pm$	0.10	&	13.9	&   10.4 &	111.5$^{*}$	&	1.14	&	0	&	AR	\\
5358.01	&	3.41	$\pm$	0.35	&2.659913   $\pm$	1.1$\times10^{-5}$	&	0.74	$\pm$	0.08	&	12.4	&	10.6	&	56.8	&	1.06	&	0	&	OT	\\
5375 b	&	11.48	$\pm$	0.35	&1.7215575   $\pm$	1.4$\times10^{-6}$	&	0.61	$\pm$	0.02	&	14.1	&	11.2	&	269.6	&	1.06	&	0	&	LP	\\
5388 b	&	1.89	$\pm$	0.56	&2.5946750   $\pm$	9.2$\times10^{-6}$	&	0.29	$\pm$	0.02	&	12.7	&	9.2	&	186.7$^{*}$	&	1.16	&	0	&	LP	\\
5416.01	&	10.99	$\pm$	8.59	&13.0087  $\pm$	1.1$\times10^{-3}$	    &	0.52	$\pm$	0.02	&	14.6	&	11.7	&	139.2$^{*}$	&	1.00	&	0	&	FP	\\
5423.01	&	4.28	$\pm$	4.42	&5.7188   $\pm$	1.1$\times10^{-3}$	    &	0.82	$\pm$	0.09	&	13.4	&	11.8	&	35.2	&	0.99	&	0	&	OT	\\
5507.01	&	0.97	$\pm$	1.85	&1.506817   $\pm$	1.1$\times10^{-5}$	&	0.34	$\pm$	0.02	&	13.0	&   9.4     &	15.4$^{*}$	&	1.62	&	0	&	AR	\\
5572.01	&	2.36	$\pm$	0.22	&5.858682   $\pm$	3.0$\times10^{-5}$	&	0.72	$\pm$	0.08	&	11.1	&	9.5	&	51.6	&	1.03	&	0	&	OT	\\
5581.01	&	9.85	$\pm$	0.65	&33.538689  $\pm$	7.9$\times10^{-5}$	&	0.81	$\pm$	0.10	&	12.6	&	11.1	&	74.9	&	1.08	&	0	&	LP	\\
5584.01	&	3.01	$\pm$	0.31	&18.63008  $\pm$	1.9$\times10^{-4}$	&	0.64	$\pm$	0.08	&	11.8	&	9.6	&	38.9	&	1.08	&	0	&	OT	\\
5706.01	&	3.41	$\pm$	2.25	&0.357656   $\pm$	4.8$\times10^{-5}$	&	0.69	$\pm$	0.08	&	11.6	&	9.8	&	102.1$^{*}$	&	3.17	&	0	&	LP	\\
5720.01	&	1.39	$\pm$	0.28	&1.434460   $\pm$	1.5$\times10^{-5}$	&	0.36	$\pm$	0.02	&	13.9	&   10.1 &	10.6$^{*}$	&	1.19	&    0	&	AR	\\
5734 b	&	2.41	$\pm$	0.31	&6.1841688   $\pm$	9.5$\times10^{-6}$	&	0.74	$\pm$	0.09	&	9.6	    &	7.6	&	132.1$^{*}$	&	0.93	&	0	&	VP	\\
5747 b	&	1.50	$\pm$	0.13	&0.573477   $\pm$	1.4$\times10^{-5}$	&	0.41	$\pm$	0.02	&	13.0	&	9.3	&	18.8$^{*}$	&	1.33	&	1	&	OT	\\
5800 b	&	2.83	$\pm$	0.27	&2.62826   $\pm$	3.5$\times10^{-4}$	&	0.78	$\pm$	0.10	&	9.6	    &	7.8	&	162.8$^{*}$	&	1.09	&	0	&	OT	\\
5814.01	&	2.74	$\pm$	0.22	&6.254395   $\pm$	2.2$\times10^{-5}$	&	0.83	$\pm$	0.10	&	11.3	&	9.6	&	48.8	&	1.04	&	0	&	OT	\\
5955.01	&	1.68	$\pm$	0.18	&0.590239   $\pm$	8.4$\times10^{-5}$	&	0.58	$\pm$	0.02	&	12.0	&	9.1	&	99.6$^{*}$	&	1.40	&	0	&	OT	\\
5974.01	&	1.21	$\pm$	1.58	&1.7266943   $\pm$	7.1$\times10^{-6}$	&	0.36	$\pm$	0.02	&	13.0	&	9.4	&	12.0$^{*}$	&	1.25	&	0	&	FP	\\
5981.01	&	2.40	$\pm$	0.75	&7.014246   $\pm$	8.1$\times10^{-5}$	&	0.54	$\pm$	0.02	&	14.4	&	10.5	&	40.7	&	1.18	&	0	&	OT	\\
5998.01	&	3.09	$\pm$	1.57	&9.33803   $\pm$	1.2$\times10^{-4}$	&	0.64	$\pm$	0.08	&	12.8	&	10.4	&	38.3	&	1.10	&	0	&	OT	\\
6008 b	&	1.25	$\pm$	0.16	&0.857373   $\pm$	1.4$\times10^{-5}$	&	0.20	$\pm$	0.02	&	16.0	&	10.4	&	18.2$^{*}$	&	1.28	&	0	&	FP	\\
6077.01	&	2.99	$\pm$	1.12	&2.211110   $\pm$	1.2$\times10^{-5}$	&	0.63	$\pm$	0.02	&	14.1	&	11.1	&	45.6	&	3.28	&	0	&	OT	\\
6086 b	&	0.97	$\pm$	1.72	&1.388858   $\pm$	1.4$\times10^{-5}$	&	0.23	$\pm$	0.02	&	15.0	&	10.8	&	13.5$^{*}$	&	1.20	&	0	&	OT	\\
6110 b	&	8.67	$\pm$	0.34	&1.2541235   $\pm$	1.7$\times10^{-6}$	&	0.54	$\pm$	0.02	&	14.7	&	11.7	&	201.2$^{*}$	&	1.01	&	1	&	LP	\\
6245.01	&	0.90	$\pm$	1.30	&3.214994   $\pm$	1.0$\times10^{-5}$	&	0.21	$\pm$	0.02	&	15.1	&	10.3	&	14.4$^{*}$	&	1.24	&	0	&	OT	\\
6255.01	&	0.96	$\pm$	0.29	&0.2381823   $\pm$	8.0$\times10^{-6}$	&	0.36	$\pm$	0.02	&	12.8	& 8.9  &	33.1$^{*}$	&	1.39	&	0	&	AR	\\
6258.01	&	1.22	$\pm$	2.68	&0.557087   $\pm$	2.5$\times10^{-5}$	&	0.37	$\pm$	0.02	&	14.4	&	10.5	&	10.0$^{*}$	&	1.37	&	0	&	FP	\\
6276 b	&	1.02	$\pm$	0.57	&6.20316   $\pm$	4.5$\times10^{-4}$	&	0.74	$\pm$	0.09	&	7.8	    &	5.8	&	24.5$^{*}$	&	0.99	&	0	&	OT	\\
6276 c	&	0.98	$\pm$	0.41	&4.46607   $\pm$	3.2$\times10^{-4}$	&	0.74	$\pm$	0.09	&	7.8	    &	5.8	&	28.0$^{*}$	&	0.99	&	0	&	OT	\\
6284.01	&	0.83	$\pm$	0.05	&3.447594   $\pm$	1.3$\times10^{-5}$	&	0.47	$\pm$	0.02	&	11.8	&	8.3	&	13.4$^{*}$	&	1.19	&	1	&	OT	\\
6324 b	&	1.06	$\pm$	0.07	&0.27922110   $\pm$	2.0$\times10^{-7}$	&	0.27	$\pm$	0.02	&	13.4	&	9.4	&	33.6$^{*}$	&	1.09	&	0	&	VP	\\
6599.01	&	1.04	$\pm$	0.31	&2.20592   $\pm$	2.1$\times10^{-4}$	&	0.17	$\pm$	0.02	&	14.9	&	10.1	&	22.0$^{*}$	&	1.27	&	0	&	OT	\\
6678.01	&	2.13	$\pm$	0.21	&3.42684   $\pm$	7.6$\times10^{-4}$	&	0.77	$\pm$	0.09	&	9.8	    &	8.0	&	98.8$^{*}$	&	1.13	&	0	&	OT	\\
6688.01	&	4.63	$\pm$	0.96	&4.95950   $\pm$	3.7$\times10^{-4}$	&	0.80	$\pm$	0.09	&	11.9	&	9.8	&	80.4	&	1.69	&	29	&	OT	\\
6709.01	&	1.09	$\pm$	0.34	&2.347734   $\pm$	1.2$\times10^{-5}$	&	0.78	$\pm$	0.10	&	9.8	    &	7.9	&	11.2$^{*}$	&	0.96	&	0	&	OT	\\
6716 b	&	1.01	$\pm$	0.51	&4.718569   $\pm$	1.6$\times10^{-5}$	&	0.20	$\pm$	0.02	&	14.8	&	10.1	&	14.5$^{*}$	&	1.13	&	0	&	OT	\\
6806 b	&	15.75	$\pm$	1.10	&23.24908  $\pm$	2.0$\times10^{-4}$	&	0.79	$\pm$	0.10	&	12.0	&	10.4	&	127.5	&	1.16	&	5	&	LP	\\
6898.01	&	3.36	$\pm$	0.43	&18.1071 $\pm$	3.5$\times10^{-3}$	    &	0.63	$\pm$	0.08	&	12.8	&	9.4	&	56.9	&	1.05	&	0	&	OT	\\
7049.01	&	2.03	$\pm$	0.09	&40.38339  $\pm$	1.1$\times10^{-4}$	&	0.23	$\pm$	0.02	&	14.2	&	9.9	&	79.1	&	1.11	&	0	&	LP	\\
7054 b	&	7.35	$\pm$	0.64	&14.7596  $\pm$	2.0$\times10^{-3}$	    &	0.75	$\pm$	0.10	&	13.0	&	11.0	&	111.0$^{*}$	&	1.05	&	88	&	VP$^{**}$	\\
7170.01	&	3.57	$\pm$	0.71	&1.91710   $\pm$	1.2$\times10^{-4}$	&	0.52	$\pm$	0.02	&	14.8	&	11.5	&	71.1	&	1.06	&	0	&	OT	\\
7336.01	&	0.92	$\pm$	0.08	&0.5418556   $\pm$	1.1$\times10^{-6}$	&	0.82	$\pm$	0.09	&	9.5	    &	7.9	&	17.4$^{*}$	&	0.87	&	0	&	LP	\\
7464 b	&	11.14	$\pm$	0.36	&20.041705  $\pm$	2.9$\times10^{-5}$	&	0.52	$\pm$	0.02	&	15.2	&	11.7	&	115.1	&	1.09	&	0	&	LP	\\
\hline
\end{longtable}
\tablefoot{{VP = Validated Planet, LP = Likely Planet, FP = False Positive, OT = Other, AR = Ambiguous Result. The results in parentheses indicate the most probable scenario for the EB candidates. Planetary radius, period, stellar mass, Vmag, Jmag and TSM values are from ExoFOP website. RUWE and $\mathrm{IPD_{FMP}}$ values are from \citet{Gaia2023}. 
\tablefoottext{*}{Candidates exceeding the TSM threshold recommended by \citet{Kempton_2018} for atmospheric characterisation.}
\tablefoottext{**}{Candidates that satisfy the statistical validation criteria but fail the stellar binarity check.}}
}
}

\renewcommand\arraystretch{1.2}
\begin{table*}
\centering

\caption{Properties and disposition of the vetted systems with derived stellar parameters.}
\label{info_vetted_derived}

\begin{tabular}{ccccccccc}

\hline
\hline
TOI & $R$ [$\mathrm{R_{\oplus}}$] & $P$ [d] & $V$ (mag) & $J$ (mag) & RUWE & $\mathrm{IPD_{FMP}}$ & Result \\
\hline
579.01  &  14.07   $\pm$   0.48 & 1.5498373  $\pm$	2.6$\times10^{-6}$ & 10.4 & 8.8 & 6.13 & 45 & LP \\
885.01  &  17.61   $\pm$   1.73 & 2.56876   $\pm$   0.00214 & 11.5 & 9.1 & 0.96 & 0 & FP \\
1215.01 &  0.84    $\pm$   0.05 & 1.2124460   $\pm$   1.5$\times10^{-6}$ & 12.3 & 8.4 & 1.74 & 8 & FP \\
1490.01 &   11.98 & 3.9943   $\pm$   3.4$\times10^{-3}$ & 9.3 & 7.5 & 0.95 & 0 & OT \\
1976.01 & - & 7.6943096   $\pm$   8.0$\times10^{-6}$  & 9.8 & 7.9 & 5.23 & 85 & OT \\
3583.01 & - & 3.12134    $\pm$    1.9$\times10^{-4}$  & 13.6 & 10.5 & 9.39 & 54 & FP \\
3606.01 & - & 9.312167   $\pm$   3.2$\times10^{-5}$ & 14.1 & 11.3 & 9.98 & 0 & FP \\
4276.01 & - & 11.85116    $\pm$    1.7$\times10^{-4}$ & 13.0 & 11.1 & 18.30 & 5 & FP \\
4625.01 & 2.30   $\pm$   1.13 & 2.1913420    $\pm$    6.7$\times10^{-6}$ & 11.6 & 9.5 & 2.73 & 66 & LP \\
4801.01 & - & 1.470060   $\pm$    1.0$\times10^{-5}$   & 13.9 & 12.0 & 6.61 & 48 & FP \\
4839.01 & - & 13.34419   $\pm$   2.4$\times10^{-4}$ & 13.4 & 11.3 & 26.43 & 73 & OT \\
4893.01 & - & 7.326261   $\pm$ 2.7$\times10^{-5}$ & 13.4 & 11.6 & 31.25 & 22 & (EBx2P) \\
5809.01 & 2.63   $\pm$   0.18 & 9.210319   $\pm$   1.9$\times10^{-5}$ & 10.8 & 8.1 & 1.12 & 0 & LP \\
6645.01 & - & 3.952573   $\pm$   3.0$\times10^{-5}$ & 11.5 & 9.3 & 9.15 & 36 & FP \\
6827.01 & 11.21 & 4.473844   $\pm$   1.9$\times10^{-5}$ & 13.0 & 11.3 & 0.88 & 12 & LP \\
\hline
\end{tabular}
\\
\footnotesize\raggedright{\textbf{Note.} VP = Validated Planet, LP = Likely Planet, FP = False Positive, OT = Other, AR = Ambiguous Result. The results in parentheses indicate the most probable scenario for the EB candidates. Planetary radius, period, Vmag, and Jmag values are from ExoFOP website. RUWE and $\mathrm{IPD_{FMP}}$ values are from \citet{Gaia2023}.}
\end{table*}

\end{appendix}

\end{document}